\documentclass[twocolumn,a4paper,reprint,amssymb,aps,prd,groupedaddress,superscriptaddress,nofootinbib]{revtex4-2}
\pdfoutput=1
\usepackage{booktabs}
\usepackage{multirow}
\usepackage{physics}
\usepackage{dcolumn}
\usepackage{bm}
\usepackage{amsmath}
\usepackage{mathtools}
\usepackage{amsfonts}
\usepackage{pbox}
\usepackage{color}
\usepackage[dvipsnames]{xcolor}
\usepackage{tabularx}
\usepackage{subfigure}
\usepackage{graphicx}
\usepackage{psfrag}
\usepackage{lipsum}
\usepackage{enumitem}
\usepackage{ragged2e}
\usepackage{changes}
\usepackage{array}
\usepackage{CJK}
\usepackage[english]{babel}

\usepackage[colorlinks = true,
            linkcolor = blue,
            urlcolor  = blue,
            citecolor = blue,
            anchorcolor = blue]{hyperref}
\usepackage{blindtext}
\usepackage[utf8]{inputenc}
\usepackage{graphicx}
\usepackage{placeins}
\usepackage{siunitx}
\usepackage{amssymb}
\usepackage{mathtools}
\usepackage{physics}
\usepackage{tikz}
\usepackage[titletoc]{appendix}
\usepackage{listings}
\usepackage{float}
\usepackage{braket}

\usepackage[capitalize]{cleveref}
\usepackage{soul}
\usepackage{array}
\begin{document}

\title{Relaxing cosmological tensions with a sign switching cosmological constant:\\
Improved results with Planck, BAO, and Pantheon data}

\author{\"{O}zg\"{u}r Akarsu}
\email{akarsuo@itu.edu.tr}
\affiliation{Department of Physics, Istanbul Technical University, Maslak 34469 Istanbul, Turkey}

\author{Suresh Kumar}
\email{suresh.math@igu.ac.in}
\affiliation{Department of Mathematics, Indira Gandhi University, Meerpur, Haryana-122502, India}

\author{Emre \"{O}z\"{u}lker}
\email{ozulker17@itu.edu.tr}
\affiliation{Department of Physics, Istanbul Technical University, Maslak 34469 Istanbul, Turkey}

\author{J. Alberto Vazquez}
\email{javazquez@icf.unam.mx}
\affiliation{Instituto de Ciencias F\'isicas, Universidad Nacional Aut\'onoma de M\'exico, Cuernavaca, Morelos 62210, Mexico}

\author{Anita Yadav}
\email{anita.math.rs@igu.ac.in }
\affiliation{Department of Mathematics, Indira Gandhi University, Meerpur, Haryana-122502, India}

\begin{abstract}
We present a further observational analysis of the $\Lambda_{\rm s}$CDM model proposed in Akarsu \textit{et al.} [\href{https://doi.org/10.1103/PhysRevD.104.123512}{Phys. Rev. D 104, 123512 (2021)}]. This model is based on the recent conjecture suggesting the Universe has transitioned from anti-de Sitter vacua to de Sitter vacua (viz., the cosmological constant switches sign from negative to positive), at redshift ${z_\dagger\sim2}$, inspired by the graduated dark energy model proposed in Akarsu \textit{et al.} [\href{https://doi.org/10.1103/PhysRevD.101.063528}{Phys. Rev. D 101, 063528 (2020)}]. $\Lambda_{\rm s}$CDM was previously claimed to simultaneously relax five cosmological discrepancies, namely, the $H_0$, $S_8$, and $M_B$ tensions along with the Ly-$\alpha$ and $\omega_{\rm b}$ anomalies, which prevail within the standard $\Lambda$CDM model as well as its canonical/simple extensions. In the present work, we extend the previous analysis by constraining the model using the Pantheon data (with and without the SH0ES $M_B$ prior) and/or the \textit{completed} BAO data along with the full \textit{Planck} CMB data. We find that $\Lambda_{\rm s}$CDM exhibits a better fit to the data compared to $\Lambda$CDM, and simultaneously relaxes the six discrepancies of $\Lambda$CDM, viz., the $H_0$, $M_B$, $S_8$, Ly-$\alpha$, $t_0$, and $\omega_{\rm b}$ discrepancies, all of which are discussed in detail. When the $M_B$ prior is included in the analyses, $\Lambda_{\rm s}$CDM performs significantly better in relaxing the $H_0$, $M_B$, and $S_8$ tensions with the constraint ${z_\dagger\sim1.8}$ even when the Ly-$\alpha$ data (which imposed the $z_\dagger\sim2$ constraint in the previous studies) are excluded. In contrast, the presence of the $M_B$ prior causes only negligible improvements for $\Lambda$CDM. Thus, the $\Lambda_{\rm s}$CDM model provides remedy to various cosmological tensions simultaneously, only that the galaxy BAO data hinder its success to some extent.
\end{abstract}

\maketitle

\section{Introduction}
In the past decade, there has been a growing consensus that today's standard model of cosmology, namely, the Lambda Cold Dark Matter ($\Lambda$CDM) model, is actually an approximation to a more realistic new cosmological model which is yet to be understood. This new model, which may be conceptually very different, is expected to show slight but probably nontrivial phenomenological deviations from $\Lambda$CDM, because, despite being in very good agreement with a wide range of astrophysical and cosmological data~\cite{Riess:1998cb,SupernovaCosmologyProject:1998vns,Planck:2018vyg,Alam:2020sor,DES:2021wwk,DES:2022ygi,ACT:2020gnv,SPT-3G:2021wgf}, $\Lambda$CDM leads to discordances between various cosmological probes increased in diversity and precision over the past decade, e.g., the $H_0$ and $S_8$ tensions, and other statistically less significant anomalies~\cite{DiValentino:2020vhf,DiValentino:2020zio,DiValentino:2020vvd,DiValentino:2020srs,DiValentino:2021izs,Perivolaropoulos:2021jda,Abdalla:2022yfr,DiValentino:2022oon,Annis:2022xgg}. While these discordances can still be in part the result of systematic errors, the fact that they survived (and in some cases are even exacerbated) after several years of accurate analyses, points to cracks in $\Lambda$CDM, and suggests searching for new physics beyond the well-established fundamental theories that underpin, and even extend, the $\Lambda$CDM model. In particular, the $H_{0}$ (Hubble constant) tension exceeds $5\sigma$ with the recent SH0ES measurement~\cite{Riess:2021jrx} which led it to be called a crisis by many. Moreover, these tensions have turned out to be more challenging than originally thought. For instance, the $H_0$ tension worsens when the cosmological constant ($\Lambda$) is replaced by generic quintessence models of dark energy (DE), and is only partially relaxed when replaced by the simplest phantom (or quintom) models, and troublingly, many of the compelling models that suggest an amelioration in the $H_0$ tension---such as early dark energy (EDE)~\cite{Poulin:2018cxd,Smith:2019ihp,Herold:2022iib,Kamionkowski:2022pkx}, new-EDE~\cite{Niedermann:2019olb,Niedermann:2020dwg}, and nonminimally interacting dark energy (IDE)~\cite{Suresh:2019nrs,Eleonora:2020ao,Suresh:2020h0t,Bernui:2023byc}---result in worsening of others, e.g., the $S_8$ (weighted amplitude of matter fluctuations) tension, and they can even exacerbate less important anomalies to significant levels~\cite{Vagnozzi:2018jhn,DiValentino:2019exe,DiValentino:2019dzu,Vazquez:2020ani,Banerjee:2020xcn,Heisenberg:2022gqk,Lee:2022cyh,Goldstein:2023gnw,Poulin:2023lkg} (see also Refs.~\cite{DiValentino:2020vhf,DiValentino:2020zio,DiValentino:2020vvd,DiValentino:2020srs,DiValentino:2021izs,Perivolaropoulos:2021jda,Abdalla:2022yfr,DiValentino:2022oon,Annis:2022xgg}).
One may see Refs.~\cite{FrancoAbellan:2020xnr,Davari:2019tni,SolaPeracaula:2020vpg,SolaPeracaula:2019zsl,Camera:2017tws,DiValentino:2018gcu,Murgia:2020ryi,Archidiacono:2019wdp,Chudaykin:2022rnl,Lucca:2021dxo,SolaPeracaula:2021gxi,Khosravi:2021csn,Benevento:2022cql,Heimersheim:2020aoc} suggesting solutions to $S_8$ tension, some of which suggest relaxing the $H_0$ tension as well. We refer the reader to Refs.~\cite{DiValentino:2020vhf,DiValentino:2020zio,DiValentino:2020vvd,DiValentino:2020srs,DiValentino:2021izs,Perivolaropoulos:2021jda,Abdalla:2022yfr,DiValentino:2022oon,Annis:2022xgg} for a comprehensive list of references and recent reviews on the cosmological tensions, including discussions on the cosmological model-independent estimations of parameters such as $H_0$ and $S_8$, and a summary of proposed solutions.

It has been reported that the $H_{0}$ tension---as well as a number of other discrepancies---could be alleviated by a dynamical DE (as an effective or actual source) that achieves negative (could be persistent or temporary) or rapidly vanishing energy density values in the near or far past; and, this has recently increased interest in the phenomenological and theoretical realization/investigation of such models, see Refs.~\cite{Delubac:2014aqe,Aubourg:2014yra,Sahni:2014ooa,DiValentino:2017rcr,Mortsell:2018mfj,Poulin:2018zxs,Capozziello:2018jya,Wang:2018fng,Dutta:2018vmq,Banihashemi:2018oxo,Banihashemi:2018has,Farhang:2020sij,Banihashemi:2020wtb,Visinelli:2019qqu,Akarsu:2019hmw,Ye:2020btb,Ye:2020oix,Perez:2020cwa,Calderon:2020hoc,Paliathanasis:2020sfe,Bonilla:2020wbn,Vazquez:2012ag,Akarsu:2020yqa,LinaresCedeno:2021aqk,Zhou:2021xov,LinaresCedeno:2020uxx,Akarsu:2021fol,Sen:2021wld,Escamilla:2021uoj,Moshafi:2022mva,Wang:2022nap,Bag:2021cqm,Akarsu:2019ygx,Saharian:2022mft,Chavda:2020tfh,Acquaviva:2021jov,Akarsu:2020vii,Ozulker:2022slu,Acquaviva:2022bju,DiGennaro:2022ykp,Akarsu:2022lhx,Wang:2022jpo,Jiang:2022uyg,Li:2020ybr,Yang:2021eud,Colgain:2022tql,Xue:2022sfv,Cruz:2021knz,Krishnan:2020vaf,DiValentino:2020vnx,Vagnozzi:2019ezj,Ong:2022wrs,Malekjani:2023dky,Omni,Akarsu:2019pvi}. In fact, the simplest example of this type of scenario is the spatially closed $\Lambda$CDM model; positive spatial curvature (analogous to cosmic strings with negative energy density\footnote{They are analogous in the sense that both contribute to the Friedmann equation as a negative energy source with an equation of state parameter equal to $-1/3$. However, the presence of spatial curvature also has the effect of modifying the interrelations of cosmological distance measures (e.g., the comoving angular diameter distance is no longer proportional to the line-of-sight comoving distance for nonzero spatial curvature), rendering these two scenarios quite different. A similar distinction also arises when considering the combination of $\Lambda$ and spatial curvature as a single effective source in the Friedmann equation.}) and positive cosmological constant together can be interpreted as a single effective source that attains negative energy densities in the past and this scenario is significantly preferred over spatially flat $\Lambda$CDM by CMB data alone~\cite{Planck:2018vyg}---this preference that can be referred as the curvature, $\Omega_k$, anomaly~\cite{Handley:2019tkm,DiValentino:2019qzk,DiValentino:2020hov,Semenaite:2022unt,Yang:2022kho,Vagnozzi:2020zrh,Akarsu:2021max,Colgain:2022rxy,Dhawan:2021mel,Gonzalez:2021ojp}, is closely related to the lensing amplitude, $A_{\rm L}$, anomaly~\cite{Calabrese:2008rt,DiValentino:2019dzu,DiValentino:2020hov} since these two parameters are degenerate. However, the fact that this scenario (also its canonical/simple extensions) worsens the $H_0$ and $S_8$ tensions and is no more preferred when the CMB data is combined with other astrophysical data~\cite{Planck:2018vyg,Handley:2019tkm,DiValentino:2019qzk,DiValentino:2020hov,Semenaite:2022unt,Calabrese:2008rt,DiValentino:2019dzu,Yang:2022kho,Vagnozzi:2020zrh,Akarsu:2021max,Colgain:2022rxy,Acquaviva:2021jov,Dhawan:2021mel,Gonzalez:2021ojp}, may be signaling the need for a source of negative energy density that contributes more unexpectedly to the evolution of the Universe. In particular, it was recently conjectured in Ref.~\cite{Akarsu:2019hmw} that the Universe underwent a rapid anti-de Sitter (AdS) to de Sitter (dS) vacua transition at redshift ${z\sim2}$. This conjecture was based on the fact that observational analyses of the graduated dark energy (gDE) favored its sign-switching cosmological constantlike ($\Lambda_{\rm s}$-like) behavior, and this behavior simultaneously ameliorated the $H_0$ and Ly-$\alpha$ (Lyman-$\alpha$) discrepancies; the conjecture was further motivated by some theoretical advantages of $\Lambda_{\rm s}$ over gDE's $\Lambda_{\rm s}$-like behavior. In a later paper~\cite{Akarsu:2021fol}, the $\Lambda_{\rm s}$CDM model (which simply replaces the usual positive cosmological constant of $\Lambda$CDM with $\Lambda_{\rm s}$)
was studied in detail in the context of cosmological tensions. In particular, it was explained how this model can simultaneously address the $H_0$, $M_B$~[Type Ia supernovae (SNIa) absolute magnitude, closely related to the $H_0$ measurements], and Ly-$\alpha$ discrepancies, and, its observational analyses using the full Planck cosmic microwave background (CMB) and baryon acoustic oscillation (BAO) data were carried out. It was found that $\Lambda_{\rm s}$CDM is able to ameliorate the $H_0$, $M_B$, and $S_8$ tensions along with the Ly-$\alpha$ and $\omega_{\rm b}$ (physical baryon density) anomalies.

In this paper, we expand the investigations in Ref.~\cite{Akarsu:2021fol}, extend the observational analyses, by using the \textit{completed} BAO data and the Pantheon SNIa sample (with and without an $M_B$ prior) along with the full Planck CMB data, extend the previous discussions on the $H_0$, $M_B$, $S_8$, Ly-$\alpha$, and $\omega_{\rm b}$ discrepancies within $\Lambda_{\rm s}$CDM, and further add the $t_0$ (present-day age of the Universe) discrepancy and a theoretical explanation of how the $S_8$ tension can be alleviated in this model. In~\cref{sec:ss}, we briefly present the $\Lambda_{\rm s}$CDM model and motivate it starting from the gDE and by discussing its behavior with respect to tensions of $\Lambda$CDM. 
In~\cref{sec:method}, we first present the methodology and data sets used in the observational analyses and then discuss the results. In~\cref{sec:tensions}, we briefly explain six discrepancies of $\Lambda$CDM, viz., the $H_0$, $M_B$, $S_8$, Ly-$\alpha$, $t_0$, and $\omega_{\rm b}$, and assess their situation within $\Lambda_{\rm s}$CDM for our data sets, and we conclude in~\cref{sec:conc}.

\section{The $\Lambda_{\rm s}$CDM model}
\label{sec:ss}
The standard $\Lambda$CDM model relies on the presence of a constant energy density term, $\Lambda$---such as the usual vacuum energy of QFT and/or an effective energy density of a geometrical cosmological constant---to drive the present-day acceleration of the Universe; this constant energy density corresponds to zero inertial mass density $\varrho=0$, where $\varrho\equiv\rho+p$ with $\rho$ and $p$ being energy density and pressure, respectively. A minimal dynamical deviation from the zero inertial mass density assumption in the form of $\varrho\propto \rho^{\lambda}$, called graduated dark energy, was first investigated in Ref.~\cite{Akarsu:2019hmw}. Having almost constant negative energy density values at large redshifts, gDE settles into a positive value in the late Universe after a continuous transition whose rapidity is controlled by the parameter $\lambda$. During the transition, its energy density vanishes at a redshift, $z_\dagger$, and exhibits a pole in its equation of state (EoS) parameter that is characteristic of the DE models with sign-changing density~\cite{Ozulker:2022slu}. The parameter space of the gDE was well-constrained in its observational analysis (see Ref.~\cite{Akarsu:2019hmw}) with a preference of $z_\dagger\approx2.3$ and large negative values of $\lambda$, in which case the gDE resembles (becomes exact for $\lambda\to-\infty$) a negative cosmological constant, $\Lambda_-<0$, that instantaneously switches sign at $z\approx2.3$ and attains its present-day positive value $\Lambda_+=\abs{\Lambda_-}$. Compared to the usual cosmological constant, the gDE shows better agreement with multitude of data. In particular, when analysed with a combined data set from CMB, BAO, SNIa, and cosmic chronometers (CCs),  the gDE model had a significantly better fit with a nonmonotonic behavior of $H(z)$ around $z_\dagger\approx2.3$ that allowed the model to bring Ly-$\alpha$ BAO (BOSS DR11) data~\cite{Delubac:2014aqe} in concordance with the rest of the observations. Moreover, it yielded a value of $H_0=69.7\pm0.9{\rm \,km\, s^{-1}\, Mpc^{-1}}$ which is in perfect agreement with the local $H_{0}=69.8\pm0.8$ km $\rm{s^{-1}}$ $\rm{Mpc^{-1}}$ measurement from the tip of the red giant branch (TRGB)~\cite{Freedman:2019cchp}. In the gDE framework, these two simultaneous improvements in $H_0$ and Ly-$\alpha$ are interrelated in the following sense. A $z_\dagger$ value smaller than the effective redshift of the Ly-$\alpha$ data leads the model to have negative DE density at that effective redshift and beyond (towards early universe). Such a negative DE is in line with the lesser $H(z)$ value of the Ly-$\alpha$ data (less than the prediction of $\Lambda$CDM when constrained by the CMB). And since the comoving angular diameter distance to last scattering, $D_M(z_*)$, which is directly related to the integral of $H^{-1}(z)$, is strictly constrained by observations almost model independently, the lesser value of $H(z)$ at the effective redshift of the Ly-$\alpha$ data should be compensated by a higher $H(z)$ value somewhere else, which, for the gDE, results in a higher $H_0$ value~\cite{Akarsu:2019hmw} (see also Refs.~\cite{Akarsu:2021fol,Akarsu:2022lhx}, for a detailed discussion).

Inspired by the observational findings, and the fact that a sign-switching cosmological constant corresponding to the ${\lambda\to-\infty}$ limit of the gDE, unlike gDE with a finite $\lambda$, evades violating the weak energy condition and bounds on the speed of sound, the authors conjectured in Ref.~\cite{Akarsu:2019hmw} that the cosmological constant has spontaneously switched sign, i.e., the Universe has transitioned from AdS vacuum with $\Lambda_-$ to dS vacuum with $\Lambda_+$. The simplest sign-switching cosmological constant model, $\Lambda_{\rm s}$CDM, can be phenomenologically constructed by promoting the usual cosmological constant ($\Lambda$) of the standard $\Lambda$CDM model to an abruptly sign-switching (switches at a redshift $z_\dagger$ which is the only extra free parameter on top of the standard $\Lambda$CDM) cosmological constant ($\Lambda_{\rm s}$) with a present-day value of $\Lambda_{\rm s0}>0$;
\begin{equation}
    \Lambda\quad\rightarrow\quad\Lambda_{\rm s}\equiv\Lambda_{\rm s0}\,{\rm sgn}[z_\dagger-z],
    \label{eq:ssdef}
\end{equation}
where, the sign-switch feature is realized by the signum function, ``sgn," that reads ${\rm sgn}[x]=-1,0,1$ for $x < 0$, $x = 0$ and $x > 0$, respectively~\cite{Akarsu:2021fol}. Before moving on to the cosmological implications of the $\Lambda_{\rm s}$CDM model in the light of observational data, it may be helpful to comment on a few subtleties to gain a clear understanding of this model. The sign-switching transition of $\Lambda_{\rm s}$ described here by the signum function (implying an abrupt transition) should be understood as an idealized description of a rapid transition (may or may not be smooth) from an AdS vacuum provided by $\Lambda_{\rm s}=-\Lambda_{\rm s0}$ to a dS vacuum provided by $\Lambda_{\rm s}=\Lambda_{\rm s0}$, or DE models such as gDE, that can mimic this behavior. Such transitions that are also smooth, can easily be constructed/described phenomenologically using sigmoid functions, e.g., the hyperbolic tangent, $\tanh \qty[x]$, and the logistic function, $1/(1+e^{-x})$. Accordingly, one can replace~\cref{eq:ssdef} with, for example, $\Lambda_{\rm s}\equiv\Lambda_{\rm s0}\tanh{\qty[\eta(z_\dagger-z)]}$ which comes with two extra free parameters on top of $\Lambda$CDM, namely, $\eta$ and $z_\dagger$.\footnote{Another $\Lambda_{\rm s}$, extending the usual $\Lambda$ with two extra parameters ($z_\dagger$, $\gamma$), can be defined: $\Lambda_{\rm s}\equiv\Lambda_{\rm s0}\qty(\gamma \,{\rm sgn}{[z_\dagger-z]}-\gamma+1)$, in which case the new parameter $\gamma>1/2$ determines the depth of the AdS vacuum, $\Lambda_-=\Lambda_{\rm s0}(1-2\gamma)$. Further, once again ${\rm sgn}$ function can be replaced with continuous sigmoid functions, e.g., ${\Lambda_{\rm s}\equiv\Lambda_{\rm s0}\qty(\gamma \tanh{\qty[\eta(z_\dagger-z)+{\rm arctanh}\qty[1-1/\gamma]]}-\gamma+1)}$ defines a $\Lambda_{\rm s}$ with three extra parameters ($z_\dagger$, $\eta$, $\gamma$) that smoothly transitions from $\Lambda_-=\Lambda_{\rm s0}(1-2\gamma)$ to $\Lambda_+=\Lambda_{\rm s0}$ with a rapidity controlled by $\eta>0$, and vanishes at $z=z_\dagger$.} Of these two, $\eta>0$ determines the rapidity of the transition from $-\Lambda_{\rm s0}$ to $\Lambda_{\rm s0}$ around $z=z_\dagger$ and the limit $\eta\to+\infty$ leads to the abrupt sign-switch behavior considered in~\cref{eq:ssdef}. In $\Lambda_{\rm s}$CDM, we simply replace the $\Lambda$ of $\Lambda$CDM with $\Lambda_{\rm s}$, so that all material constituents of the Universe are locally conserved separately, and thus $\Lambda_{\rm s}$ also submits to the usual continuity equation due to the twice-contracted Bianchi identity in general relativity (GR). Accordingly, the corresponding EoS parameter reads $w_{\Lambda_{\rm s}}=-1-\eta(1+z)(1-\tanh^2{\qty[\eta(z_\dagger-z)]})/3\tanh{\qty[\eta(z_\dagger-z)]}$, which exhibits a pole at $z=z_\dagger$ (viz., yields ${\lim_{z\to z_\dagger^{\pm}}w_{\Lambda_{\rm s}}(z)=\pm\infty}$; such a singularity\footnote{For some examples with EoS parameters presenting singularities of the same type, see Refs.~\cite{Sahni:2014ooa,Wang:2018fng,Akarsu:2019hmw,Escamilla:2021uoj,Akarsu:2019ygx,Acquaviva:2021jov,Akarsu:2022lhx,DiValentino:2017rcr,Ong:2022wrs,Omni,Akarsu:2019pvi} where they are studied within the context of cosmological tensions; and see Refs.~\cite{Sahni:2004fb,Tsujikawa:2008uc,Zhou:2009cy,Bauer:2010wj,Sahni:2002dx} for some earlier examples.} is necessary for the energy density to change sign~\cite{Ozulker:2022slu}) and, approaching minus unity more and more with increasing $|z_\dagger-z|$, becomes indistinguishable from $w_{\Lambda_{\rm s}}=-1$ for all redshifts far enough away from $z_\dagger$.
We note that for a given definition of $\Lambda_{\rm s}$, the corresponding EoS parameter $w_{\Lambda_{\rm s}}$ is free to behave as necessary to ensure that the $\Lambda_{\rm s}$ satisfies the continuity equation, and when the limit $\eta\rightarrow+\infty$ corresponding to the abrupt sign-switch behavior in~\cref{eq:ssdef} is taken at face value, $w_{\Lambda_{\rm s}}(z\neq z_\dagger)=-1$ would be satisfied and the deviation of $w_{\Lambda_{\rm s}}$ from minus unity would be squeezed into the single redshift\footnote{Discontinuity of the signum function results in mild complications in familiar notions, e.g., the spacetime metric is no longer differentiable at $z=z_\dagger$ (though, it is continuous; see the solution of the metric for $\Lambda_{\rm s}$CDM in Ref.~\cite{Akarsu:2021fol}) and imposing that the $\Lambda_{\rm s}$ is conserved requires making use of generalized functions (distributions) to express its corresponding EoS parameter. We do not concern ourselves with such mathematical intricacies in the present study because the discontinuous $\Lambda_{\rm s}$ as defined in \cref{eq:ssdef} can be treated as an idealized parametrization/limiting case as stated in the main text.} $z=z_\dagger$; see \cref{sec:conc} for more comments on the abrupt sign-switch scenario and potential mechanisms underlying it. On the other hand, from a phenomenological point of view, for sufficiently large values of $\eta$, the ${\rm sgn}[x]$ and $\tanh[x]$ parametrizations become barely distinguishable; however, working with the abrupt AdS-dS transition as defined in~\cref{eq:ssdef} is much more convenient thanks to its simplicity, particularly for observational analyses. For instance, for $\eta=100$, we have $|\Lambda_{\rm s}|=\Lambda_{\rm s0}$ with $10^{-2}$ percent precision and $w_{\Lambda_{\rm s}}=-1$ with one percent precision at $z=z_\dagger\pm0.05$, improving to $10^{-6}$ percent precision and $10^{-4}$ percent precision, respectively, at $z=z_\dagger\pm0.1$. Thus, the abrupt sign-switching $\Lambda_{\rm s}$ we consider in this study can also be taken as an approximation for the more general, but rapidly sign-switching $\Lambda_{\rm s}$ models using, for instance, continuous sigmoid functions.

In Ref.~\cite{Akarsu:2021fol}, $\Lambda_{\rm s}$CDM was analyzed both theoretically and observationally; when the consistency of the model with the CMB is ensured, \textbf{(i)} $H_0$ and $M_B$ values are inversely correlated with $z_\dagger$ and reach $H_0\approx73.4~{\rm km\, s^{-1}\, Mpc^{-1}}$ and $M_B\approx-19.25\,{\rm mag}$ for $z_\dagger=1.6$ in remarkable agreement with the measurements from SH0ES~\cite{Riess:2021jrx,Camarena:2021dcvm}, and \textbf{(ii)} the model inherently presents an excellent fit to the Ly-$\alpha$ data provided that $z_\dagger\lesssim 2.34$. Since $\Lambda_{\rm s}$CDM is equivalent to $\Lambda$CDM for $z<z_\dagger$ except for the values of its parameters, it respects the internal consistency of the methodology used by local $H_0$ measurements that infer it from $M_B$ by assuming a $\Lambda$CDM-like cosmography~\cite{Riess:2016jrr,Freedman:2019cchp} such as SH0ES and TRGB;  thus, resolving the $H_0$ tension within $\Lambda_{\rm s}$CDM is almost equivalent to resolving the $M_B$ tension~\cite{Akarsu:2021fol}. To see if the model can achieve these promising features, it was confronted with observational data in Ref.~\cite{Akarsu:2021fol}; when only the CMB data set from \textit{Planck} 2018 is used, the model yields to $H_0 = 70.22\pm 1.78{\rm \,km\, s^{-1}\, Mpc^{-1}}$ with weak constraints on $z_\dagger$, and when BAO are also included with the CMB data set, it yields to $H_0 = 68.82\pm 0.55{\rm \,km\, s^{-1}\, Mpc^{-1}}$, fully consistent with the TRGB measurement ${H_0=69.8\pm0.8~{\rm km\, s^{-1}\, Mpc^{-1}}}$~\cite{Freedman:2019cchp} (or ${H_0=69.8\pm0.6~{\rm km\, s^{-1}\, Mpc^{-1}}}$~\cite{Freedman:2021ahq}), and a well-constrained $z_\dagger = 2.44\pm 0.29$, removing the ${\sim2\sigma}$ discrepancy with the Ly-$\alpha$ DR14~\cite{Blomqvist:2019rah} measurements that arises within $\Lambda$CDM. The lower and upper limits of $z_\dagger$ are controlled by the Galaxy and Ly-$\alpha$ BAO data, respectively, and the larger $z_{\dagger}$ values imposed by the Galaxy BAO data prevent the model from agreeing perfectly with the SH0ES measurements of $H_0=73.04\pm1.04~{\rm km\, s^{-1}\, Mpc^{-1}}$~\cite{Riess:2021jrx} and $M_B=-19.244\pm0.037\,\rm mag$~\cite{Camarena:2021dcvm}. Furthermore, the observational analyses of Ref.~\cite{Akarsu:2021fol} show that lower values of $z_\dagger$ also alleviate the $S_8$ tension despite having larger $\sigma_8$ (amplitude of mass fluctuations on scales of $8\, h^{-1}$ Mpc with $h\equiv H_0/100\, {\rm km\,s}^{-1}{\rm Mpc}^{-1}$ being the dimensionless reduced Hubble constant), i.e., more structures, and also in the case of CMB+BAO data, $\Lambda_{\rm s}$CDM accommodates a physical baryon density lower than that of $\Lambda$CDM in better agreement with its recent estimations from BBN constraints on the abundance of light elements such as $100\omega_{\rm b}=2.233\pm 0.036$~\cite{Mossa:2020gjc}. In summary, as $z_\dagger$ gets smaller, four discrepancies of $\Lambda$CDM, viz., the $H_0$, $M_B$, $S_8$, and Ly-$\alpha$ discrepancies, are better alleviated with potential improvements in the $\omega_{\rm b}$ discrepancy; and for $z_\dagger\sim1.6$ which is not preferred by the Galaxy BAO data, $\Lambda_{\rm s}$CDM can have remarkable agreement with multitude of observational data including the above four that $\Lambda$CDM is discordant with.

Besides all these superior phenomenological aspects of $\Lambda_{\rm s}$CDM over $\Lambda$CDM, $\Lambda_{\rm s}$CDM is also one of the simplest one-parameter extensions of $\Lambda$CDM. In fact, it is identical to $\Lambda$CDM  for both $z<z_\dagger$ and $z>z_\dagger$ except for the values of its parameters, in the sense that the Friedmann equations restricted to either one of these intervals have the same functional form; for other equivalently simple models inspired by $\Lambda_{\rm s}$CDM, see Ref.~\cite{Akarsu:2022jerk}. Thus, it is highly tempting to further explore how $\Lambda_{\rm s}$CDM is a good candidate to replace $\Lambda$CDM by extending the work of Ref.~\cite{Akarsu:2021fol} both theoretically and observationally. In Ref.~\cite{Akarsu:2021fol}, a detailed discussion was given on how it can alleviate the discrepancies with $H_0$, $M_B$, and BAO Ly-$\alpha$ measurements, and how these discrepancies are affected by the extra free parameter $z_\dagger$; in the observational analyses, CMB alone was found to be consistent with any value of $z_{\dagger}\gtrsim 1.5$, and it was constrained to $z_{\dagger}\gtrsim 2.3$ when the BAO data was included in the analysis. In what follows, we first expand the observational analysis of Ref.~\cite{Akarsu:2021fol}, by using the CMB data combined with the Pantheon data (with and without the $M_B$ prior from the SH0ES measurements), and also along with either the latest full BAO data set, only Ly-$\alpha$ BAO data, or without BAO data. Then, in the light of the results we have obtained, we extend the discussions on the $H_0$, $M_B$, $S_8$, Ly-$\alpha$, and $\omega_{\rm b}$ discrepancies made in Ref.~\cite{Akarsu:2021fol}; furthermore, we add the $t_0$ tension to our discussions and give a theoretical explanation of how the $S_8$ tension is alleviated in $\Lambda_s$CDM.

\section{Observational Analysis}
\label{sec:method}

Considering the background and perturbation dynamics, in what follows we explore the full parameter space of the $\Lambda_{\rm s}$CDM model and, for comparison, that of the standard $\Lambda$CDM model. The baseline seven free parameters of the $\Lambda_{\rm s}$CDM model are given by
\begin{equation}
\label{baseline1}
\mathcal{P}= \left\{ \omega_{\rm b}, \, \omega_{\rm c}, \, \theta_s, \,  A_{\rm s}, \, n_s, \, \tau_{\rm reio},
\,   z_\dagger \right\}.
\end{equation}
Here, the first six parameters are the common ones with the standard $\Lambda$CDM model, viz., $\omega_{\rm b}=\Omega_{\rm b} h^2$ and $\omega_{\rm c}=\Omega_{\rm c}h^2$ ($\Omega$ being the present-day density parameter) are, respectively, the present-day physical density parameters of baryons and cold dark matter, $\theta_{\rm s}$ is the ratio of the sound horizon to the angular diameter distance at decoupling, $A_{\rm s}$ is the initial super-horizon amplitude of curvature perturbations at $k=0.05$ Mpc$^{-1}$, $n_{\rm s}$ is the primordial spectral index, and $\tau_{\rm reio}$ is the reionization optical depth. We assume three neutrino species, approximated as two massless states and a single massive neutrino of mass $m_{\nu}=0.06\,\rm eV$. We use uniform priors $\omega_{\rm b}\in[0.018,0.024]$, $\omega_{\rm c}\in[0.10,0.14]$, $100\,\theta_{\rm s}\in[1.03,1.05]$, $\ln(10^{10}A_{\rm s})\in[3.0,3.18]$, $n_{\rm s}\in[0.9,1.1]$, and $\tau_{\rm reio}\in[0.04,0.125]$ for the common free parameters of the models, and $z_\dagger\in[1,3]$ for the additional free parameter characterizing the $\Lambda_{\rm s}$CDM model.

\begin{table}[b!]
\caption{\rm Clustering measurements for each of the BAO samples from Ref.~\cite{Alam:2020sor}.}
 \scalebox{0.9}{
\centering
\begin{tabular}{l|c|c|c|c}
\toprule
\hline
\textbf{Parameter} & $\bm{\,\,z_{\rm eff}\,\,}$  &  $\bm{\,\,D_V(z)/r_{\rm d}\,\,}$ & $\bm{\,\,D_M(z)/r_{\rm d}\,\,}$ & $\bm{\,\,D_H(z)/r_{\rm d}\,\,}$  \\
\hline

\hline
MGS & 0.15 & $4.47 \pm 0.17$ & ... & ... \\

BOSS Galaxy & $0.38$ & ... & $10.23 \pm 0.17$ & $25.00 \pm 0.76$ \\

BOSS Galaxy & $0.51$ & ... & $13.36 \pm 0.21$ & $22.33 \pm 0.58$ \\

eBOSS LRG & $0.70$ & ... & $17.86 \pm 0.33$ & $19.33 \pm 0.53$ \\

eBOSS ELG & $0.85$ & $18.33_{-0.62}^{+0.57}$ & ... & ... \\

eBOSS Quasar & $1.48$ & ... & $30.69 \pm 0.80$ & $13.26 \pm 0.55$ \\

Ly$\alpha$-Ly$\alpha$ & $2.33$ & ... & $37.6 \pm 1.9$ & $8.93 \pm 0.28$ \\

Ly$\alpha$-Quasar & $2.33$ & ... &$37.3 \pm 1.7$ & $9.08 \pm 0.34$ \\
\hline
\bottomrule
\hline
\end{tabular} }
\label{tab:BAO_measurements}
\end{table}

In order to constrain the models, we use the latest \textit{Planck} CMB data combined with other data sets from independent observations. From the \textit{Planck} 2018 legacy data release~\cite{Planck:2018lbu,Planck:2019nip}, we use measurements of CMB temperature anisotropy and polarization power spectra, their cross-spectra, and lensing power spectrum, viz., (i) the high-$\ell$ \texttt{Plik} likelihood for TT (in the multipole range $30 \leq \ell \leq 2508$), (ii) TE and EE (in the multipole range $30 \leq \ell \leq 1996$), (iii) the low-$\ell$ TT-only ($2 \leq \ell \leq 29$) likelihood based on the \texttt{Commander} component-separation algorithm in pixel space, (iv) the low-$\ell$ EE-only ($2 \leq \ell \leq 29$) \texttt{SimAll} likelihood, and (v) the CMB lensing power spectrum measurements reconstructed from the temperature 4-point function. Along with the \textit{Planck} CMB data, we use the high-precision BAO measurements at different redshifts up to $z=3.5$, viz., the BAO measurements compiled in Table~\ref{tab:BAO_measurements}, from final measurements of clustering using galaxies, quasars, and Lyman-$\alpha$ (Ly-$\alpha$) forests from the \textit{completed} Sloan Digital Sky Survey (SDSS) lineage of experiments in large-scale structure~\cite{Alam:2020sor}. It is worth noting that we include the Ly-$\alpha$ measurements in our BAO compilation as these have a substantial impact on the parameters of $\Lambda_{\rm s}$CDM, whereas these have a minor impact on the parameters of $\Lambda$CDM, which is why the Ly-$\alpha$ measurements were excluded from the default BAO compilation by the \textit{Planck} (2018) Collaboration~\cite{Planck:2018vyg}.  In our analyses, we first consider only the Ly-$\alpha$ data and then the full set of BAO data. We use the Pantheon~\cite{scolnic:2018dm} distance moduli measurements for Type Ia Supernovae which provide the constraints on the slope of the late-time expansion rate $H_{0}d_{L}(z)$, i.e., the noncalibrated light distance. The theoretical apparent magnitude $m_B$ of an SNIa at redshift $z$ reads $m_{B}(z)= 5 \log_{10} \left[d_{L}(z)/1\, {\rm Mpc}\right]+25+M_{B}$, where $M_{B}$ is the absolute magnitude. The distance modulus is then given by $\mu(z)=m_{B}-M_{B}$. We constrain the models also by using a Gaussian prior on $M_{B}$, viz., $M_B=-19.2435\pm 0.0373\,\rm mag$ that corresponds to the SH0ES SNIa measurements~\cite{Camarena:2021dcvm}---alternatively, one could prefer using an $H_0$ prior; see~\cref{footnote5} in~\cref{subsec:h0} for advantages and disadvantages of using $M_{B}$ or $H_0$ as a prior. We use the publicly available Boltzmann code \texttt{CLASS}~\cite{Blas:2011jt} with the parameter inference code  \texttt{Monte Python}~\cite{Audren:2013ea} to obtain correlated Monte Carlo Markov Chain (MCMC) samples. We analyze the MCMC samples using the python package \texttt{GetDist}; and use the \texttt{MCEvidence}~\cite{MCEvidence} algorithm to approximate the Bayesian evidence, used to perform a model comparison through the Jeffreys' scale~\cite{Vazquez:2011xa}. See Ref.~\cite{Padilla:2019mgi}, and references therein, for an extended review of the cosmological parameter inference and model selection procedure. In general, for a data set $D$ and a given model $\mathcal{M}_a$ with a set of parameters $\Theta$, Bayes' theorem results in
\begin{equation}\label{eq:bayes}
P(\Theta|D, \mathcal{M}_a) = \frac{\mathcal{L}(D|\Theta, \mathcal{M}_a) \pi(\Theta|\mathcal{M}_a)}{\mathcal{E}(D|\mathcal{M}_a)},
\end{equation}
where $P(\Theta|D, \mathcal{M}_a)$ is the posterior probability distribution function of the parameters, $\pi(\Theta|\mathcal{M}_a)$ is the prior for the parameters, $\mathcal{L}(D|\Theta, \mathcal{M}_a)$ is the likelihood function, and $\mathcal{E}(D|\mathcal{M}_a)$ is the Bayesian evidence given by
\begin{equation}\label{eq:evidence}
\mathcal{E}(D|\mathcal{M}_a) = \int_{\mathcal{M}_a} \mathcal{L}(D|\Theta, \mathcal{M}_a) \pi(\Theta|\mathcal{M}_a) \text{d}\Theta.
\end{equation}
To make a comparison of the model $\mathcal{M}_a$ with some other model $\mathcal{M}_b$, we compute the ratio of the posterior probabilities of the models, given by
\begin{equation}
\frac{P(\mathcal{M}_a|D)}{P(\mathcal{M}_b|D)} = B_{ab}\frac{P(\mathcal{M}_a)}{P(\mathcal{M}_b)},
\end{equation}
where $B_{ab}$ is the Bayes' factor given by
\begin{equation}
    B_{ab}=\frac{\mathcal{E}(D|\mathcal{M}_a)}{\mathcal{E}(D|\mathcal{M}_b)}\equiv\frac{\mathcal{Z}_a}{\mathcal{Z}_b}.
\end{equation}
So the relative log-Bayesian evidence reads as 
\begin{equation}\label{eq:evidence}
    \ln B_{ab}=\ln \mathcal{Z}_a-\ln \mathcal{Z}_b\equiv\Delta\ln \mathcal{Z}.
\end{equation}
The model with smaller $|\ln \mathcal{Z}|$ is the preferred model, and therefore considered as the reference model in model comparison. To interpret the results, we refer to the revised Jeffreys' scale as given in Ref.~\cite{Kass:1995loi}. Accordingly, a weak evidence is indicated by $0 \leq |\Delta\ln \mathcal{Z}|  < 1$, a definite evidence $1 \leq | \Delta\ln \mathcal{Z}|  < 3$, a strong evidence by $3 \leq | \Delta\ln \mathcal{Z}|  < 5$, and a very strong evidence by $| \Delta\ln \mathcal{Z} | \geq 5$,  in favor of the reference model.

In Ref.~\cite{Akarsu:2021fol}, the authors investigated the observational constraints on the parameters of the models, $\Lambda_{\rm s}$CDM and $\Lambda$CDM, with the CMB and CMB+BAO data. In the present study, we obtain the observational constraints on the parameters of these models by using the data combinations of CMB+Pan, CMB+Pan+Ly-$\alpha$, and CMB+Pan+BAO without and with $M_B$ prior separately present in~\cref{tab:withoutMB,tab:withMB}, respectively. Also, see~\cref{fig:nomb,fig:mb,fig:nomblya,fig:mblya,fig:nombbao,fig:mbbao} in~\cref{sec:Appendix} for the corresponding one- and two-dimensional [at 68\% and 95\% confidence levels (CLs)] marginalized distributions of the model parameters.~\footnote{Note that the BAO data used in the present study is an updated and extended version of that in Ref.~\cite{Akarsu:2021fol}, hence the results are not directly comparable.} In the last three rows of these tables, we list the best fit ($-2\ln{\mathcal{L}_{\rm max}}$), the $\log$-Bayesian evidence ($\ln \mathcal{Z}$), and the $\log$-Bayesian evidence relative to the reference model ($\Delta\ln \mathcal{Z}$).

\begin{figure}[t!]
    \centering
    \includegraphics[width=4.25cm]{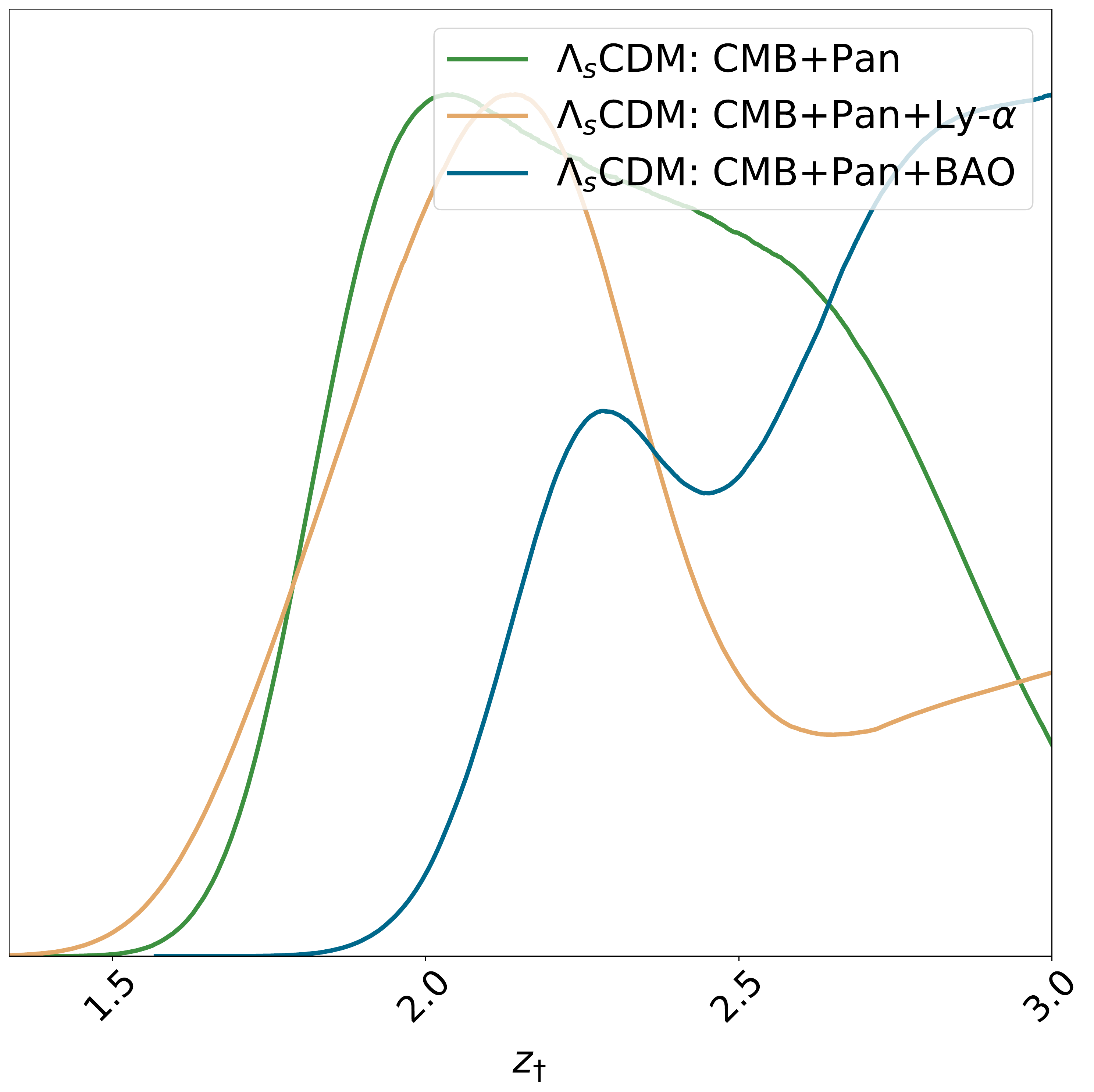}
    \includegraphics[width=4.25cm]{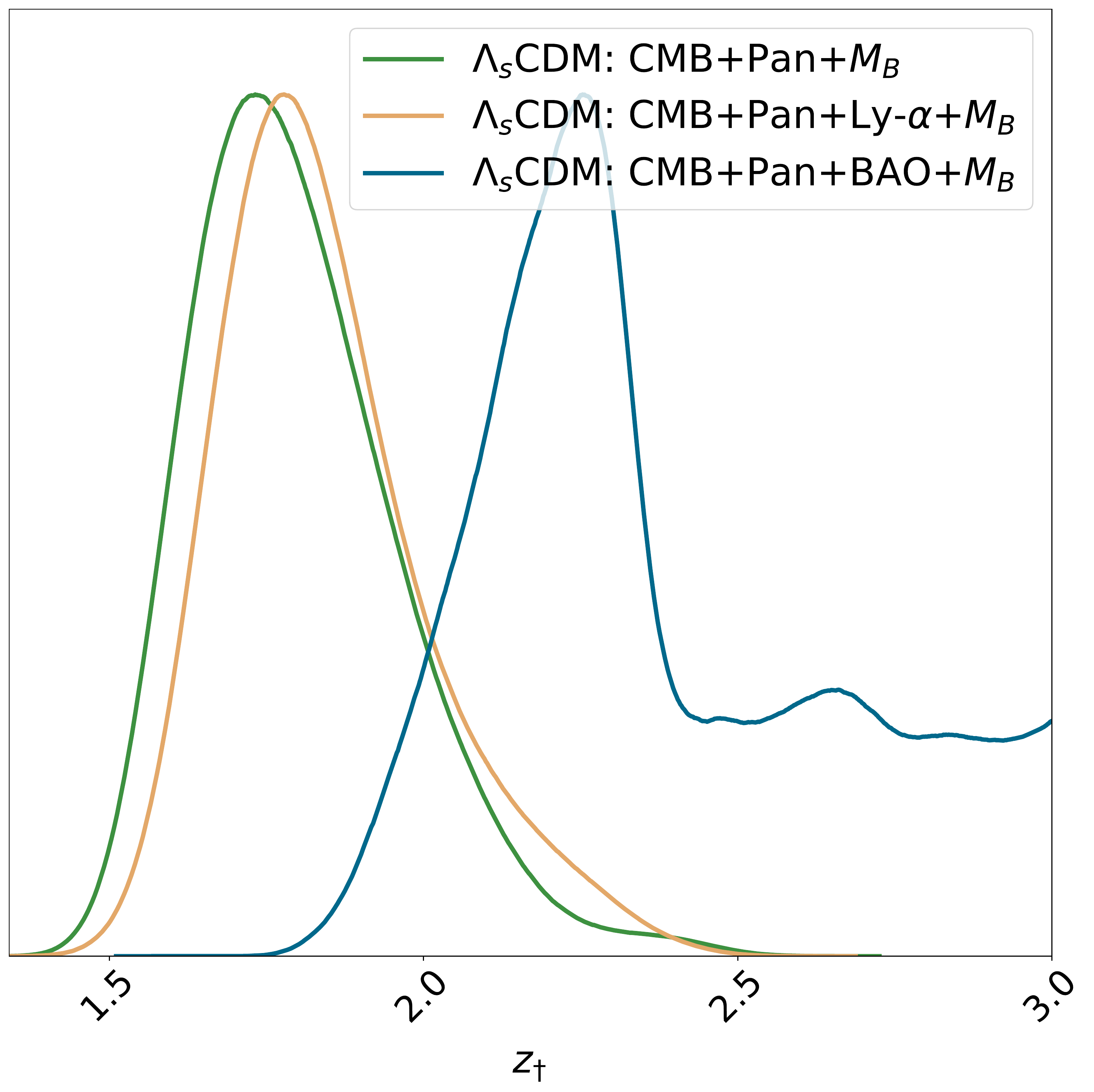}
   \caption{One-dimensional marginalized posterior distributions of the parameter $z_\dagger$ of the $\Lambda_{\rm s}$CDM model, the redshift at which the cosmological constant ($\Lambda_{\rm s}$) changes sign, for various data set combinations.}
    \label{fig:zdagger1d}
\end{figure}

The distinctive free parameter of the $\Lambda_{\rm s}$CDM model is $z_\dagger$, the redshift at which the cosmological constant ($\Lambda_{\rm s}$) changes sign. In~\cref{fig:zdagger1d}, we present the one-dimensional marginalized distributions of the parameter $z_\dagger$ for various data set combinations. From Ref.~\cite{Akarsu:2021fol}, we know that the CMB data alone is not able to constrain $z_{\dagger}$, implying that any $z_{\dagger}\gtrsim1.5$ ($1.5$ is the lower limit of the prior used in Ref.~\cite{Akarsu:2021fol}), i.e., a negative cosmological constant $\Lambda_{\rm s}(z>z_\dagger)=-\Lambda_{\rm s0}\sim-2.9\times 10^{-122}\,l_{\rm Planck}^{-2} $ is consistent with the CMB data. But when the SNIa data are included in the analysis with CMB (see the green curve in~\cref{fig:zdagger1d}), the shape of the distribution changes, and we find lower bound of $z_{\dagger}>1.77$ and with the inclusion of the Ly-$\alpha$ data (which favor $z_\dagger$ values less than $\sim2.33$) as well, we see a clear peak at $z_\dagger\sim2.2$ with a plateaulike tail for $z_{\dagger}\gtrsim2.5$, the region where the model approaches $\Lambda$CDM. However, with the inclusion of the full BAO data, rather than only the Ly-$\alpha$ data, again we find only a lower bound, $z_{\dagger}>2.13$. This is because the low-redshift BAO data tend to push $z_{\dagger}$ to larger values, despite the opposition of the Ly-$\alpha$; this point was discussed in Ref.~\cite{Akarsu:2021fol} thoroughly, also see~\cref{subsec:bao}. We notice that including the $M_{B}$ prior in the analysis has important consequences in the results. When the $M_B$ prior is present, whether the Ly-$\alpha$ data are included or not on top of CMB+Pan data, $z_{\dagger}$ is very well constrained at $z\approx1.8$ with $\sim10\%$ precision at $\%68$ CL. While the CMB+Pan+BAO data combination without the $M_B$ prior is able to provide only a lower bound on $z_\dagger$, with the $M_B$ prior it leads to a clear peak at $z\approx2.3$ with $\sim10\%$ precision at $\%68$ CL, with a flat tail for $z\gtrsim 2.4$ seems to have arisen from the preference of higher $z_\dagger$ values of the low-redshift BAO data.

\begin{table*}[tb]
\caption{Constraints (68\% CL) on  the free and some derived parameters of the $\Lambda_{\rm s}$CDM and standard $\Lambda$CDM models for CMB+Pan, CMB+Pan+Ly-$\alpha$ and CMB+Pan+BAO data. In the last three rows, the best fit ($-2\ln{\mathcal{L}_{\rm max}}$), the $\log$-Bayesian evidence ($\ln \mathcal{Z}$), and the relative $\log$-Bayesian evidence $\Delta\ln \mathcal{Z}$ [see \cref{eq:evidence}] are listed. For each combination of data sets, the model with $\Delta\ln \mathcal{Z}=0$ is the reference (preferred) model.}\label{tab:withoutMB}
{
 \scalebox{0.96}{
\begin{centering}
	  \begin{tabular}{lcccccc}
  	\hline
    \toprule
    \multicolumn{1}{l}{\textbf{Data set}} & \multicolumn{2}{c}{\textbf{CMB+Pan}} & \multicolumn{2}{c}{\textbf{CMB+Pan+Ly-$\bm{\alpha}$}}& \multicolumn{2}{c}{\textbf{CMB+Pan+BAO}} \\  \hline
      & \textbf{{$\bm{\Lambda}$CDM}} & \textbf{$\bm{\Lambda}_{\textbf{s}}$CDM} & \textbf{{$\bm{\Lambda}$CDM}}  & \textbf{$\bm{\Lambda}_{\textbf{s}}$CDM}  & \textbf{{$\bm{\Lambda}$CDM}}  & \textbf{$\bm{\Lambda}_{\textbf{s}}$CDM} \\ 
      \midrule
      \vspace{0.1cm}
$10^{2}\omega_{\rm b }$ & $2.240\pm0.015$ &  $2.241\pm 0.014$ & $2.242\pm 0.013$ &   $2.241\pm 0.015 $ 
    & $2.242 \pm 0.013$    & $2.235\pm 0.014$  
    \\ 
    
\vspace{0.1cm}
$\omega_{\rm c }$  &$0.1197\pm 0.0012$ & $0.1196 \pm 0.0011$ & $0.1193\pm 0.0009 $ & $0.1196\pm 0.0011$ 
 & $0.1193 \pm 0.0009$ & $0.1206 \pm 0.0010$  
 \\ 
 
\vspace{0.1cm}
$100 \theta_{s } $  & $1.04191 \pm 0.00029$    &$1.04190\pm 0.00028$  & $1.04191\pm 0.00029$ & $1.04190\pm 0.00029 $  
&    $1.04194\pm0.00028$ & $1.04180\pm0.00030$ 
\\ 

\vspace{0.1cm}
$\ln(10^{10}A_{\rm s})$ &     $3.047\pm0.015$  & $3.041\pm0.014$ & $3.047\pm0.014$ & $3.040\pm 0.015$ 
& $3.047\pm0.015$  &  $3.040\pm 0.014$ 
\\ 

\vspace{0.1cm}
$n_{s } $  & $0.9662\pm0.0042$   & $0.9668\pm 0.0040$  & $0.9669^{+0.0039}_{-0.0036}$ & $0.9668\pm0.0041$ 
&     $0.9665\pm 0.0037$ & $0.9644\pm0.0037$ 
\\ 

\vspace{0.1cm}
$\tau_{\rm reio } $  &     $0.0556\pm0.0075$  & $0.0533\pm0.0075$ & $0.0560\pm0.0069$ & $0.0528\pm 0.0077 $ 
& $0.0561\pm0.0076$  &  $0.0515 \pm 0.0073$ 
\\

\vspace{0.1cm}
$z_\dagger$  &   --- & $>1.80$ (95\% CL) & --- & $2.21^{+0.16}_{-0.38} $ 
& --- & $>2.13$ (95\% CL)  
\\

\vspace{0.1cm}
$M_{B}$ [mag]  &   $-19.421\pm0.014$ & $-19.363^{+0.021}_{-0.037}$ & $-19.418\pm 0.011$ & $-19.349\pm0.028$ 
& $-19.418\pm0.012$ & $-19.387\pm0.015$   
\\

\hline

\vspace{0.10cm} 
$\Omega_{\rm{m} }$ &   $0.3129 \pm 0.0071$   & $0.2940^{+0.0120}_{-0.0093}$  & $0.3110\pm 0.0053$ & $0.2899\pm 0.0097 $   & $0.3109\pm 0.0056$  & $0.3039 \pm 0.0058$
\\

\vspace{0.10cm} 
$\omega_{\rm{m} }$ &   $0.1427\pm 0.0011$   & $0.1427\pm 0.0010$  & $0.1424\pm 0.0008$ & $0.1426\pm0.0010$   & $0.1424\pm 0.0009$  & $0.1436\pm 0.0010$ 
\\

\vspace{0.10cm} 
$H_{\rm 0}$ [km/s/Mpc] &    $67.55\pm0.53$  & $69.68^{+0.77}_{-1.40}$   & $ 67.68\pm 0.40 $ & $70.17^{+0.96}_{-1.10}$ 
& $67.69^{+0.38}_{-0.43}$   & $68.74^{+0.49}_{-0.55}$ 
\\

\vspace{0.10cm} 
$t_{\rm 0}$ [Gyr] &    $13.79\pm0.02$  & $13.65^{+0.06}_{-0.04}$   & $ 13.79\pm 0.02 $ & $13.62^{+0.09}_{-0.03}$ & $13.79\pm0.02$   & $13.71^{+0.03}_{-0.02}$  \\

\vspace{0.10cm} 
$\sigma_{8}$ &    $0.8111^{+0.0056}_{-0.0063}$  & $0.8167^{+0.0059}_{-0.0067}$   & $0.8104\pm0.0060$ & $0.8182\pm 0.0066$ 
& $0.8101\pm0.0063$   & $0.8167\pm 0.0062$  
\\

  \vspace{0.10cm} 
$S_{8}$ &    $0.828\pm 0.013$  & $0.809\pm 0.015$   & $0.825\pm 0.010$ & $0.804\pm 0.014 $ & $0.825\pm 0.011$   & $0.822\pm 0.010$ 
\\
  \hline

 \vspace{0.10cm}    
$-2\ln{\mathcal{L}_{\rm max}}$ & $3807.24$ & $3805.00$ & $3819.36$ & $3806.88$ 
& $3819.26$ & $3819.06$
\\

\vspace{0.10cm} 
 $\ln \mathcal{Z}$ & $-1937.82$ & $-1938.02$ & $-1944.53$ & $-1939.75$ & $-1944.51$ & $-1944.76$
 \\

\vspace{0.10cm} 
 $\Delta\ln \mathcal{Z}$ & $0$ & $0.20$ & $4.78$ & $0$ & $0$ & $0.25$\\ 
\hline
  \bottomrule
    \hline 
  \end{tabular}
  \end{centering}
  
  }}
\end{table*}

\begin{table*}[ht!]
  \caption{Constraints (68\% CL) on  the free and some derived parameters of the $\Lambda_{\rm s}$CDM and standard $\Lambda$CDM models for CMB+Pan, CMB+Pan+Ly-$\alpha$ and CMB+Pan+BAO data with the SH0ES $M_{B}$ prior. In the last three rows, the best fit ($-2\ln{\mathcal{L}_{\rm max}}$), the $\log$-Bayesian evidence ($\ln \mathcal{Z}$), and the relative $\log$-Bayesian evidence $\Delta\ln \mathcal{Z}$ [see \cref{eq:evidence}] are listed. For each combination of data sets, the model with $\Delta\ln \mathcal{Z}=0$ is the reference (preferred) model.}
  \label{tab:withMB}
	\scalebox{0.96}{
	\setlength\extrarowheight{2pt}
	\begin{centering}
	  \begin{tabular}{lcccccc}
  	\hline
    \toprule
    \multicolumn{1}{l}{\textbf{Data set}} & \multicolumn{2}{c}{\textbf{CMB+Pan+$\bm{M_B}$}} & \multicolumn{2}{c}{\textbf{CMB+Pan+Ly-$\bm{\alpha}$+$\bm{M_B}$}}& \multicolumn{2}{c}{\textbf{CMB+Pan+BAO+$\bm{M_B}$}} \\  \hline
      & \textbf{{$\bm{\Lambda}$CDM}} & \textbf{$\bm{\Lambda}_{\textbf{s}}$CDM} & \textbf{{$\bm{\Lambda}$CDM}}  & \textbf{$\bm{\Lambda}_{\textbf{s}}$CDM}  & \textbf{{$\bm{\Lambda}$CDM}}  & \textbf{$\bm{\Lambda}_{\textbf{s}}$CDM} \\ 
      \midrule
      \vspace{0.1cm}
$10^{2}\omega_{\rm b }$ & $2.256\pm0.015$ &  $2.248\pm 0.014$ & $2.253\pm 0.013$ &   $2.247^{+0.014}_{-0.013} $ 
    & $2.255 \pm 0.013$    & $2.242\pm 0.014$  \\ 
    
\vspace{0.1cm}
$\omega_{\rm c }$  &$0.1181\pm 0.0011$ & $0.1191 \pm 0.0011$ & $0.1183\pm 0.0008 $ & $0.1191\pm 0.0011$ 
 & $0.1181 \pm 0.0009$ & $0.1200^{+0.0010}_{-0.0011}$  \\ 
 
\vspace{0.1cm}
$100 \theta_{s } $  & $1.04208 \pm 0.00029$    &$1.04197\pm 0.00031$  & $1.04204\pm 0.00028$ & $1.04196\pm 0.00028 $  
&    $1.04207^{+0.00029}_{-0.00026}$& $1.04186 \pm 0.00028$  \\ 

\vspace{0.1cm}
$\ln(10^{10}A_{\rm s})$ &     $3.053^{+0.014}_{-0.017}$  & $3.039\pm0.014$ & $3.052^{+0.013}_{-0.016}$ & $3.041\pm 0.015$ 
& $3.053^{+0.014}_{-0.016}$  &  $3.041\pm 0.015$  \\ 

\vspace{0.1cm}
$n_{s } $  & $0.9701\pm0.0040$   & $0.9687^{ +0.0043}_{-0.0038}$  & $0.9697\pm 0.0035$ & $0.9684\pm 0.0041$ 
&     $0.9702\pm 0.0035$ & $0.9661\pm0.0037$  \\ 

\vspace{0.1cm}
$\tau_{\rm reio } $  &     $0.0601^{+0.0072}_{-0.0085}$  & $0.0526\pm0.0074$ & $0.0593^{+0.0064}_{-0.0079}$ & $0.0535\pm 0.0077 $ 
& $0.0603^{+0.0070}_{-0.0078}$  &  $0.0524 \pm 0.0076$   \\

\vspace{0.1cm}
$z_\dagger$  &   --- & $1.78^{+0.14}_{-0.18}$  & --- & $1.84^{+0.13}_{-0.21} $ 
& --- & $2.36\pm0.28$    \\

\vspace{0.1cm}
$M_{B}$ [mag]  &   $-19.399\pm0.014$ & $-19.290^{+0.026}_{-0.029}$ & $-19.402\pm 0.011$ & $-19.299\pm0.028$ 
& $-19.399\pm0.011$ & $-19.366^{+0.013}_{-0.015}$    \\

\hline

\vspace{0.10cm} 
$\Omega_{\rm m }$ &   $0.3028 \pm 0.0068$   & $0.2716 \pm 0.0084$  & $0.3043\pm 0.0050$ & $0.2743^{+0.0086}_{-0.0097} $   & $0.3030\pm 0.0051$  & $0.2965 \pm 0.0055$ \\

\vspace{0.10cm} 
$\omega_{\rm m }$ &   $0.1413\pm 0.0011$   & $0.1422\pm 0.0010$  & $0.1415\pm 0.0008$ & $0.1422\pm 0.0011$   & $0.1413\pm 0.0008$  & $0.1431\pm 0.0010$ \\

\vspace{0.10cm} 
$H_{\rm 0}$ [km/s/Mpc] &    $68.31\pm0.52$  & $72.38^{+0.98}_{-1.10}$   & $ 68.19\pm 0.38 $ & $72.0\pm1.1$ 
& $68.29\pm0.39$   & $69.48^{+0.48}_{-0.55}$  \\

\vspace{0.10cm} 
$t_{\rm 0}$ [Gyr] &    $13.76\pm0.02$  & $13.55\pm0.05$   & $ 13.76\pm 0.02 $ & $13.56^{+0.04}_{-0.04}$ & $13.76\pm0.02$   & $13.67\pm0.03$  \\

\vspace{0.10cm} 
$\sigma_{8}$ &    $0.8090\pm 0.0064$  & $0.8255^{+0.0072}_{-0.0081}$   & $0.8091^{+0.0054}_{-0.0063}$ & $0.8243\pm 0.0076$ 
& $0.8092^{+0.0057}_{-0.0061}$   & $0.8176 \pm 0.0063$  \\

  \vspace{0.10cm} 
$S_{8}$ &    $0.813\pm 0.012$  & $0.785\pm 0.012$   & $0.815\pm 0.010$ & $0.788^{+ 0.012}_{-0.014} $ & $0.813\pm 0.010$   & $0.813\pm 0.010$  \\
  \hline

    \vspace{0.10cm}    
$-2\ln{\mathcal{L}_{\rm max}}$ & $3826.56$ & $3808.58$ & $3837.36$ & $3811.76$ 
    & $3839.70$ & $3831.14$
\\

\vspace{0.10cm} 
 $\ln \mathcal{Z}$ & $-1947.83$ & $-1940.06$ & $-1954.17$ & $-1941.85$ & $-1955.02$ & $-1951.79$\\
 
\vspace{0.10cm} 
  $\Delta\ln \mathcal{Z}$ & $7.77$ & $0$ & $12.32$ & $0$ & $3.23$ & $0$\\ 
  \hline
  \bottomrule
    \hline 
  \end{tabular}
  \end{centering}}
\end{table*}

In Ref.~\cite{Akarsu:2021fol}, no strong statistical evidence was found to discriminate between the $\Lambda_{\rm s}$CDM and $\Lambda$CDM models in the analyses with neither the CMB data nor the CMB+BAO data (estimates $z_\dagger\sim 2.4$). We see in the current work that, without the $M_B$ prior, this picture does not change for the cases CMB+Pan (estimates $z_\dagger\gtrsim1.8$) and CMB+Pan+BAO (estimates $z_\dagger\gtrsim2.1$), while the $\Lambda_{\rm s}$CDM model finds a strong evidence ($\Delta\ln \mathcal{Z}\sim5$) against the standard $\Lambda$CDM model for the case CMB+Pan+Ly-$\alpha$ (estimates $z_\dagger\sim2.2$); see~\cref{tab:withoutMB}. On the other hand, when we analyze the models with the same data sets by including the $M_B$ prior that corresponds to the SH0ES SNIa measurements~\cite{Camarena:2021dcvm}, it turns out that the $\Lambda_{\rm s}$CDM model (estimates $z_\dagger\sim2$) is always preferred over the standard $\Lambda$CDM model; namely, $\Lambda_{\rm s}$CDM finds very strong evidence (reaching $\Delta\ln \mathcal{Z}\sim12$) against $\Lambda$CDM by predicting $z_\dagger\sim1.8$ for both the CMB+Pan+$M_B$ and CMB+Pan+Ly-$\alpha$+$M_B$ cases, and finds strong evidence ($\Delta\ln \mathcal{Z}\sim3$) by predicting $z_\dagger\sim2.4$ for the CMB+Pan+BAO+$M_B$ case; see~\cref{tab:withMB}. Hence, the relative log-Bayesian evidences are significantly strengthened in favor of $\Lambda_{\rm s}$CDM in all cases with the inclusion of $M_B$ prior. Regarding the best fits ($-2\ln\mathcal{L}_{\rm max}$), the inclusion of the $M_B$ prior results in a substantial worsening ($-2\Delta\ln\mathcal{L}_{\rm max}\sim20$) of $\Lambda$CDM's fit to the data for all three data compilations; compare $-2\ln\mathcal{L}_{\rm max}$ of~\cref{tab:withoutMB,tab:withMB}. On the other hand, for $\Lambda_{\rm s}$CDM, there is no significant worsening ($-2\Delta\ln\mathcal{L}_{\rm max}\sim5$) without the full BAO data, and while it becomes noticeable ($-2\Delta\ln\mathcal{L}_{\rm max}\sim12$) when the full BAO data is included, it still is milder compared to $\Lambda$CDM. This implies that $\Lambda_{\rm s}$CDM has much better consistency with the $M_B$ prior than $\Lambda$CDM and signals $\Lambda_{\rm s}$CDM relaxes the $M_B$ tension and thus the closely related $H_0$ tension as well. Also, in both tables (\cref{tab:withoutMB,tab:withMB}), we see that the expansion of CMB+Pan and CMB+Pan+$M_B$ analyzes by including the Ly-$\alpha$ data makes a significant improvement ($\sim5$) in the relative log-Bayesian evidence in favor of $\Lambda_{\rm s}$CDM, which indicates that $\Lambda_{\rm s}$CDM is also highly compatible with the Ly-$\alpha$ data. On the other hand, when we expand the CMB+Pan and CMB+Pan+$M_B$ analyzes by including the full BAO data listed in~\cref{tab:BAO_measurements} (equivalent to expanding the cases CMB+Pan+Ly-$\alpha$ and CMB+Pan+Ly-$\alpha$+$M_B$ by adding the low-redshift BAO data) we compromise on this improvement; namely, the strong evidence ($\Delta\ln \mathcal{Z}\sim5$) from the CMB+Pan+Ly-$\alpha$ data set in favor of $\Lambda_{\rm s}$CDM is lost ($\Delta\ln \mathcal{Z}\sim0$) in the CMB+Pan+BAO case, and  the very strong evidence ($\Delta\ln \mathcal{Z}\sim12$) from the CMB+Pan+Ly-$\alpha$+$M_B$ data set in favor of $\Lambda_{\rm s}$CDM is reduced to strong evidence ($\Delta\ln \mathcal{Z}\sim3$) in the CMB+Pan+BAO+$M_B$ case. It is worth mentioning here that the Ly-$\alpha$ data support $z_\dagger$ values less than $\sim2.3$, whereas some low-redshift BAO data prefer $z_\dagger$ values greater than $\sim2.3$, forcing the $\Lambda_{\rm s}$CDM model to its $\Lambda$CDM limit ($z_{\dagger}\rightarrow\infty$). 

\section{Relaxing cosmological tensions}
\label{sec:tensions}

\begin{figure*}
    \centering
    \includegraphics[width=18.0cm]{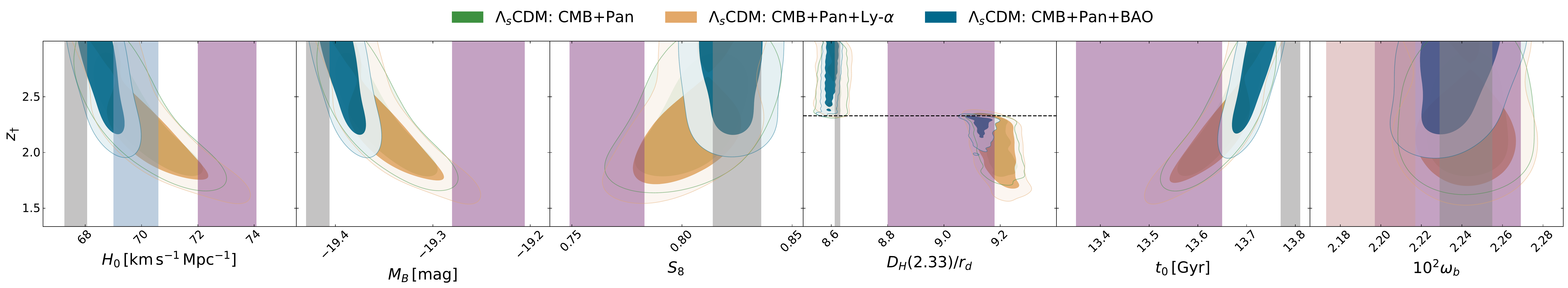}
    \includegraphics[width=18.0cm]{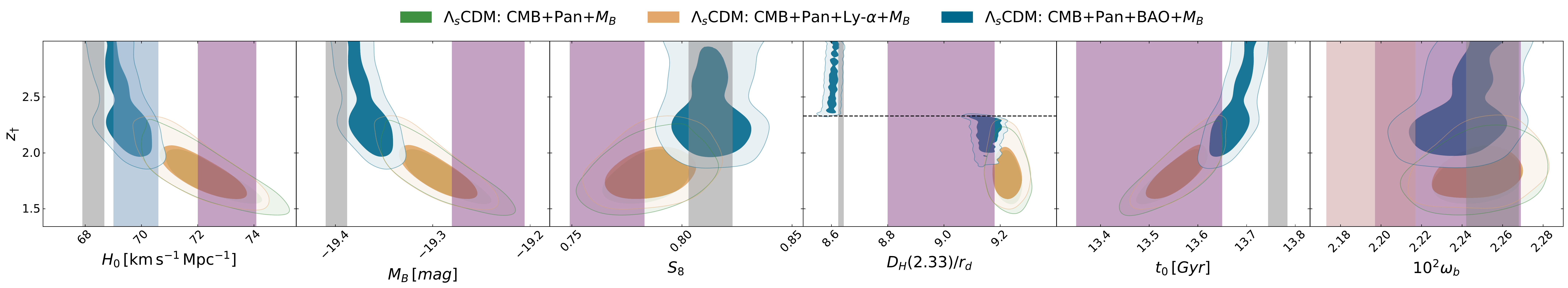}
    \caption{Two-dimensional marginalized probability posteriors of $z_{\dagger}$ versus $H_0$, $M_B$, $S_8$, $D_H(2.33)/r_{\rm d}$ ($D_H/r_{\rm d}$ at $z_{\rm eff}=2.33$ relevant to the Ly-$\alpha$ measurements), $t_0$, and $\omega_{\rm b}$ in the $\Lambda_{\rm s}$CDM model from various combinations of the data sets. The vertical gray bands are the constraints (68\% CL) for the $\Lambda$CDM model, where in the upper panels we consider only CMB+Pan+BAO and in the lower panels CMB+Pan+BAO+$M_B$ since the vertical gray bands obtained for other combinations of data sets do not differ visually. The vertical purple bands stand for the theoretical/direct observational estimations (at 68\% CL) of the corresponding parameters commonly used in the literature: $H_{\rm 0}^{\rm R21}=73.04\pm1.04~{\rm km\, s^{-1}\, Mpc^{-1}}$~\cite{Riess:2021jrx}; $M_B=-19.244\pm 0.037\,\rm mag$ (SH0ES)~\cite{Camarena:2021dcvm}; $S_8=0.766^{+0.020}_{-0.014}$ (WL+GC KiDS-1000 $3\times2$pt)~\cite{Heymans:2020gsg}; $D_H(2.33)/r_{\rm d}=8.99\pm0.19$ (for combined Ly-$\alpha$ data)~\cite{duMasdesBourboux:2020pck}; $t_{\rm u}=13.50\pm0.15\,\rm Gyr$ (systematic uncertainties are not included)~\cite{Valcin:2021jcg}; $10^2\omega_{\rm b}^{\rm LUNA}=2.233\pm0.036$~\cite{Mossa:2020gjc}. In addition, we show vertical blue and brown bands for $H_{\rm 0}^{\rm TRGB}=69.8\pm0.8~{\rm km\, s^{-1}\, Mpc^{-1}}$ ~\cite{Freedman:2019cchp} and $10^2\omega_{\rm b}^{\rm PCUV21}=2.195\pm0.022$ ~\cite{Pitrou:2020etk}, respectively. Note that the disjoint contours (around the horizontal $z_\dagger=2.33$ dashed line) of $\Lambda_{\rm s}$CDM for $D_H(2.33)/r_{\rm d}$ are as expected since $\Lambda_{\rm s}$ at $z=2.33$ is negative for $z_\dagger<2.33$ and positive for $z_\dagger>2.33$.}
    \label{fig:zdagger2d}
\end{figure*}

As we discussed in the previous section, the $\Lambda_{\rm s}$CDM model generically finds better fit to the data compared to the $\Lambda$CDM model. Since the inclusion of the $M_B$ prior and/or the Ly-$\alpha$ data in the data sets causes $\Lambda_{\rm s}$CDM to perform even better compared to $\Lambda$CDM, we expect it to resolve, or at least relax, the $M_B$ and the closely related $H_0$ tensions along with the Ly-$\alpha$ discrepancy. In~\cref{fig:zdagger2d}, we show the two-dimensional marginalized probability posteriors of $z_{\dagger}$ versus $H_0$, $M_B$, $S_8$, $D_H(2.33)/r_{\rm d}$ (viz., the $D_H/r_{\rm d}$ at $z_{\rm eff}=2.33$ relevant to the Ly-$\alpha$ measurements), $t_0$, and $\omega_{\rm b}$ in the $\Lambda_{\rm s}$CDM model from various combinations of the data sets and in~\cref{tab:tensionswithoutMB} we quantify the concordances/discordances between the $\Lambda$CDM and $\Lambda_{\rm s}$CDM models and the theoretical/direct observational estimations, viz., $H_{\rm 0}^{\rm R21}=73.04\pm1.04~{\rm km\, s^{-1}\, Mpc^{-1}}$~\cite{Riess:2021jrx} and $H_{\rm 0}^{\rm TRGB}=69.8\pm0.8~{\rm km\, s^{-1}\, Mpc^{-1}}$~\cite{Freedman:2019cchp}; $M_B=-19.244\pm 0.037\,\rm mag$ (SH0ES)~\cite{Camarena:2021dcvm}; $S_8=0.766^{+0.020}_{-0.014}$ (WL+GC KiDS-1000 $3\times2$pt)~\cite{Heymans:2020gsg}; $D_H(2.33)/r_{\rm d}=8.99\pm0.19$ (for the combined Ly-$\alpha$ data)~\cite{duMasdesBourboux:2020pck}; $t_{0}=13.50\pm0.27\,\rm Gyr$~\cite{Valcin:2021jcg}; $10^2\omega_{\rm b}^{\rm LUNA}=2.233\pm0.036$~\cite{Mossa:2020gjc} and $10^2\omega_{\rm b}^{\rm PCUV21}=2.195\pm0.022$~\cite{Pitrou:2020etk}. In what follows, we discuss these tensions and how they are relaxed within the $\Lambda_{\rm s}$CDM model compared to the $\Lambda$CDM model.

\subsection{$H_0$ discrepancy}
\label{subsec:h0}
The most statistically significant and pressing tension is in $H_0$, between its direct local distance ladder measurements and its estimations from the CMB data assuming the standard $\Lambda$CDM model. More precisely, there exists approximately $5\sigma$ tension between its $\Lambda$CDM value $H_0 = 67.27\pm 0.60{\rm \,km\, s^{-1}\, Mpc^{-1}}$ (68\% CL)~\cite{Planck:2018vyg} inferred from \textit{Planck} 2018 and the SH0ES measurement $H_0 = 73.04\pm 1.04{\rm \,km\, s^{-1}\, Mpc^{-1}}$ (68\% CL)~\cite{Riess:2021jrx} based on the SNIa calibrated by Cepheid variables. This tension reduces to a mild discrepancy of $2.5\sigma$ (or $2.7\sigma$) when the TRGB measurement $H_0=69.8\pm0.8{\rm \,km\, s^{-1}\, Mpc^{-1}}$ (68\% CL)~\cite{Freedman:2019cchp} (or $H_0=69.8\pm0.6{\rm \,km\, s^{-1}\, Mpc^{-1}}$~\cite{Freedman:2021ahq}, at 68\% CL), which is $2.5\sigma$ (or $2.7\sigma$) tension with the SH0ES measurement, is considered. There are in fact plenty of other independent (at least partially) and direct $H_0$ measurements relying on different methods and astrophysical observations~\cite{Huang:2019yhh,Blakeslee:2021rqi,Garnavich:2022hef,deJaeger:2022lit,Pesce:2020xfe,Kourkchi:2020iyz,FernandezArenas:2017dux,Birrer:2022chj,Moresco:2022phi} (see Ref.~\cite{Abdalla:2022yfr} for a further list of direct $H_0$ measurements). Almost all of these are statistically consistent with the latest SH0ES measurement, but their error percentages are large compared to those of SH0ES and TRGB measurements. Among these alternatives, the time-delay related measurements (based on Ref.~\cite{Refsdal:1964nw}) stand out as they are independent of the distance ladders on which SH0ES and TRGB $H_0$ measurements rely, and as they can provide error percentages comparable to those of SH0ES and TRGB $H_0$ measurements; namely, $H_0 = 73.3^{+1.7}_{-1.8}$ of H0LiCOW~\cite{Wong:2019kwg} and $H_0=74.2\pm1.6$ of TDCOSMO~\cite{Millon:2019slk}---though, their low error percentages require assumptions on the mass density profiles of the deflector galaxies to break the so-called mass-sheet degeneracy, leaving the method prone to systematics; relaxed assumptions on the mass density profile result in looser constraints, e.g., the TDCOSMO results $H_0=73.3\pm5.8$~\cite{Birrer:2020tax}, and the recent $H_0=77.1^{+7.3}_{-7.1}$~\cite{Shajib:2023uig} from the analysis of a single system.\footnote{\label{footnote5}Note that, since some of these $H_0$ measurements based on the alternative methods are independent of the calibration of supernova absolute magnitudes, deciding to use an $H_0$ prior instead of an $M_B$ prior allows the usage of a wider variety of measurements related to the present-day expansion of the Universe. 
We use the SH0ES measurement due to its robustness and reliability, and we chose their $M_B$ estimation as our prior instead of $H_0$ since $M_B$ is the more direct estimation whereas their inference of $H_0$ require some minimal assumptions related to low-redshift cosmography (See \cref{subsec:mb}). 
It is possible that a cosmological model agrees with one of these quantities ($H_0$, $M_B$) without agreeing with both~\cite{Alestas:2020zol,Alestas:2021luu,Marra:2021fvf,Camarena:2019moy,Camarena:2021jlr,Efstathiou:2021ocp,Benevento:2020fev,Cai:2021weh,Greene:2021shv,Benisty:2022psx}. Thus, if one decides to use an $H_0$ prior from a certain measurement, they should also compare their results against independent $M_B$ measurements---if the used $H_0$ prior is inferred from an $M_B$ value, a comparison with that value is also required. Similarly, if one decides to use an $M_B$ prior from a certain measurement, they should also compare their results against independent $H_0$ measurements---but not necessarily against the $H_0$ value inferred from that $M_B$ prior. Since almost all direct measurements of $H_0$ independent of the SH0ES measurement are statistically consistent with the SH0ES value, instead of discussing other independent measurements, we compare our results again with the SH0ES $H_0$ estimation. Also, due to the discrepancy between the SH0ES and TRGB measurements (note, however, the recent work Ref.~\cite{Anderson:2023aga}), we include comparisons of our results with the TRGB $H_0$ measurement.}

\begin{table*}[ht!]
\caption{Concordance/discordance between the $\Lambda$CDM/$\Lambda_{\rm s}$CDM models and the theoretical/direct observational estimations, viz., $H_{\rm 0}^{\rm R21}=73.04\pm1.04~{\rm km\, s^{-1}\, Mpc^{-1}}$~\cite{Riess:2021jrx} and $H_{\rm 0}^{\rm TRGB}=69.8\pm0.8~{\rm km\, s^{-1}\, Mpc^{-1}}$~\cite{Freedman:2019cchp}; $M_B=-19.244\pm 0.037\,\rm mag$ (SH0ES)~\cite{Camarena:2021dcvm}; $S_8=0.766^{+0.020}_{-0.014}$ (WL+GC KiDS-1000 $3\times2$pt)~\cite{Heymans:2020gsg}; $D_H(2.33)/r_{\rm d}=8.99\pm0.19$ (for the combined Ly-$\alpha$ data)~\cite{duMasdesBourboux:2020pck}; $t_{0}=13.50\pm0.15\,\rm Gyr$ (systematic uncertainties are not included)~\cite{Valcin:2021jcg}; $10^2\omega_{\rm b}^{\rm LUNA}=2.233\pm0.036$~\cite{Mossa:2020gjc} and $10^2\omega_{\rm b}^{\rm PCUV21}=2.195\pm0.022$~\cite{Pitrou:2020etk}.  The results marked with ${}^*$ should be interpreted with caution since the SH0ES $M_B$ prior is not fully consistent with the TRGB measurements.}
\label{tab:tensionswithoutMB}
	\scalebox{0.82}{
\begin{centering}
	  \begin{tabular}{l|cc|cc|cc|cc|cc|cc}
  	\hline
    \toprule
    \multicolumn{1}{l}{\textbf{Data set}} & \multicolumn{2}{c}{\textbf{CMB+Pan}} & \multicolumn{2}{c}{\textbf{CMB+Pan+Ly-$\bm{\alpha}$}}& \multicolumn{2}{c}{\textbf{CMB+Pan+BAO}} & \multicolumn{2}{c}{\textbf{CMB+Pan+$\bm{M_B}$}} & \multicolumn{2}{c}{\textbf{CMB+Pan+Ly-$\bm{\alpha}$+$\bm{M_B}$}}& \multicolumn{2}{c}{\textbf{CMB+Pan+BAO+$\bm{M_B}$}}  \\  \hline
      & \textbf{{$\bm{\Lambda}$CDM}} & \textbf{$\bm{\Lambda}_{\textbf{s}}$CDM} & \textbf{{$\bm{\Lambda}$CDM}}  & \textbf{$\bm{\Lambda}_{\textbf{s}}$CDM}  & \textbf{{$\bm{\Lambda}$CDM}}  & \textbf{$\bm{\Lambda}_{\textbf{s}}$CDM} & \textbf{{$\bm{\Lambda}$CDM}}  & \textbf{$\bm{\Lambda}_{\textbf{s}}$CDM} & \hphantom{,,,}\textbf{{$\bm{\Lambda}$CDM}}\hphantom{,,}  & \hphantom{,,}\textbf{$\bm{\Lambda}_{\textbf{s}}$CDM}\hphantom{,,,} & \hphantom{,,,}\textbf{{$\bm{\Lambda}$CDM}}\hphantom{,,}  & \hphantom{,,}\textbf{$\bm{\Lambda}_{\textbf{s}}$CDM}\hphantom{,,,}\\ 
      \hline
    $H_{\rm 0}^{\rm R21}$ & $4.7\sigma$ & $2.2\sigma$ & $4.8\sigma$ & $2.0\sigma$ & $4.8\sigma$ & $3.7\sigma$ & $4.1\sigma$ & $0.4\sigma$ & $4.4\sigma$ & $0.7\sigma$ & $4.3\sigma$ & $3.1\sigma$   
\\

$H_{\rm 0}^{\rm TRGB}$ & $2.3\sigma$ & $0.1\sigma$ & $2.8\sigma$ & $0.3\sigma$ & $2.4\sigma$ & $1.1\sigma$ & $1.6\sigma$* & $2.0\sigma$* & $1.8\sigma$* & $1.6\sigma$* & $1.7\sigma$*  & $0.3\sigma$* 
\\

$M_{B}$  &   $4.5\sigma$ & $2.5\sigma$ & $4.5\sigma$ & $2.3\sigma$ & $4.5\sigma$ & $3.6\sigma$ &$3.9\sigma$ & $1.0\sigma$ & $4.1\sigma$ & $1.2\sigma$ & $4.0\sigma$ & $3.1\sigma$  
\\

$S_{8}$ &   $2.9\sigma$ & $1.9\sigma$ & $3.0\sigma$ & $1.7\sigma$ & $2.9\sigma$ & $2.8\sigma$ & $2.3\sigma$ & $0.9\sigma$ & $2.5\sigma$ & $1.0\sigma$ & $2.4\sigma$ & $2.4\sigma$  
\\ 

$D_H(2.33)/r_{\rm d}$  & $2.0\sigma$ & $0.2\sigma$ & $1.9\sigma$ & $0.1\sigma$ & $1.9\sigma$ & $1.1\sigma$ & $1.9\sigma$ & $1.2\sigma$ & $1.8\sigma$ & $1.2\sigma$ & $1.9\sigma$ & $0.1\sigma$ 
   \\

       $t_{\rm u}$ & $1.9\sigma$ & $1.0\sigma$ & $1.9\sigma$ & $0.8\sigma$ & $1.9\sigma$ & $1.4\sigma$ & $1.7\sigma$ & $0.3\sigma$ & $1.7\sigma$ & $0.4\sigma$ & $1.7\sigma$ & $1.1\sigma$ 
    \\

$\omega_{\rm b }^{\rm PCUV21}$ & $1.7\sigma$ & $1.8\sigma$ & $1.8\sigma$ & $1.7\sigma$ & $1.8\sigma$ & $1.5\sigma$ & $2.3\sigma$ & $2.0\sigma$ & $2.3\sigma$ & $2.0\sigma$ & $2.3\sigma$ & $1.8\sigma$ 
    \\

$\omega_{\rm b}^{\rm LUNA}$ & $0.2\sigma$ & $0.2\sigma$ & $0.2\sigma$ & $0.2\sigma$ & $0.2\sigma$ & $0.1\sigma$ & $0.6\sigma$ & $0.4\sigma$ & $0.5\sigma$ & $0.4\sigma$ & $0.6\sigma$ & $0.2\sigma$

\\
 
\hline
  \bottomrule
    \hline
  \end{tabular}
  \end{centering}
 }
\end{table*}

Consistency with CMB requires that the presence of a sign-switching cosmological constant instead of a regular one always results in a higher $H_0$ value inversely correlated with $z_\dagger$; this behavior is visible in the leftmost panels of~\cref{fig:zdagger2d} (for a detailed explanation, see Ref.~\cite{Akarsu:2021fol} and particularly Figs.~2~and~8 therein). Hence, the higher $H_0$ values of $\Lambda_{\rm s}$CDM compared to $\Lambda$CDM
in~\cref{tab:withMB,tab:withoutMB} for all six data sets are no surprise; and, as seen from~\cref{tab:tensionswithoutMB}, for all six data sets, $\Lambda_{\rm s}$CDM is in better agreement with the SH0ES $H_0$ measurement (so also with the H0LiCOW and TDCOSMO $H_0$ measurements) and is compatible (i.e., discrepancy is less than $2\sigma$) with the TRGB $H_0$ measurement having at most a $2\sigma$ discrepancy in the case of CMB+Pan+$M_B$ and only because it predicts too high of an $H_0$ value compared to TRGB. Also, note that, the $M_B$ prior we use is that of SH0ES and this must be kept in mind when the constraints on $H_0$ in its presence are compared with the TRGB $H_0$ measurement. As seen from~\cref{fig:zdagger1d}, the $M_B$ prior clearly prefers a sign switch at lower redshifts $1.6\lesssim z_\dagger\lesssim2$; thus, when the $M_B$ prior is included in the data sets, the estimations of $H_0$ within $\Lambda_{\rm s}$CDM are higher compared to the same data sets without the $M_B$ prior due to the inverse correlation of $z_\dagger$ and $H_0$. This results in removal of the SH0ES $H_0$ tension for the CMB+Pan+$M_B$ and CMB+Pan+Ly-$\alpha$+$M_B$ cases. In fact, for these cases, $H_0$ predictions of $\Lambda_{\rm s}$CDM are high enough that they start introducing mild discrepancies with the TRGB $H_0$ measurement. In contrast, addition of the $M_B$ prior makes little to no difference for the $\Lambda$CDM model in amelioration of the SH0ES $H_0$ tension.

However, for the CMB+Pan+BAO cases with or without the $M_B$ prior, the preference of high $z_\dagger$ values by the lower redshift BAO hinders the success of $\Lambda_{\rm s}$CDM in ameliorating the discrepancies displayed in~\cref{tab:tensionswithoutMB} including the SH0ES $H_0$ tension---the opposition of the low-redshift BAO data (viz., consensus Galaxy BAO from $z_{\rm eff} = 0.38,\, 0.51,\, 0.61$) to lower $z_\dagger$ values and hence to higher $H_0$ values was discussed in Ref.~\cite{Akarsu:2021fol}. Closely related to this, the $H_0$ tension within $\Lambda$CDM not only exists between the local $H_0$ measurements and the inference of $H_0$ from CMB, but also between the local $H_0$ measurements and the BAO data set (combined with a BBN prior) when CMB data set is not used~\cite{Cuceu:2019for,Schoneberg:2022ggi,Schoneberg:2019wmt,Alam:2020sor,Aubourg:2014yra}. It is worth mentioning that this tension with the BAO does not originate from any particular BAO measurement, rather, it is due to the different degeneracy directions of the constraints from BAO at high redshifts ($z>1$) and galaxy BAO at low redshifts ($z<1$) in the $\Omega_{\rm m}-H_0$ plane; see Refs.~\cite{Cuceu:2019for,Schoneberg:2022ggi,Schoneberg:2019wmt,Alam:2020sor,Aubourg:2014yra}, for instance, Fig.~5 of Ref.~\cite{Alam:2020sor}. Here, $\Omega_{\rm m}\equiv8\pi G\rho_{\rm m0}/(3H_0^2)$ is the present-day ($z=0$) matter density parameter with $\rho_{\rm m0}$ being the present-day matter energy density. Note that the CMB agrees very well with the BBN constraints used in Refs.~\cite{Aubourg:2014yra,Alam:2020sor} and the degeneracy direction of the constraints from high-redshift BAO data agrees with that of the CMB within both $\Lambda$CDM and $\Lambda_{\rm s}$CDM with contours for $\Lambda_{\rm s}$CDM being shifted to higher $H_0$ values as indicated by the analyses in Ref.~\cite{Akarsu:2021fol}. While $\Lambda_{\rm s}$CDM is able to address the $H_0$ tension with the CMB, the different degeneracy direction of the galaxy BAO will still introduce problems. That is because, since $z_\dagger>1$ is satisfied for any reasonable expansion history within $\Lambda_{\rm s}$CDM (see Fig.~5 and the relevant discussion in Ref.~\cite{Akarsu:2021fol}), both models are equivalent for the whole range of the galaxy BAO and would yield the same contours in a BBN+galaxy BAO analysis as in Refs.~\cite{Aubourg:2014yra,Alam:2020sor} and the shift to higher $H_0$ values within $\Lambda_{\rm s}$CDM in the $\Omega_{\rm m}-H_0$ plane by itself is not adequate for a full resolution of the BAO-based $H_0$ tension but only an amelioration. This inadequacy manifests itself in the impairing of $\Lambda_{\rm s}$CDM in alleviating the $H_0$ tension when the full BAO data is included in our analyses as can be seen from, in addition to~\cref{tab:tensionswithoutMB}, the blue contours in the $H_0$ panels of~\cref{fig:zdagger2d}, and particularly clearly by comparing the rightmost panels of~\cref{fig:h3D0mb,fig:3dh0s8} showing the analyses including the full BAO data to the rest of their panels showing the cases without the low-redshift BAO data.

Another point of interest is the relation of the $H_0$ tension with the $M_B$ and $S_8$ tensions (the two other prominent discrepancies of $\Lambda$CDM) within $\Lambda_{\rm s}$CDM. The two-dimensional marginalized posterior distributions of $M_B$ versus $H_0$, and $S_8$ versus $H_0$ are given in~\cref{fig:h3D0mb,fig:3dh0s8}, respectively, both color coded by $z_\dagger$. These two figures have some striking common features: (i) there is a strong correlation with $H_0$ and the other two parameters; (ii) lower $z_\dagger$ values are preferred by all three discrepancies; (iii) the presence of the full BAO data set hinders the alleviation of the tensions by preferring higher $z_\dagger$ values that blur the phenomenological differences between the two models; (iv) the presence of the $M_B$ prior results in better alleviation of the tensions and greater differentiation between the two models by preferring lower $z_\dagger$ values. The correlation is particularly pronounced between $M_B$ and $H_0$; the analyses without the full BAO data yield a correlation of $\sim0.99$ and the ones with the full BAO data yield $\sim0.96$.

\subsection{$M_B$ discrepancy} 
\label{subsec:mb}

\begin{figure*}[ht!]
    \centering
    \includegraphics[width=5.9cm]{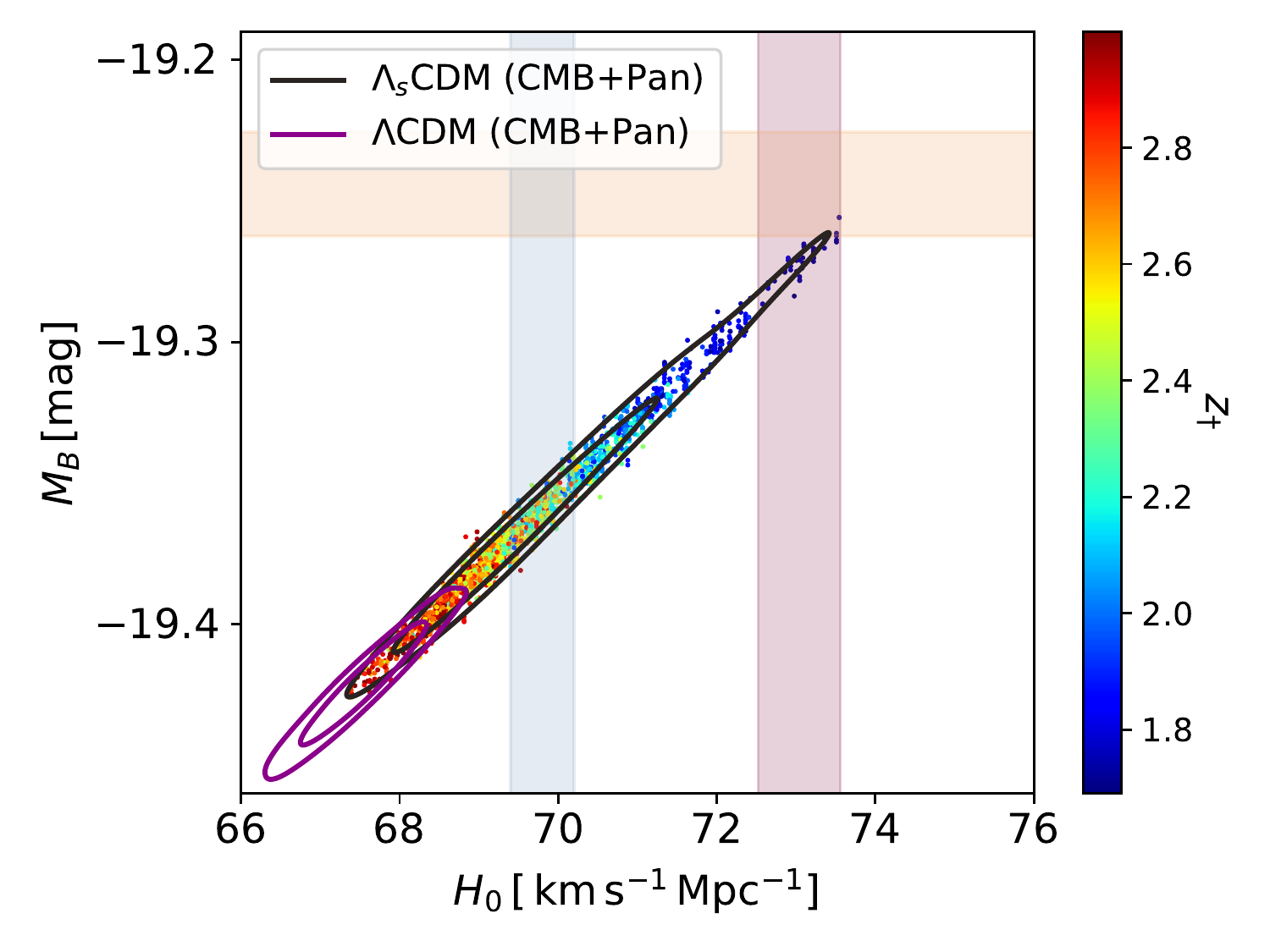}
    \includegraphics[width=5.9cm]{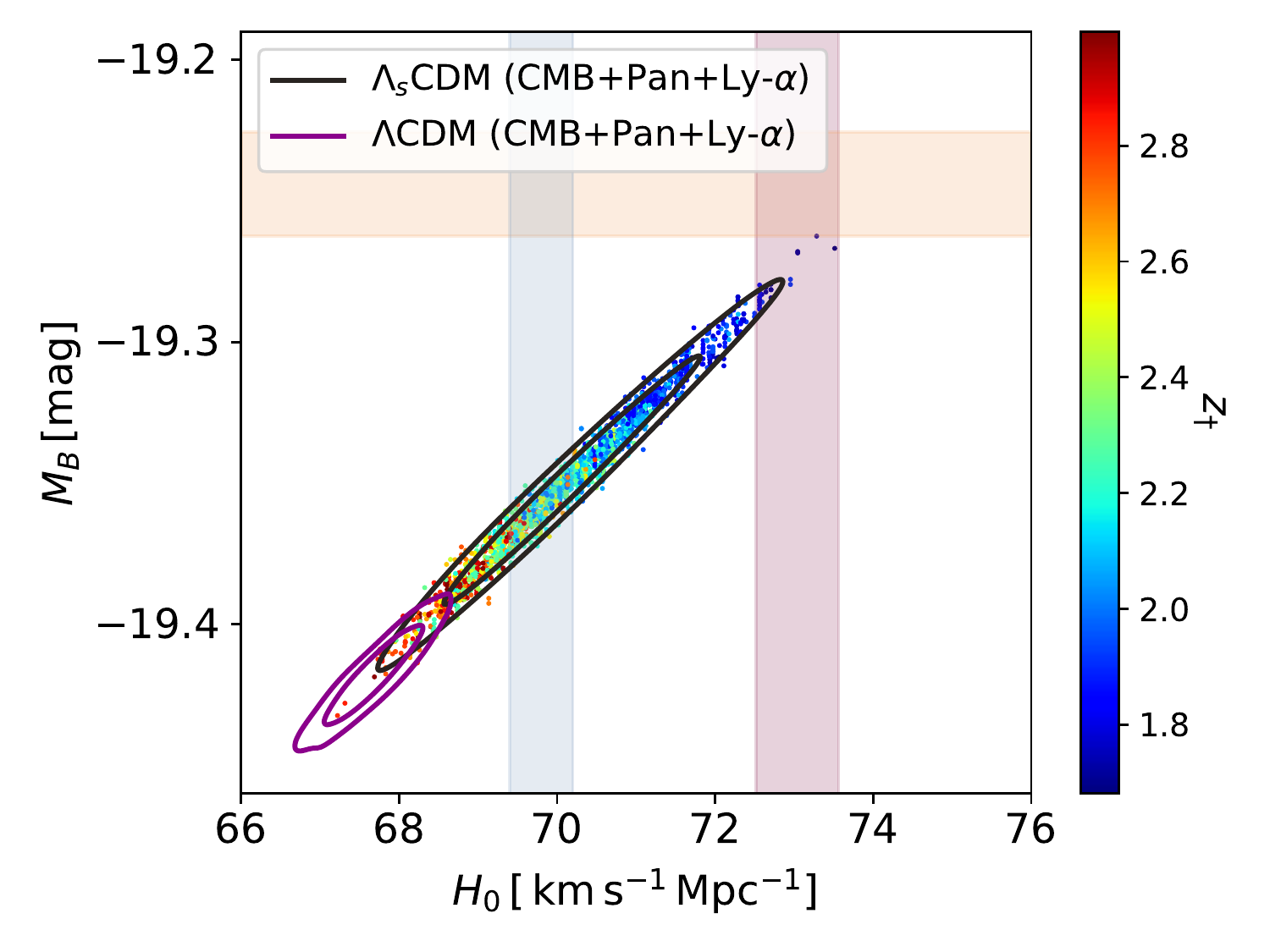}
    \includegraphics[width=5.9cm]{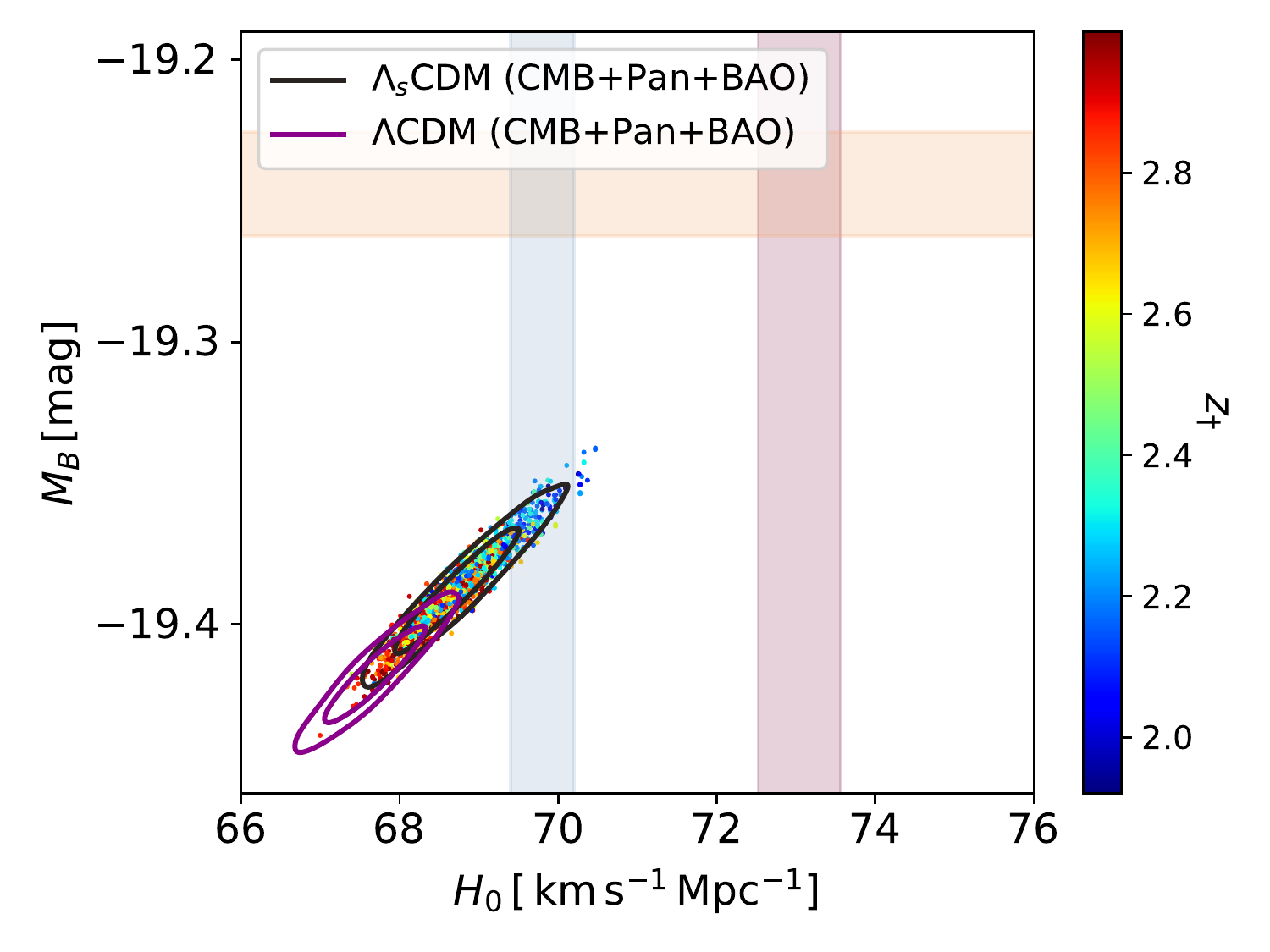}
    \includegraphics[width=5.9cm]{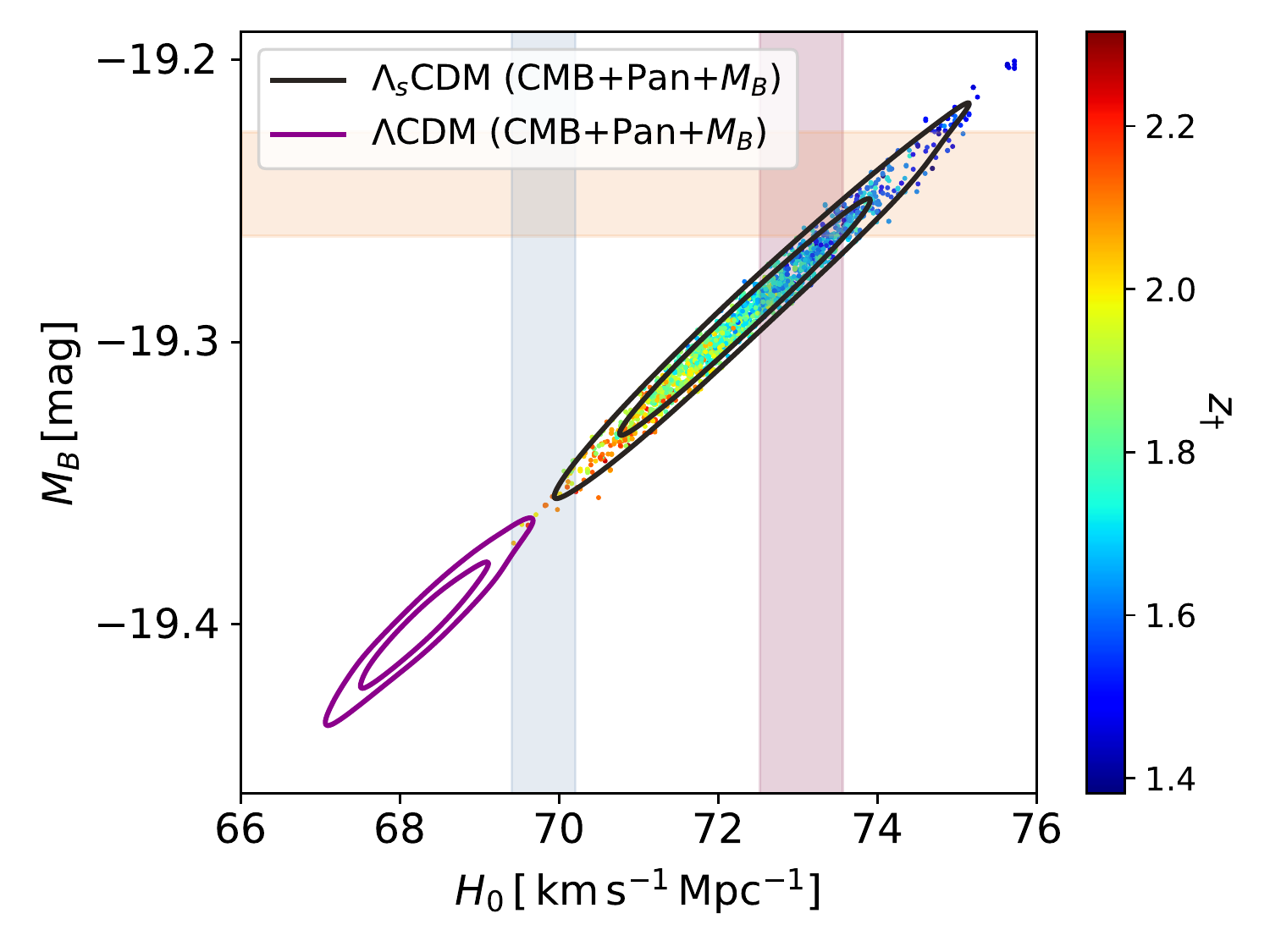}
    \includegraphics[width=5.9cm]{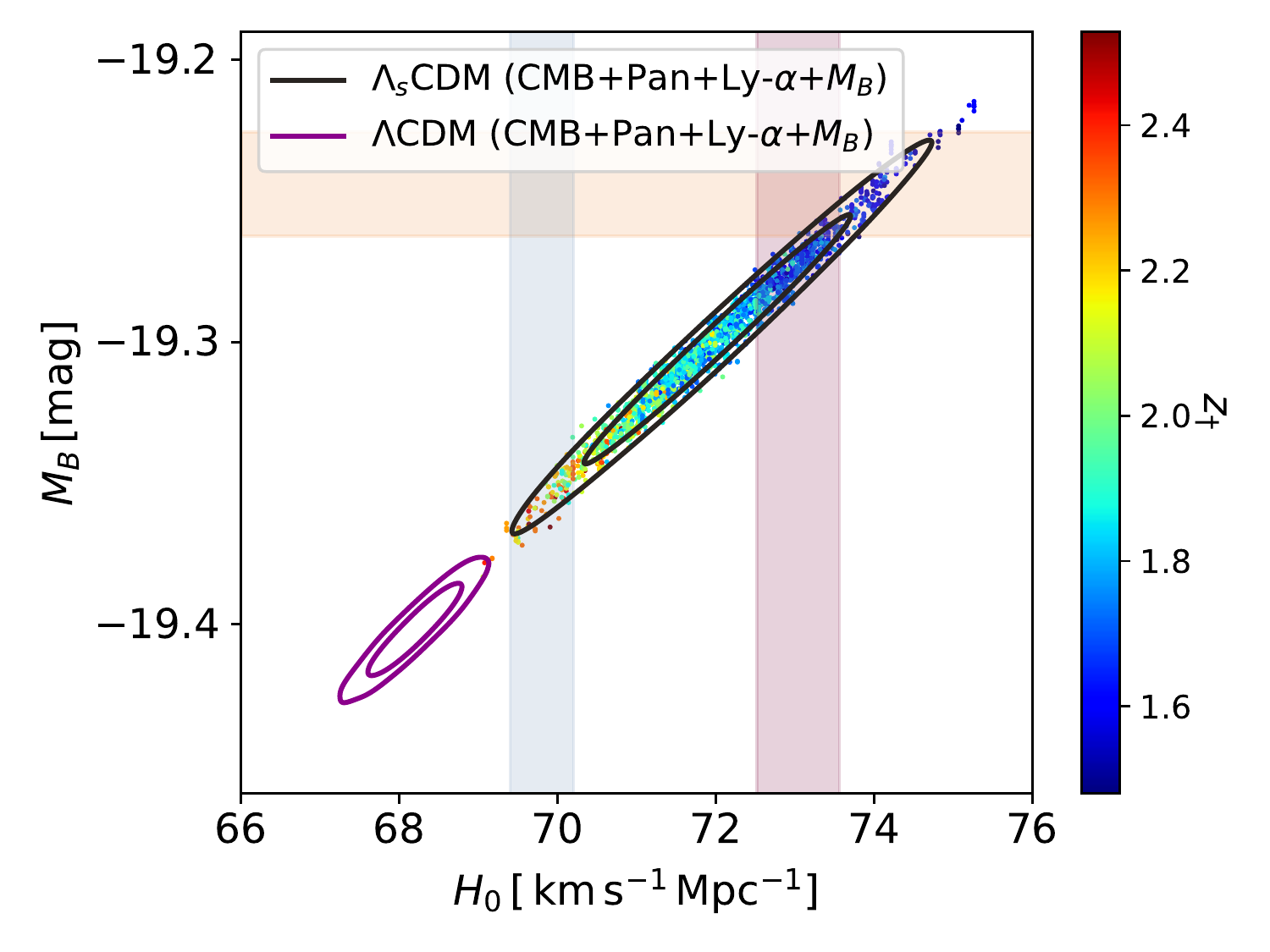}
    \includegraphics[width=5.9cm]{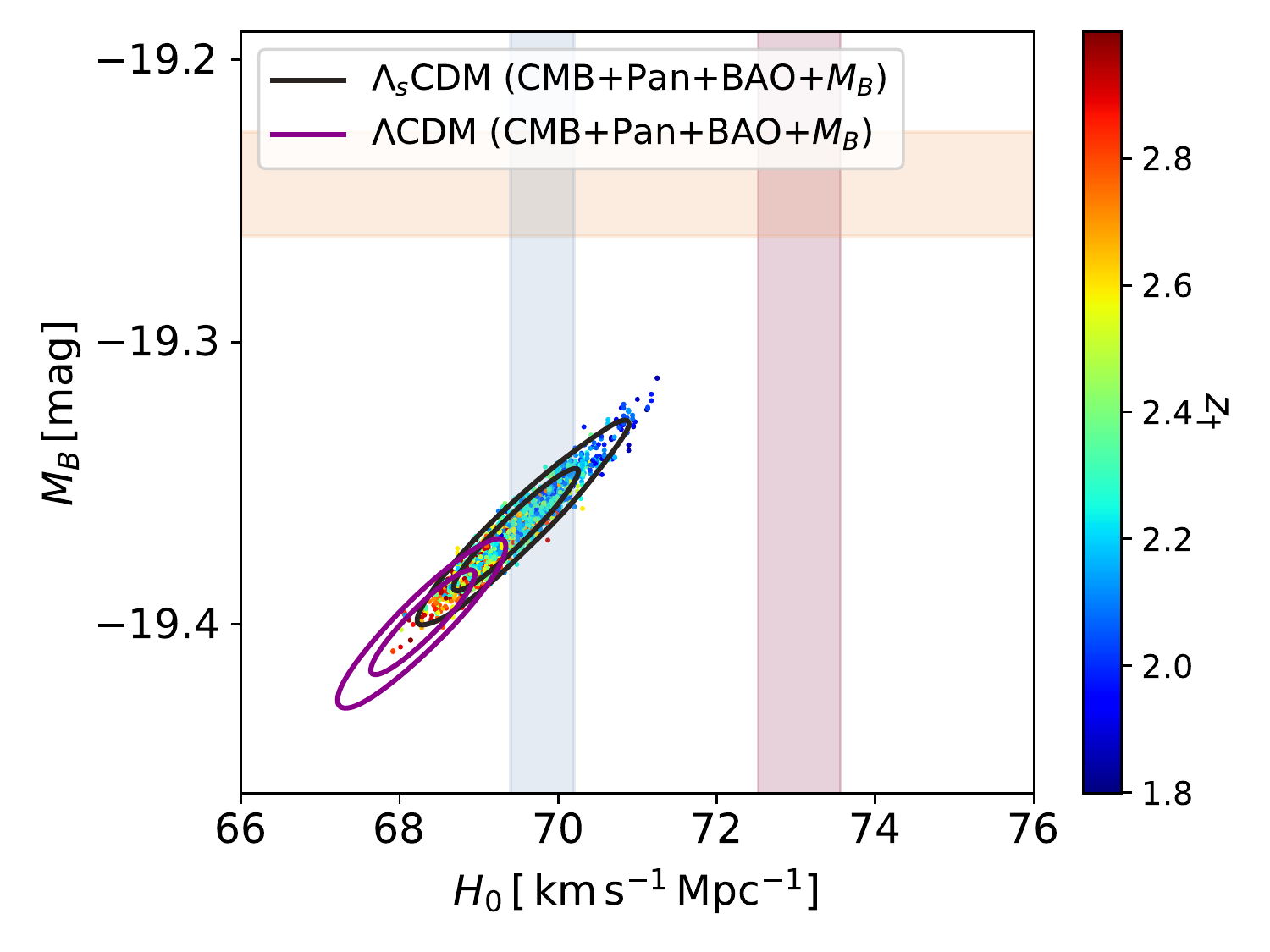}
    \caption{Two-dimensional marginalized posterior distributions (68\% and 95\% CLs) in the $M_B$-$H_0$ plane for the $\Lambda_{\rm s}$CDM (color coded by $z_\dagger$) and $\Lambda_{\rm}$CDM for different data combinations. We overlay $1\sigma$ bands for the local measurements $H_{0}^{\rm R21}=73.04\pm1.04$ km $\rm{s^{-1}}$ $\rm{Mpc^{-1}}$~\cite{Riess:2021jrx}, $H_{\rm 0}^{\rm TRGB}=69.8\pm0.8~{\rm km\, s^{-1}\, Mpc^{-1}}$ ~\cite{Freedman:2019cchp}, and $M_B=-19.244\pm 0.037\,\rm mag$ (SH0ES)~\cite{Camarena:2021dcvm}.  The larger $z_\dagger$ is, the closer the $\Lambda_{\rm s}$CDM model is to the standard $\Lambda$CDM model.}
    \label{fig:h3D0mb}
\end{figure*}

The $M_B$ tension is closely related to the $H_0$ tension~\cite{Abdalla:2022yfr,Camarena:2019rmj,Camarena:2021dcvm} (see also Refs.~\cite{Alestas:2020zol,Marra:2021fvf,Alestas:2021luu}). The local $H_0$ measurements rely on observations of astrophysical objects that extend into redshifts where the Hubble flow dominates over peculiar velocities. Particularly, the two most quoted measurements of $H_0$, namely the TRGB and SH0ES values, are based on the calibration (using Cepheid variables for SH0ES, and tip of the red giant branch for TRGB) of the SNIa absolute magnitude. From the calibrated absolute magnitude, using the apparent magnitudes of SNIa that extend up to $z=0.15$, the value of $H_0$ is then inferred by assuming a $\Lambda$CDM-like cosmography for which the distance modulus $\mu(z)$ depends only on $H_0$. This $H_0$ value, as discussed in the previous subsection, is in substantial tension with the one inferred from the CMB assuming $\Lambda$CDM cosmology. This implies a serious inconsistency between the CMB and local measurements. This inconsistency is also present if, instead of propagating the local calibration of $M_B$ to an $H_0$ value, one propagates the constraints on $r_{\rm d}$ from CMB to constraints on $M_B$ through the inverse distance ladder as in Ref.~\cite{Camarena:2019rmj} utilizing BAO measurements. The CMB calibration yields $M_B^{\rm CMB}=-19.401\pm0.027\,\rm mag$~\cite{Camarena:2019rmj} while the SH0ES calibration yields $M_B=-19.2435\pm 0.0373\,\rm mag$~\cite{Camarena:2021dcvm}. Alternatively, instead of comparing the local $H_0$ value (inferred from the Cepheid or TRGB calibrated $M_B$ value) with the $H_0$ value inferred by constraining the parameters of a model (most often making use of the CMB), one can calculate the distance modulus for the constrained model which can be used to infer the SNIa absolute magnitude ($M_B$) from their apparent magnitudes, and then directly compare this $M_B$ value with the one calibrated using Cepheid variables or TRGB. Within $\Lambda$CDM, where the cosmographic assumptions used in the inference of the local $H_0$ measurements are accurate, the $M_B$ and $H_0$ tensions are almost equivalent. It is important to note that, for an arbitrary model, the direct comparison of the $M_B$ values instead of $H_0$ is advantageous as this method is not prone to finding fake resolutions of the $H_0$ tension as discussed in Refs.~\cite{Camarena:2019moy,Camarena:2021jlr,Efstathiou:2021ocp,Benevento:2020fev,Cai:2021weh,Greene:2021shv,Benisty:2022psx}. 

Since the SH0ES $H_0$ measurement is based on a $\Lambda$CDM-like cosmography to infer $H_0$ from the $M_B$ value found by the calibration of SNIa up to $z=0.15$ by Cepheid variables from $z<0.01$, within $\Lambda_{\rm s}$CDM, for which the functional form of the cosmographic parameters are exactly those of $\Lambda$CDM for $z<z_\dagger$, the resolution of the $H_0$ and $M_B$ tensions are almost equivalent just as it is within $\Lambda$CDM (note that the constraint on $z_\dagger$ is well above the redshift range of the SNIa data used by the local measurements~\cite{Freedman:2019cchp,Riess:2021jrx}, and is greater than most of the available SNIa sample~\cite{scolnic:2018dm}). In other words, the $\Lambda_{\rm s}$CDM model respects the internal consistency of the methodology used by the SH0ES collaboration. Figure \ref{fig:h3D0mb} and the almost perfect correlation between $H_0$ and $M_B$ within $\Lambda_{\rm s}$CDM (see the end of~\cref{subsec:h0}) clearly illustrate this feature; also compare the first and third rows of~\cref{tab:tensionswithoutMB}. As a result, the discussion for the $M_B$ tension follows the $H_0$ tension discussion in the previous subsection very closely. For all six data sets, $\Lambda_{\rm s}$CDM yields higher (fainter) $M_B$ values compared to $\Lambda$CDM as shown in~\cref{tab:withMB,tab:withoutMB}. Higher $M_B$ values are also what the local calibrations find, and this is reflected in~\cref{tab:tensionswithoutMB} where, compared to $\Lambda$CDM, $\Lambda_{\rm s}$CDM is always in less tension. As in the case of the $H_0$ tension, the inclusion of the $M_B$ prior reduces the $M_B$ tension significantly for $\Lambda_{\rm s}$CDM for all three data compilations (down to $1\sigma$ for the CMB+Pan+$M_B$ case) and the inclusion of the full BAO in the compilation has a hindering effect. Note that, in contrast, the addition of the $M_B$ prior makes little to no difference for the $\Lambda$CDM model in amelioration of the $M_B$ tension.

\subsection{$S_8$ discrepancy}
\label{subsec:s8}

\begin{figure*}[htbp]\centering
\includegraphics[width=5.8cm]{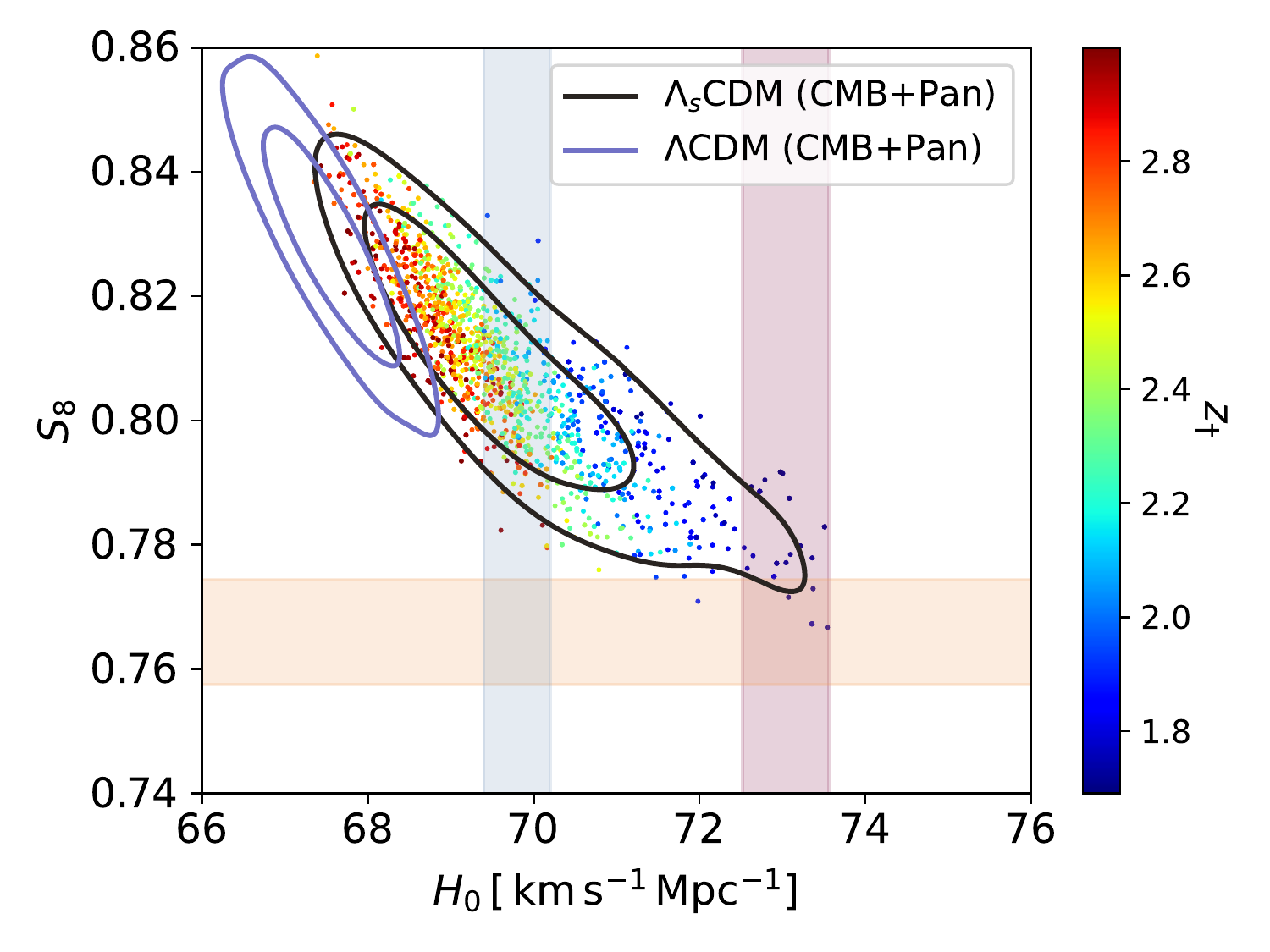}
\includegraphics[width=5.8cm]{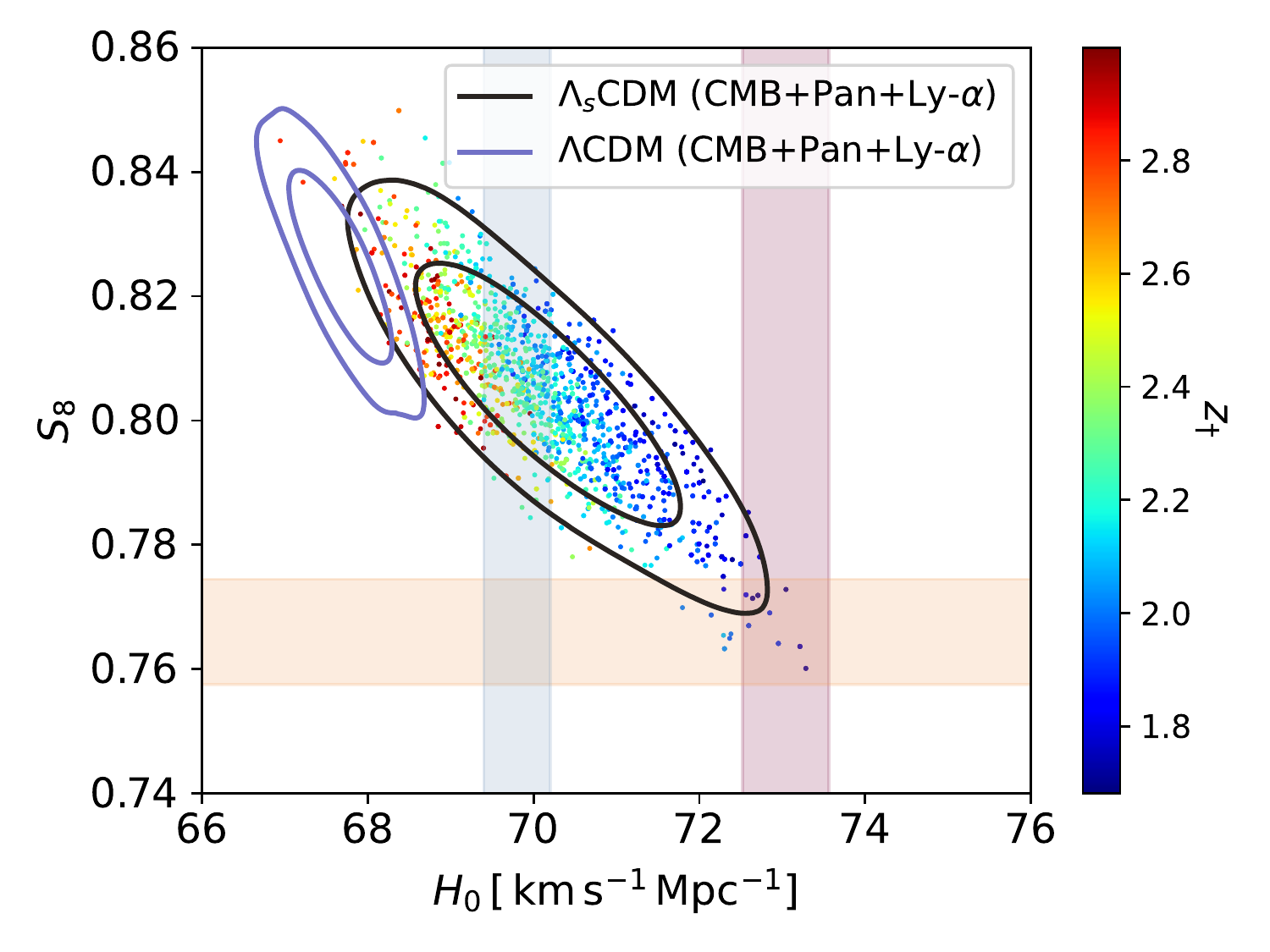}
\includegraphics[width=5.8cm]{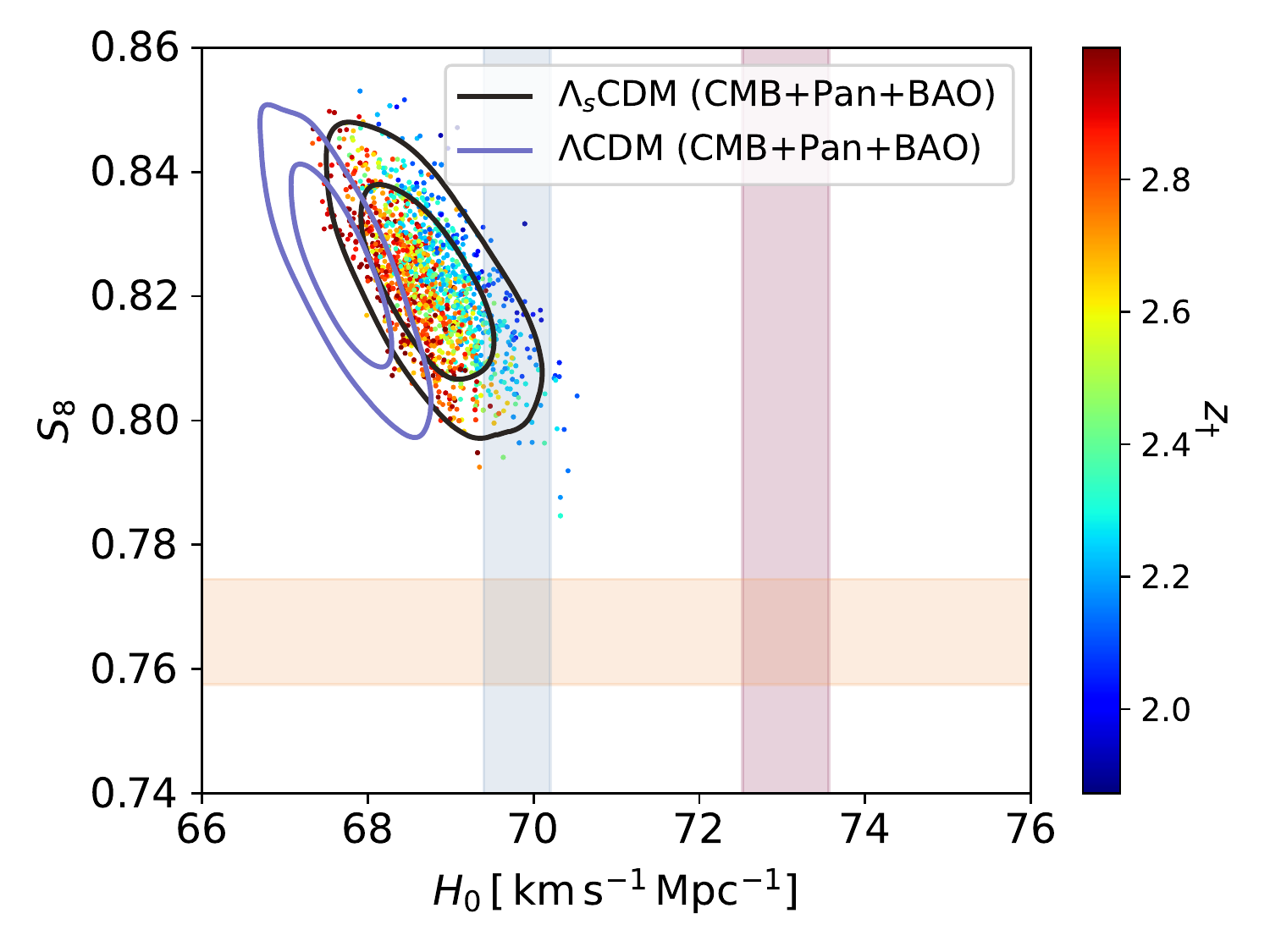}
\includegraphics[width=5.8cm]{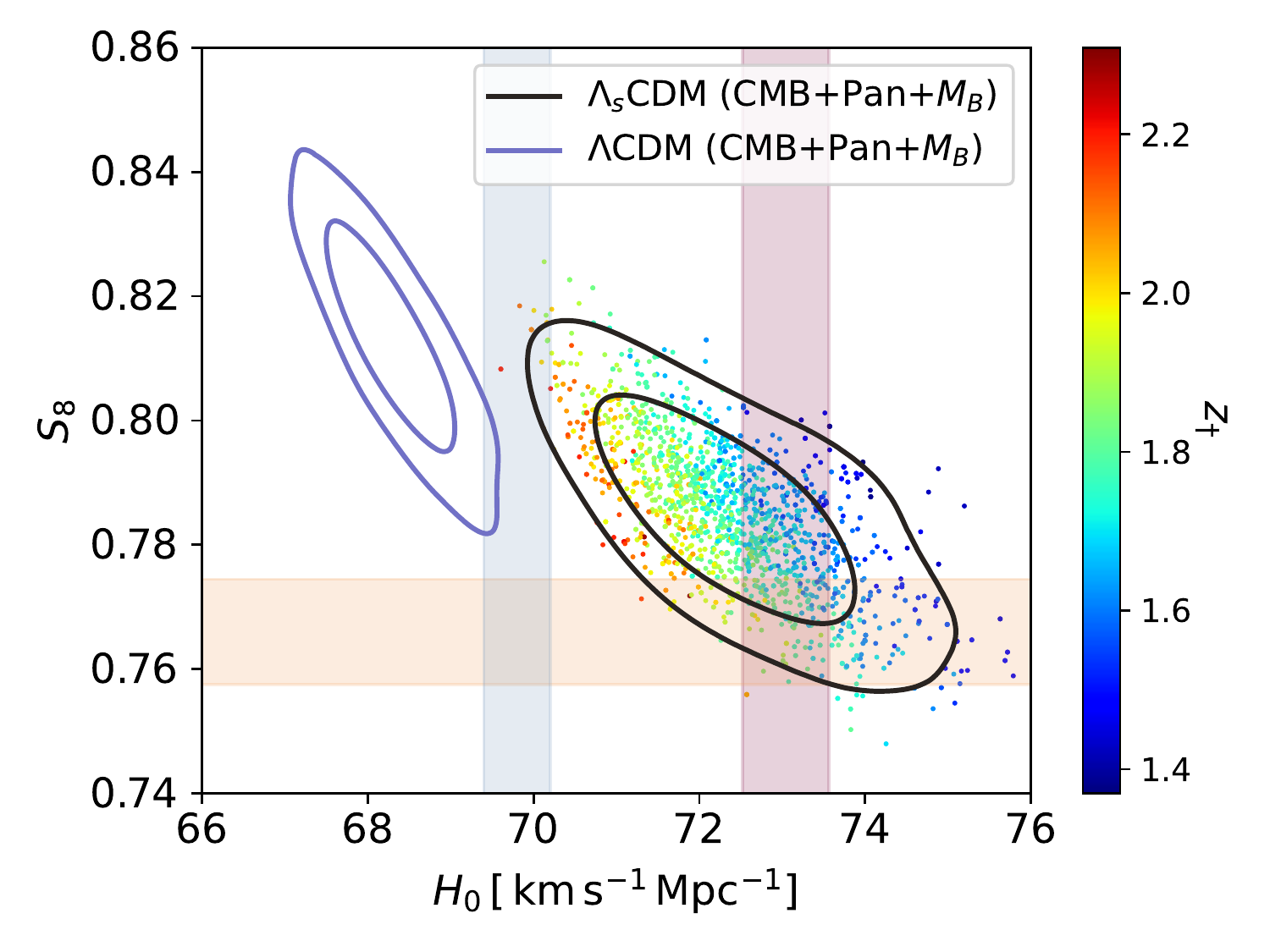}
\includegraphics[width=5.8cm]{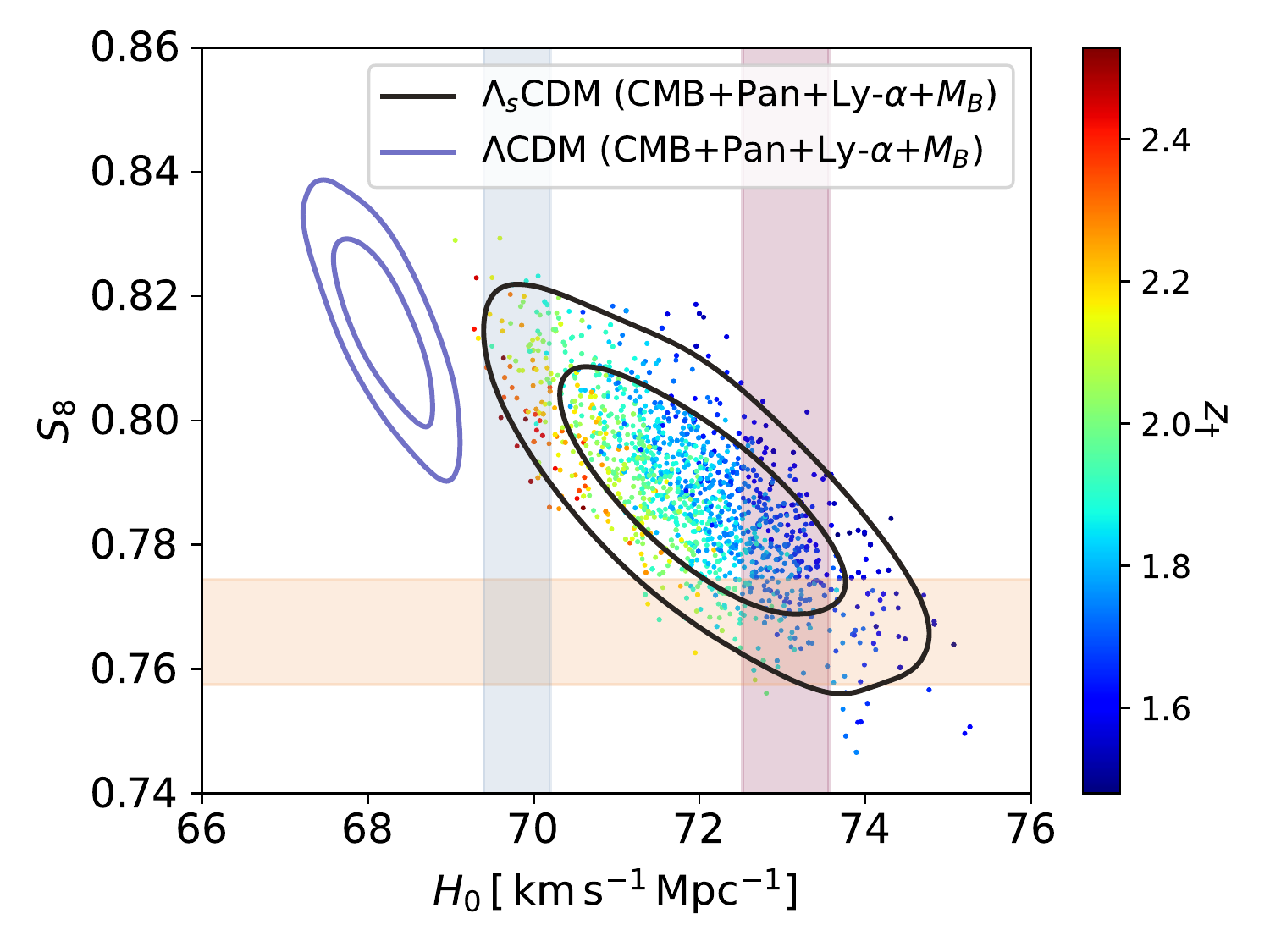}
\includegraphics[width=5.8cm]{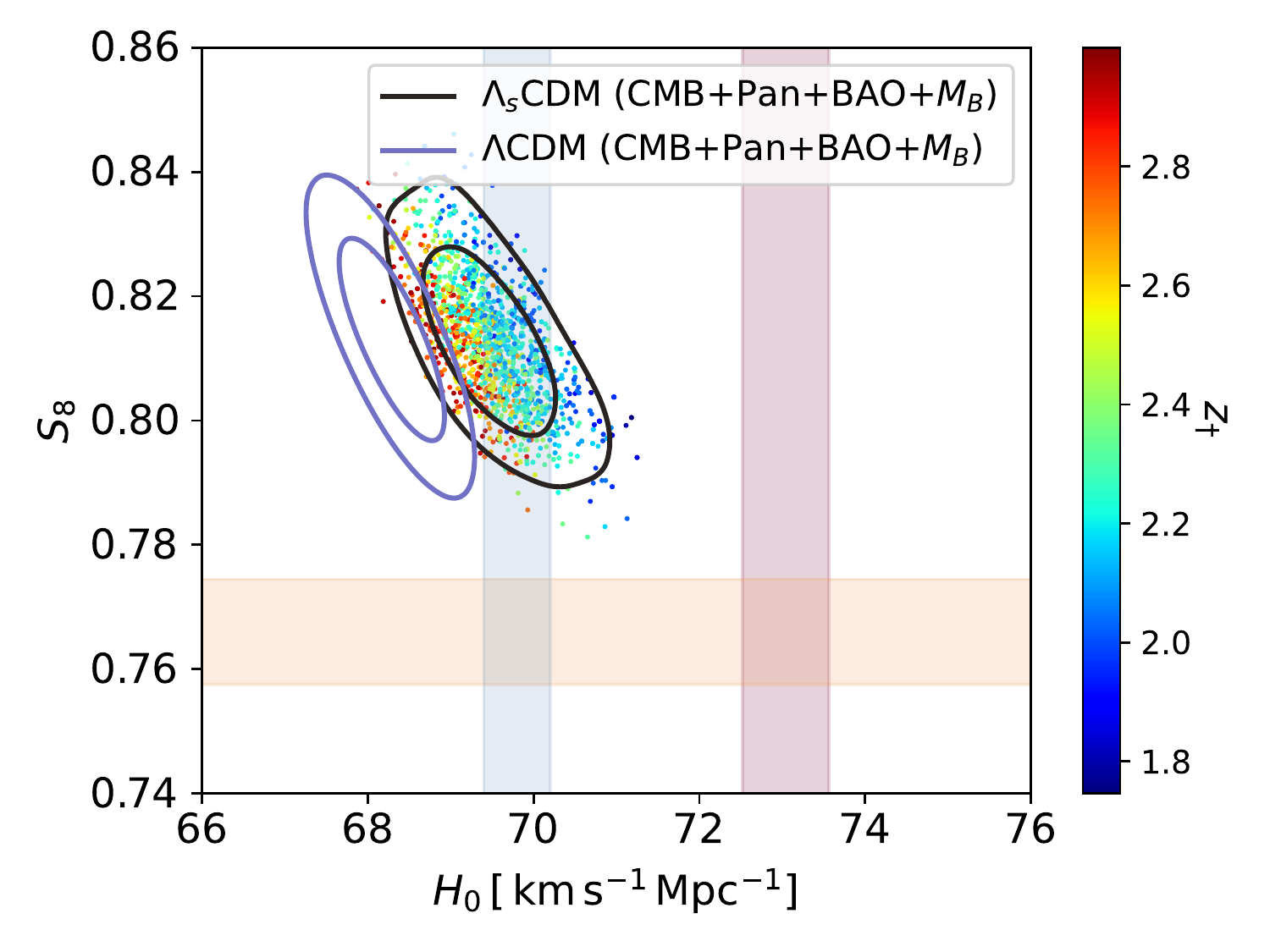}

\caption{Two-dimensional marginalized posterior distributions (68\% and 95\% CLs) in the $S_8$-$H_0$ plane for the $\Lambda_{\rm s}$CDM (color coded by $z_\dagger$) and $\Lambda_{\rm}$CDM for different data combinations. We overlay $1\sigma$ bands for the local measurements $H_{0}^{\rm R21}=73.04\pm1.04$ km $\rm{s^{-1}}$ $\rm{Mpc^{-1}}$~\cite{Riess:2021jrx}, $H_{\rm 0}^{\rm TRGB}=69.8\pm0.8~{\rm km\, s^{-1}\, Mpc^{-1}}$ ~\cite{Freedman:2019cchp}, and $S_{8}=0.766^{+0.020}_{-0.014}$ (WL+GC KiDS-1000 $3\times2$pt)~\cite{Heymans:2020gsg}.  The larger $z_\dagger$ is, the closer the $\Lambda_{\rm s}$CDM model is to the standard $\Lambda$CDM model. }
\label{fig:3dh0s8}
\end{figure*}

There is a discordance within $\Lambda$CDM between CMB and dynamical low-redshift cosmological probes (weak lensing, cluster counts, redshift-space distortion) that manifests itself in the $\sigma_8-\Omega_{\rm m}$ plane, where the $\sigma_8$ parameter quantifying the amplitude of growth of structure is the root-mean-square of the present-day matter density fluctuations within spheres of $8h^{-1}\, {\rm Mpc}$. This discordance is typically quantified by the $S_8\equiv\sigma_8\sqrt{\Omega_{\rm m}/0.3}$ parameter that characterizes the main degeneracy direction of the weak lensing measurements in the $\sigma_8-\Omega_{\rm m}$ plane. Assuming the $\Lambda$CDM model, the CMB constraints on $S_8$ from the full Planck data yield $S_8=0.834\pm0.016$~\cite{Planck:2018vyg} up to $3\sigma$ tension with the low-redshift measurements such as $S_8=0.766^{+0.020}_{-0.014}$ (WL+GC KiDS-1000 $3\times2$pt)~\cite{Heymans:2020gsg}, and $S_8=0.759\pm0.025$ (DES-Y3)~\cite{DES:2021vln}---although, note the Hyper Suprime-Cam (HSC) measurement $S_8=0.823^{+0.032}_{-0.028}$~\cite{Hamana:2019etx} that is consistent with the Planck $\Lambda$CDM value. Thus, the resolution of this discrepancy within a different model calls for a reduced $S_8$ prediction without compromising the agreement with the CMB. While this implies a reduction in the values of the parameters $\sigma_8$ and $\Omega_{\rm m}$, it is possible that a significant enough reduction in the value of either of the parameters can work just as well even if the remaining one's value is increased.
Indeed, the observational assessments of $\Lambda$CDM and $\Lambda_{\rm s}$CDM in Ref.~\cite{Akarsu:2021fol} presented higher $\sigma_8$ values for the $\Lambda_{\rm s}$CDM model, and the CMB-only data set yielded a matter density parameter value of $\Omega_{\rm m}=0.2900\pm0.0160$ for $\Lambda_{\rm s}$CDM lower compared to the $\Lambda$CDM value of $\Omega_{\rm m}=0.3162\pm 0.0084$, overcompensating the $\Lambda_{\rm s}$CDM's increased $\sigma_8$ parameter and consequently resulting in a relaxed $S_8= 0.8071 \pm 0.0210$ value compared to the $S_8=0.8332 \pm 0.0163$ of $\Lambda$CDM. Pleasantly, this amelioration of the $S_8$ tension is closely related to the amelioration of the $H_0$ tension within the $\Lambda_{\rm s}$CDM model as its reduced $\Omega_{\rm m}$ value is not due to a reduced physical matter density but its increased $H_0$ value. Note that, relaxing the $S_8$ tension is not a generic property of models that relax the $H_0$ tension, on the contrary, they often exacerbate it due to an excessively large $\sigma_8$ parameter~\cite{DiValentino:2020vvd,Abdalla:2022yfr}. For instance, amongst many, EDE~\cite{Poulin:2018cxd,Smith:2019ihp,Herold:2022iib,Kamionkowski:2022pkx}, as well as related models such as new-EDE~\cite{Niedermann:2019olb,Niedermann:2020dwg}, is one of the most popular promising ones for relaxing the $H_0$ tension, however both EDE and new-EDE exacerbate the $S_8$ tension. AdS-EDE~\cite{Ye:2020btb,Ye:2020oix,Wang:2022nap} is especially worth mentioning, because, similar to $\Lambda_{\rm s}$CDM, it is based on an AdS-dS transition. On the other hand, $\Lambda_{\rm s}$CDM considers the possibility of a rapid AdS-dS transition at redshift ${z\sim2}$, whereas AdS-EDE has an AdS phase that begins at ${z\sim2000}$ and ends shortly after recombination ($z_{\rm rec}\simeq1100$), settling down in a $\Lambda>0$ (dS) phase that still continues today. However, AdS-EDE, like other EDE models, relaxes the $H_0$ tension but worsens the $S_8$ tension~\cite{Wang:2022nap}.

To understand the structure formation within $\Lambda_{\rm s}$CDM and how it compares to $\Lambda$CDM, we start with the Newtonian equation for the growth of structure of the minimally interacting pressureless sources (baryons and CDM) after decoupling,
\begin{equation}
\partial_t^2\delta_{\rm m}=-2H\partial_t\delta_{\rm m}+4\pi G\bar\rho_{\rm m}\delta_{\rm m}, \label{eq:newtpert}
\end{equation}
where $\bar\rho_{\rm m}$ is the spatially uniform background energy density and $\delta_{\rm m}$ is the fractional overdensity of the pressureless fluid~\cite{Peebles1980Book}. We take $\delta_{\rm m}=\frac{\bar\rho_{\rm b}\delta_{\rm b}+\bar\rho_{\rm c}\delta_{\rm c}}{\bar\rho_{\rm b}+\bar\rho_{\rm c}}\approx\delta_{\rm b}\approx\delta_{\rm c}$ as quickly after recombination, the fractional overdensity in the baryons, $\delta_{\rm b}$, approaches that of the CDM, $\delta_{\rm c}$, and the matter behaves like a single pressureless fluid with total density contrast $\delta_{\rm m}$. The first term in the right-hand side, yielding negative values (assuming expanding universe, $H>0$), is antagonist to the growth of structure, and the second term, yielding positive values, endorses the growth of structure. We recall that the Hubble parameters, assuming expanding universe, are given by $H_{\Lambda\rm CDM}=\sqrt{8\pi G\bar\rho_{\rm m}/3+\Lambda/3}$ for $\Lambda$CDM, and ${H_{\Lambda_{\rm s}\rm CDM}=\sqrt{8\pi G\bar\rho_{\rm m}/3+\Lambda_{\rm s}/3}}$ for $\Lambda_{\rm s}$CDM, where we work in units such that the speed of light, $c$, equals unity. Thus, if both models have the same initial conditions for $\bar{\rho}_{\rm m}$ before the effects of the cosmological constants set in (which is what we assume in the rest of this discussion relying on it being well-constrained by the CMB power spectrum), $\Lambda_{\rm s}$CDM will have a weaker antagonist term up to the redshift $z_\dagger$ due to its negative valued cosmological constant which supports structure formation by lowering $H(z>z_\dagger)$ compared to both the $\Lambda$CDM and Einstein-de Sitter (viz., $\Lambda$CDM with $\Lambda=0$) models, consequently yielding an enhanced growth of structure at least for $z>z_\dagger$ (i.e., for $z\gtrsim2$ according to constraints we found on $z_\dagger$ in this work).\footnote{In line with this feature of the $\Lambda_{\rm s}$CDM model, the recent data from the James Webb Space Telescope seem to indicate enhanced growth of structure compared to $\Lambda$CDM at high redshifts \cite{Haslbauer:2022vnq,Boylan-Kolchin:2022kae,Lovell:2022bhx} (see also Refs.~\cite{Steinhardt:2015lqa,Behroozi:2016mne,Miyatake:2021qjr}).} If the values of the cosmological constants for both models were to be the same after the sign switch (i.e., ${\abs{\Lambda_{\rm s}}=\Lambda}$) for a given $\delta_{\rm m}(z>z_\dagger)$ value for both models, this would result in enhancement in the present-day structure for $\Lambda_{\rm s}$CDM since $H(z)$ would be the same for both models for $z<z_\dagger$ while the structure supporting nature of the negative cosmological constant of $\Lambda_{\rm s}$CDM would have resulted in a greater $\delta_{\rm m}$ value at $z=z_\dagger$. However, the observational constraints on $D_M(z_*)$ require that the lower $H(z>z_\dagger)$ values of $\Lambda_{\rm s}$CDM compared to $\Lambda$CDM should be compensated by higher $H(z<z_\dagger)$ values, i.e., ${\abs{\Lambda_{\rm s}}>\Lambda}$. Hence, for $z<z_\dagger$, the cosmological constant of $\Lambda_{\rm s}$CDM will have a stronger impact against growth of structures compared to $\Lambda$CDM. The answer to whether these two competing effects before and after $z_\dagger$ result in a greater present-day amplitude of growth of structure for $\Lambda_{\rm s}$CDM or not, can be reached by observational analysis, and is conceivably dependent on the value of $z_\dagger$, which controls both the value of $\abs{\Lambda_{\rm s}}$ and the amount of time the Universe spends in the phases with negative and positive cosmological constants. Note that, a smaller $z_{\dagger}$ results in a greater value for $\abs{\Lambda_{\rm s}}$ and an extended era with the negative cosmological constant, and in the $z_\dagger\to\infty$ limit, $\Lambda_{\rm s}$CDM approaches $\Lambda$CDM.

The results of the observational analyses in~\cref{tab:withMB,tab:withoutMB} present $S_8$ values that are lower for $\Lambda_{\rm s}$CDM compared to $\Lambda$CDM for all six data sets except for the CMB+Pan+BAO+$M_B$ case for which both models yield the same constraints. This is despite $\Lambda_{\rm s}$CDM yielding higher constraints on $\sigma_8$ for all cases in line with our theoretical discussion. Since the low-redshift probes find lower $S_8$ values compared to the predictions of $\Lambda$CDM, the tensions presented in~\cref{tab:tensionswithoutMB} are always lower for $\Lambda_{\rm s}$CDM except for the CMB+Pan+BAO+$M_B$ case for which both models have the same amount of tension. For $\Lambda_{\rm s}$CDM, the inclusion of the $M_B$ prior results in a better amelioration and the inclusion of the full BAO data has an hindering effect---note that, in contrast, addition of the $M_B$ prior makes little to no difference for the $\Lambda$CDM model in amelioration of the $S_8$ discrepancy. The similarities between this discussion on the constraints and tensions of $S_8$ and the ones in~\cref{subsec:h0,subsec:mb} on the constraints and tensions on $M_B$ and $H_0$ are unsurprising due to the strong correlations among these parameters (see~\cref{fig:3dh0s8} for the correlation between $H_0$ and $S_8$). Interestingly, \cref{fig:3dh0s8} indicates that the simultaneous alleviation of the $H_0$ and $S_8$ tensions within $\Lambda_{\rm s}$CDM is possible if the local $H_0$ measurement of SH0ES is considered but not the TRGB. Finally, note that the $S_8$ values as measured by the low-redshift probes are not model-independent, and an absolute determination of the status of the $S_8$ discrepancy within $\Lambda_{\rm s}$CDM requires the analyses of the low-redshift observations with $\Lambda_{\rm s}$CDM as the underlying cosmological model.

\subsection{BAO and Ly-$\alpha$ discrepancies}
\label{subsec:bao}
In all, the SDSS, BOSS, and eBOSS surveys provide galaxy and quasar samples from which BAO can be measured covering all redshifts $z < 2.2$, and Ly-$\alpha$ forest observations over $2 < z < 3.5$. In~\cref{tab:BAO_measurements}, we list the latest BAO measurements at seven different effective redshifts ($z_{\rm eff}$), viz., $D_H/r_{\rm d}$, $D_M/r_{\rm d}$, and $D_V/r_{\rm d}$, where $D_{H}(z)\equiv c/H(z)$ is the Hubble distance at redshift $z$, $D_{M}(z)\equiv c\int_0^z{\dd{z'}/H(z')}$ is the comoving angular diameter distance in a spatially flat Robertson-Walker (RW) spacetime, $D_V(z)=[zD_{H}(z)D_{M}^2(z)]^{1/3}$ is the spherically averaged distance, and $r_{\rm d}=\int_{z_{\rm d}}^{\infty}\dd{z}c_{\rm s}(z)/H(z)$ is the radius of sound horizon at drag epoch ($z_{\rm d}\sim1060$) with $c_{\rm s}(z)=c[3+9\rho_{\rm b}/4\rho_{\gamma}(z)]^{-1/2}$ being the speed of sound in the baryon-photon fluid.

\begin{figure}[t!]
    \centering
    \includegraphics[width=8.4cm]{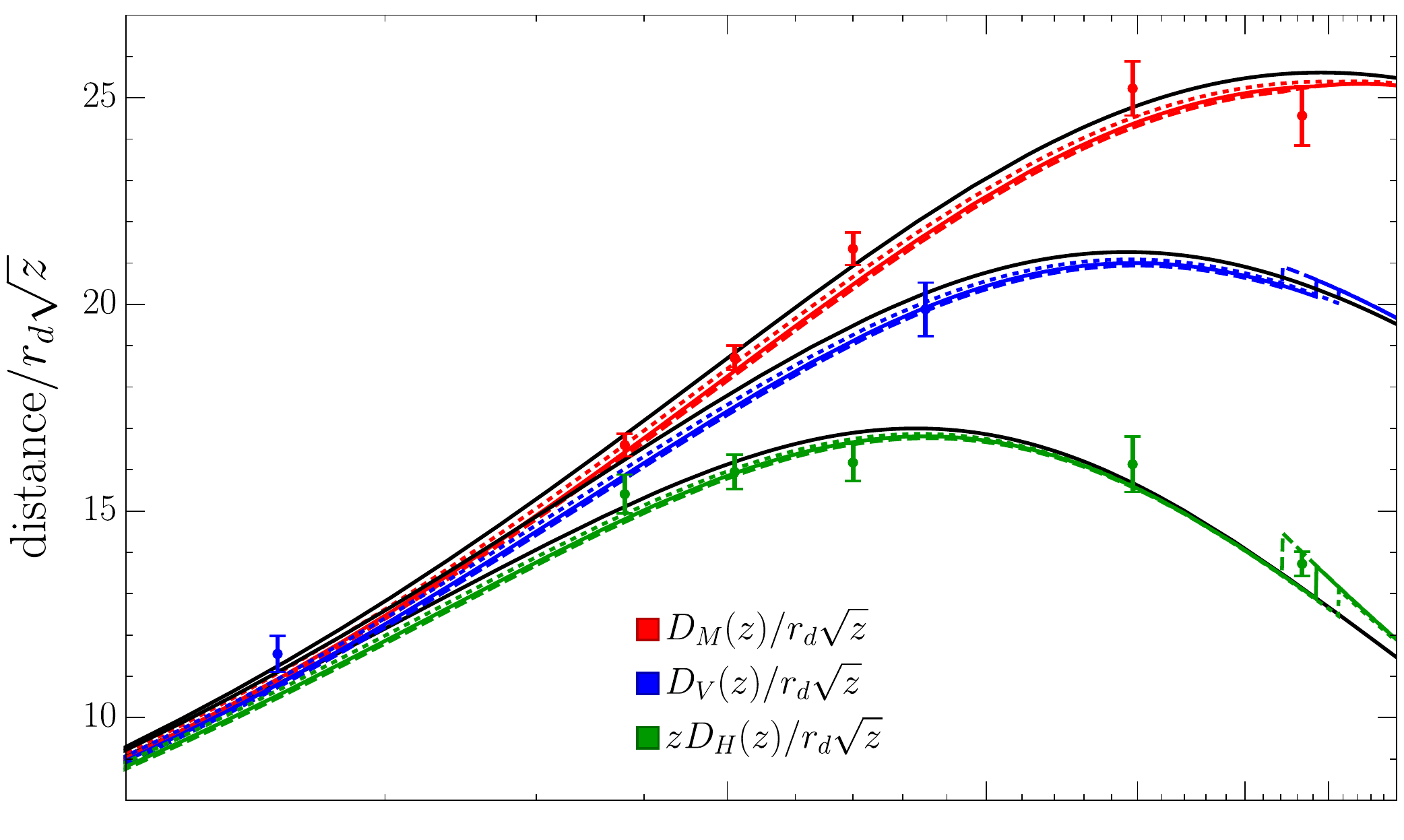}
    \includegraphics[width=8.4cm]{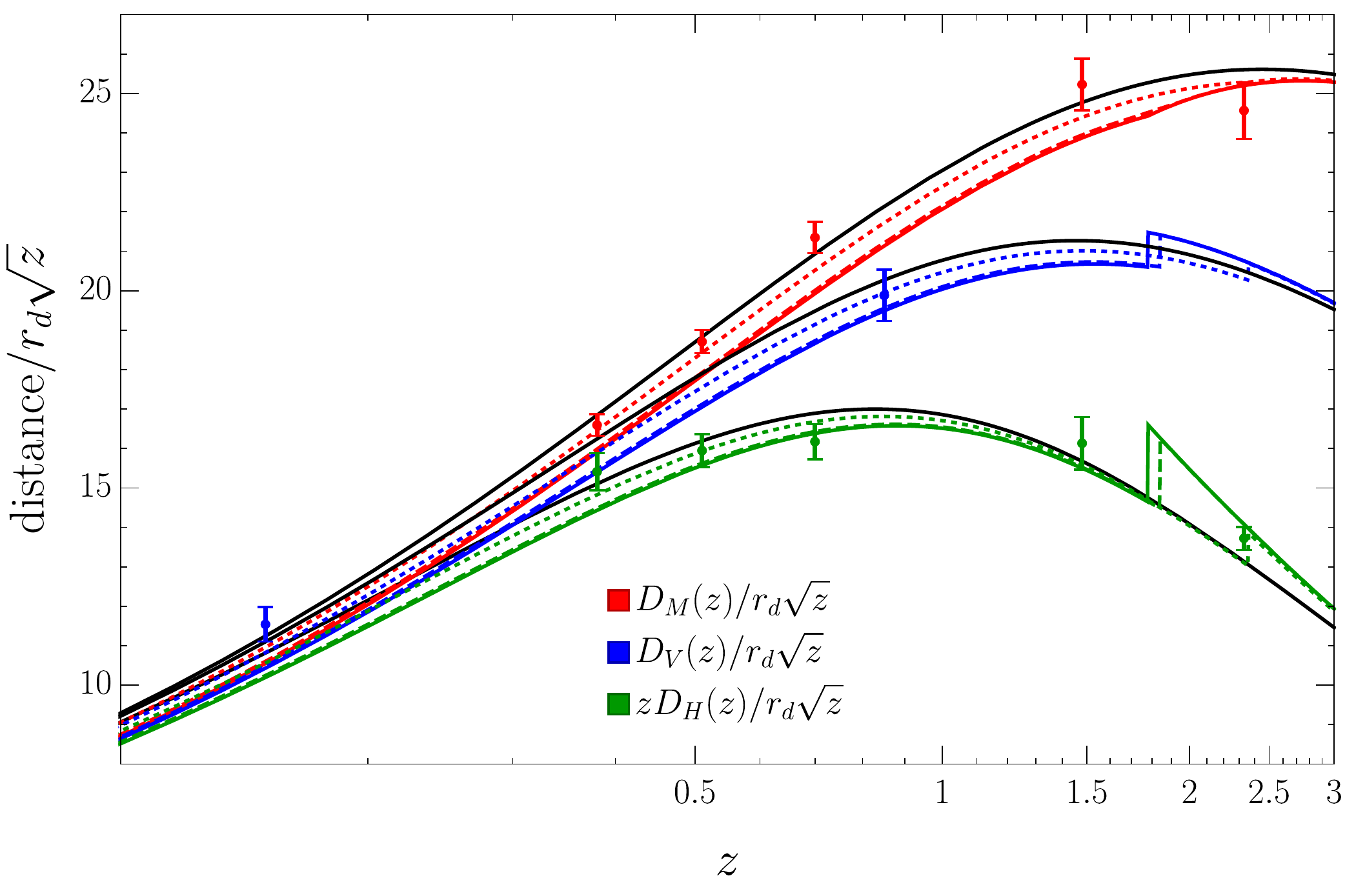}
    \caption{Expansion histories of $\Lambda_{\rm s}$CDM for the mean values of the analyses with six different data sets presented in~\cref{tab:withoutMB} and~\cref{tab:withMB}. The top and bottom panels respectively show the plots for analyses without and with the $M_B$ prior. The solid lines are for the CMB+Pan, the dashed lines are for the CMB+Pan+Ly-$\alpha$ and the dotted lines are for the CMB+Pan+BAO data sets. Both the data (from~\cref{tab:BAO_measurements} except we combine the Ly-$\alpha$ values and use $D_H(2.33)/r_{\rm d}=8.99\pm0.19$ and $D_M(2.33)/r_{\rm d}=37.5\pm1.1$) and the plots are color coded for different distance measures with red corresponding to $D_M(z)/r_{\rm d}\sqrt{z}$, blue to $D_V(z)/r_{\rm d}\sqrt{z}$ and green to $zD_H(z)/r_{\rm d}\sqrt{z}$. The plots for $\Lambda$CDM are given only for the CMB+Pantheon analysis without $M_B$ as the plots for different data sets are not visually distinguishable in the figure; $\Lambda$CDM plots are all solid black and each correspond to the obvious distance measure of the branch it is closest to.} 
    \label{fig:expansion_history}
\end{figure}

\begin{table*}[ht!]
  \caption{Concordance/discordance between the $\Lambda$CDM and $\Lambda_{\rm s}$CDM models and the BAO measurements listed in Table~\ref{tab:BAO_measurements}. For the values relevant to the Ly-$\alpha$ measurements at $z_{\rm eff}=2.33$, we have considered the combined values of $D_H(2.33)/r_{\rm d}=8.99\pm0.19$ and $D_M(2.33)/r_{\rm d}=37.5\pm1.1$~\cite{duMasdesBourboux:2020pck}.}
  \label{tab:tensionswithMB}
	\scalebox{0.82}{
	\setlength\extrarowheight{2pt}
	\begin{centering}
	  \begin{tabular}{lcc|cc|cc|cc|cc|cc}
 	\hline
    \toprule
    \multicolumn{1}{l}{\textbf{Data set}} & \multicolumn{2}{c}{\textbf{CMB+Pan}} & \multicolumn{2}{c}{\textbf{CMB+Pan+Ly-$\bm{\alpha}$}}& \multicolumn{2}{c}{\textbf{CMB+Pan+BAO}} & \multicolumn{2}{c}{\textbf{CMB+Pan+$\bm{M_B}$}} & \multicolumn{2}{c}{\textbf{CMB+Pan+Ly-$\bm{\alpha}$+$\bm{M_B}$}}& \multicolumn{2}{c}{\textbf{CMB+Pan+BAO+$\bm{M_B}$}}  \\  \hline
      & \textbf{{$\bm{\Lambda}$CDM}} & \textbf{$\bm{\Lambda}_{\textbf{s}}$CDM} & \textbf{{$\bm{\Lambda}$CDM}} & \textbf{$\bm{\Lambda}_{\textbf{s}}$CDM}  & \textbf{{$\bm{\Lambda}$CDM}} & \textbf{$\bm{\Lambda}_{\textbf{s}}$CDM} & \textbf{{$\bm{\Lambda}$CDM}} & \textbf{$\bm{\Lambda}_{\textbf{s}}$CDM} & \hphantom{,,,}\textbf{{$\bm{\Lambda}$CDM}}\hphantom{,,} & \hphantom{,,}\textbf{$\bm{\Lambda}_{\textbf{s}}$CDM}\hphantom{,,,} & \hphantom{,,,}\textbf{{$\bm{\Lambda}$CDM}} \hphantom{,,}& \hphantom{,,}\textbf{$\bm{\Lambda}_{\textbf{s}}$CDM} \hphantom{,,,}\\  \hline
$D_V(0.15)/r_{\rm d}$ & $0.9\sigma$& $1.7\sigma$ & $1.0\sigma$ & $1.8\sigma$ & $1.0\sigma$ & $0.7\sigma$ & $1.2\sigma$ & $2.4\sigma$ & $1.3\sigma$ & $2.3\sigma$ & $1.2\sigma$ & $1.6\sigma$  \\

$D_V(0.85)/r_{\rm d}$ &  $0.7\sigma$ & $0.0\sigma$ &  $0.6\sigma$ & $0.1\sigma$ & $0.6\sigma$ & $0.3\sigma$ & $0.5\sigma$ & $0.6\sigma$ & $0.4\sigma$ & $0.5\sigma$ & $0.5\sigma$ & $0.2\sigma$ \\  
 \hline
$D_M(0.38)/r_{\rm d}$ & $0.9\sigma$ & $0.6\sigma$ & $0.9\sigma$ & $0.9\sigma$ & $0.8\sigma$ & $0.2\sigma$ & $0.3\sigma$ & $2.0\sigma$ & $0.3\sigma$ & $1.8\sigma$ & $0.4\sigma$ & $0.3\sigma$  \\

$D_M(0.51)/r_{\rm d}$ &  $0.5\sigma$ & $0.9\sigma$ & $0.4\sigma$ & $1.2\sigma$ & $0.4\sigma$ & $0.3\sigma$ & $0.1\sigma$ & $2.3\sigma$ & $0.1\sigma$ & $2.2\sigma$ & $0.0\sigma$ & $0.8\sigma$  \\

$D_M(0.70)/r_{\rm d}$ & $0.9\sigma$ & $1.9\sigma$ & $0.9\sigma$ & $2.1\sigma$ & $1.0\sigma$ & $1.5\sigma$ & $1.3\sigma$ & $3.1\sigma$ & $1.4\sigma$ & $3.0\sigma$ & $1.3\sigma$ & $2.0\sigma$  \\

$D_M(1.48)/r_{\rm d}$ & $0.6\sigma$ & $1.3\sigma$ & $0.6\sigma$ & $1.5\sigma$ & $0.7\sigma$ & $1.0\sigma$ & $0.8\sigma$ & $1.9\sigma$ & $0.9\sigma$ & $1.9\sigma$ & $0.8\sigma$ & $1.2\sigma$  \\

$D_M(2.33)/r_{\rm d}$ & $1.5\sigma$ & $1.0\sigma$ & $1.5\sigma$ & $1.0\sigma$ & $1.5\sigma$ & $1.2\sigma$  & $1.3\sigma$ & $0.9\sigma$ & $1.3\sigma$ & $0.9\sigma$ & $1.4\sigma$ & $1.1\sigma$ \\ \hline

 $D_H(0.38)/r_{\rm d}$  &  $0.6\sigma$ & $1.3\sigma$ & $0.6\sigma$ & $1.4\sigma$ & $0.6\sigma$ & $0.9\sigma$ & $0.8\sigma$ & $2.0\sigma$ & $0.9\sigma$ & $1.9\sigma$ & $0.8\sigma$ & $1.2\sigma$  \\ 

 $D_H(0.51)/r_{\rm d}$  & $0.7\sigma$ & $0.1\sigma$ & $0.6\sigma$ & $0.3\sigma$ & $0.6\sigma$ & $0.3\sigma$ & $0.4\sigma$ & $0.8\sigma$ & $0.4\sigma$ & $0.8\sigma$ & $0.4\sigma$ & $0.0\sigma$  \\ 

 $D_H(0.70)/r_{\rm d}$  & $1.7\sigma$ & $1.0\sigma$ & $1.7\sigma$ & $0.9\sigma$ & $1.7\sigma$ & $1.4\sigma$ & $1.5\sigma$ & $0.5\sigma$ & $1.5\sigma$ & $0.5\sigma$ & $1.5\sigma$ & $1.2\sigma$  \\ 

 $D_H(1.48)/r_{\rm d}$  &  $0.6\sigma$ & $0.8\sigma$ & $0.6\sigma$ & $0.8\sigma$ & $0.6\sigma$ & $0.7\sigma$ & $0.6\sigma$ & $0.9\sigma$ & $0.6\sigma$ & $0.9\sigma$ & $0.6\sigma$ & $0.8\sigma$  \\

$D_H(2.33)/r_{\rm d}$ & $2.0\sigma$ & $0.2\sigma$ & $1.9\sigma$ & $0.1\sigma$ & $1.9\sigma$ & $1.1\sigma$ & $1.9\sigma$ & $1.2\sigma$ & $1.8\sigma$ & $1.2\sigma$ & $1.9\sigma$ & $0.1\sigma$  \\ 

\hline
    \bottomrule
    \hline 
  \end{tabular}
  \end{centering}
}
\end{table*}

There appears to be a discordance between the low- and high-redshift BAO data within $\Lambda$CDM. The Ly-$\alpha$ BAO measurements of $D_M(2.34)/r_{\rm d}$ and $D_H(2.34)/r_{\rm d}$ from the BOSS DR11 sample were found to be in a tension of approximately $2.5\sigma$ with the best fit predictions of Planck CMB within $\Lambda$CDM, whereas the Galaxy BAO measurements from lower redshifts including the ones from the same sample showed no significant discrepancy~\cite{Aubourg:2014yra}. Moreover, an unanchored analysis of these BAO data without the presence of additional data such as CMB, presented a tension of approximately $2.5\sigma$ with a nonevolving DE (i.e., the usual cosmological constant) for $z<2.34$~\cite{Evslin:2016gre}; and when DE was allowed to evolve in Ref.~\cite{Aubourg:2014yra}, Ly-$\alpha$ data showed a preference for negative DE density values around $z=2.34$. With the final eBOSS (SDSS DR16) measurement, this tension between the Ly-$\alpha$ BAO and Planck CMB data is reduced to approximately $1.5\sigma$~\cite{duMasdesBourboux:2020pck}---this would also correspond to a reduction of the tension in the above mentioned unanchored analysis and preference of negative DE densities, also, it is closely related to the internal tension of high and low-redshift BAO as quantified in Ref.~\cite{Cuceu:2019for} where it was also shown to diminish with updated data releases in line with the results of the recent study in Ref.~\cite{Schoneberg:2022ggi}. Despite the reduction in these discrepancies, the BAO anomalies are still important. As discussed in~\cref{subsec:h0}, the different degeneracy directions of the high- and low-redshift BAO data in the $\Omega_{\rm m}-H_0$ plane when combined with BBN constraints result in a $H_0$ value in agreement with the CMB prediction but in significant tension with local measurements~\cite{Alam:2020sor}. Moreover, parametric and nonparametric reconstructions of the DE density that utilize the BAO data keep finding negative (although usually consistent with vanishing) DE densities around the Ly-$\alpha$ data~\cite{Wang:2018fng,Escamilla:2021uoj,Bernardo:2021cxi} indicating a DE density that transits from negative to positive today. Also, note the parallelisms of the Ly-$\alpha$ and $S_8$ discrepancies that may indicate that the resolution of these two tensions are related; first, the $S_8$ constraints based on the Ly-$\alpha$ data and weak lensing surveys probing similar redshift scales as the Ly-$\alpha$ measurements agree~\cite{Palanque-Delabrouille:2019iyz}, second, the weakening of the tension with recent measurements happened also for the $S_8$ discrepancy~\cite{Heymans:2020gsg,DES:2021vln}, and third, minimal extensions of $\Lambda$CDM that relax either of these tensions tend to exacerbate the $H_0$ tension~\cite{DiValentino:2020zio,DiValentino:2020vvd}.

In the analyses of both models with six different data sets, the ones that include our full BAO data have distinctive properties from the rest. For the data sets without the full BAO, both models yield similar posterior distributions (especially for the CMB+Pan data set without the $M_B$ prior) for the baseline six free parameters of $\Lambda$CDM, whereas including the full BAO data results in slight separation of the contours (see~\cref{tab:withMB,tab:withoutMB} and Figs.~\ref{fig:nomb}-\ref{fig:mbbao} presented in the~\cref{sec:Appendix}). Regarding the derived parameters, $\Lambda_{\rm s}$CDM results in significantly lower $S_8$ values despite its higher $\sigma_8$ parameter for all data sets except when full BAO data is included in which case both models yield very similar constraints; however, $\Lambda_{\rm s}$CDM yields higher $H_0$ and $M_B$ values, and a lower $t_0$ value compared to $\Lambda$CDM whether or not full BAO is included in the data set. Expanding the BAO data set from Ly-$\alpha$ to the full BAO means inclusion of the Galaxy BAO at the $z_{\rm eff}=0.15,\,0.38,\,0.51,\,0.70,\,0.85$ and also the Quasar BAO at $z_{\rm eff}=1.48$. The effect of the Galaxy BAO at $z_{\rm eff} = 0.38,\, 0.51,\, 0.61$ on $\Lambda_{\rm s}$CDM was discussed in Ref.~\cite{Akarsu:2021fol} where it was found that the preference of the Galaxy BAO data for higher $z_\dagger$ values holds the model back from working efficiently in alleviating the tensions of $\Lambda$CDM as the phenomenological difference between the two models diminishes with the increasing values of $z_\dagger$. The same observation can be made also from the analyses of the present paper where the inclusion of the full BAO data set, majority of which is galaxy BAO, results in higher $z_\dagger$ values, and hence is accompanied with a worsening in amelioration of the tensions (cf. \cref{tab:tensionswithoutMB}). In~\cref{fig:expansion_history}, we give expansion histories of $\Lambda_{\rm s}$CDM for the mean values of the analyses with six different data sets presented in~\cref{tab:withoutMB,tab:withMB}. And in~\cref{{tab:tensionswithMB}}, we quantify the concordance/discordance between the $\Lambda$CDM and $\Lambda_{\rm s}$CDM models and the BAO measurements listed in Table~\ref{tab:BAO_measurements}. For the values relevant to the Ly-$\alpha$ measurements at $z_{\rm eff}=2.33$, we have considered the combined values of $D_H(2.33)/r_{\rm d}=8.99\pm0.19$ and $D_M(2.33)/r_{\rm d}=37.5\pm1.1$~\cite{duMasdesBourboux:2020pck}. 

We see in~\cref{tab:tensionswithoutMB} that $\Lambda$CDM is typically in approximately $2\sigma$ tension with $D_H(2.33)/r_{\rm d}=8.99\pm0.19$ in all cases. On the other hand, $\Lambda_{\rm s}$CDM is typically fully consistent with $D_H(2.33)/r_{\rm d}=8.99\pm0.19$ with the level of tension being almost zero in some cases and without exceeding $1.2\sigma$ even in the worst case. The $D_H(z)$ plots in~\cref{fig:expansion_history} show how a $z_\dagger<2.33$, i.e., a sign switch at smaller redshifts than the effective redshift of the Ly-$\alpha$ data, results in an excellent fit to the $D_H(2.33)$ measurements that is immediately lost for $z_\dagger>2.33$ (also, cf. Fig.~3 in Ref.~\cite{Akarsu:2021fol}). When we consider $D_M(2.33)/r_{\rm d}$, both $\Lambda$CDM and $\Lambda_{\rm s}$CDM models are in good consistency with $D_M(2.33)/r_{\rm d}=37.5\pm1.1$, yet $\Lambda_{\rm s}$CDM does systematically better; while the level of tension is approximately $1.5\sigma$ in $\Lambda$CDM in all cases, it is $1\sigma$ in $\Lambda_{\rm s}$CDM. The better agreement with Ly-$\alpha$ was expected by the theoretical and observational analyses in Ref.~\cite{Akarsu:2021fol}, and so was the tension with the Galaxy BAO presented in~\cref{tab:tensionswithMB}. However, a careful examination of~\cref{tab:tensionswithMB} exposes a characteristic of $\Lambda_{\rm s}$CDM that is not present in $\Lambda$CDM; that is, in certain cases, $\Lambda_{\rm s}$CDM is discrepant with the $D_M(z)/r_{\rm d}$ value of a BAO measurement while it is in agreement with its $D_H(z)/r_{\rm d}$ value. This is possible, since unlike $D_H(z)$, which gives information about a single instance of time, $D_M(z)$ relies on a cumulative effect from present-day up to a redshift, i.e., the integral $\int_0^z\dd{z'}/{H(z')}$. Thus, if the $H(z)$ of a model deviates from the actual Hubble parameter describing the Universe at low redshifts, this deviation will carry over to higher redshifts when $D_M(z)$ is considered, and can be corrected only if another deviation in the opposite direction happens (see Ref.~\cite{Akarsu:2022lhx} for the implications of this when $D_M(z_*)$ is considered). Moreover, since $1/H(z)$ decays rapidly with increasing $z$, the integral $\int_0^z\dd{z'}/{H(z')}$ gets most of its contribution from lower redshifts, and hence is more sensitive to deviations at low redshifts. It seems that~\cref{tab:tensionswithMB} and~\cref{fig:expansion_history} show imprints of this effect for $\Lambda_{\rm s}$CDM. Let us consider the CMB+Pan+$M_B$ case in~\cref{tab:tensionswithMB} as an example since it is the one where this situation is most apparent. The tension of $\Lambda_{\rm s}$CDM with the $D_M(0.70)/r_{\rm d}$ measurement is at $3.1\sigma$ level whereas it is only $0.5\sigma$ for $D_H(0.70)/r_{\rm d}$; this is likely to be caused by the tensions with the $D_H(z)/r_{\rm d}$ values for $z<0.5$, i.e., the $2\sigma$ tension with $D_H(0.38)/r_{\rm d}$ and the $2.4\sigma$ tension with the $D_V(0.15)/r_{\rm d}$ measurement, that carry over to higher redshifts for $D_M(z)/r_{\rm d}$. This effect, illustrated with the above example, seems to permeate~\cref{tab:tensionswithMB}, and indicates that $\Lambda_{\rm s}$CDM's conflict  is mainly with the BAO measurements for which $z_{\rm eff}<0.5$, and also that the model can fit both CMB and full BAO excellently if its Hubble radius is superposed with a wavelet as discussed in Ref.~\cite{Akarsu:2022lhx}.

\subsection{Age discrepancy}

The (present-day) age of the Universe can also be measured using very old astrophysical objects, such as globular clusters (GCs), in a cosmological model-agnostic way, in the sense that it does not depend in any significant way on the cosmological model adopted. It is estimated in Ref.~\cite{Valcin:2021jcg} (see also Refs.~\cite{Valcin:2020vav,Bernal:2021yli}) that the age of the oldest GCs is $t_{\rm GC}=13.32\pm0.10\,({\rm stat.})\pm0.23\,({\rm sys.})$ Gyr at 68\% CL, which is transformed to an age of the Universe $t_{\rm u}=13.50\pm0.15\,({\rm stat.})\pm0.23\,({\rm sys.})$ Gyr ($\pm0.27$ when adding statistical and systematic uncertainties in quadrature). It is in good agreement with the Planck18 $\Lambda$CDM inferred age $t_0=13.80\pm0.02$ Gyr~\cite{Planck:2018vyg}. However, this success may be due to the systematic uncertainties that are currently too large; there are ongoing efforts to reduce the impact of systematic uncertainties so that GCs' constraints on $t_0$ can potentially discriminate among different cosmological models, in particular, the models that are proposed to solve the $H_0$ tension~\cite{Bernal:2021yli,Vagnozzi:2021tjv}. When we consider the age of the Universe estimated from GCs by taking only the statistical uncertainties into account, viz., $t_{\rm u}=13.50\pm0.15$ Gyr at 68\% CL, while the Planck18 $\Lambda$CDM finds $2\sigma$ tension, the $\Lambda_{\rm s}$CDM model is expected to find an even better agreement as $\Lambda_{\rm s}$ reduces the age of the Universe~\cite{Akarsu:2021fol}. Our results for $t_0$ are summarized in~\cref{tab:tensionswithoutMB} and~\cref{fig:zdagger2d}. We see that in all three analyses without the $M_B$ prior, $\Lambda$CDM is in tension with $t_{\rm u}$ estimated from GCs mentioned above at the level of $1.9\sigma$, whereas the $\Lambda_{\rm s}$CDM model is in tension at less than $1\sigma$, except reaches $1.4\sigma$ tension for the CMB+Pan+BAO case. On the other hand, when the $M_B$ prior is included in the analysis, the tensions of $\Lambda$CDM decrease only slightly to $1.7\sigma$ for all three analyses, but $\Lambda_{\rm s}$CDM becomes fully consistent; even the largest tension for $\Lambda_{\rm s}$CDM is just $1.1\sigma$ (CMB+Pan+BAO+$M_B$). Of course, to be able to conclude whether there is a real tension within $\Lambda$CDM between the age of the Universe as predicted by CMB and the one inferred from GCs, and to use $t_{\rm u}$ as a discriminator between cosmological models, we need the systematic uncertainties in $t_{\rm u}$ to be reduced. However, it is important to notice the clear correlations of the parameter $z_\dagger$ of the $\Lambda_{\rm s}$CDM model with not only $t_0$ but also the parameters $H_0$, $M_B$, $S_8$, and $D_H(2.33)/r_{\rm d}$ in~\cref{fig:zdagger2d}. Moreover, not only the $\Lambda_{\rm s}$CDM predicted $t_0$ values find better agreement with the one predicted by GCs, but also the $\Lambda_{\rm s}$CDM predicted values of $H_0$, $M_B$, $S_8$, and $D_H(2.33)/r_{\rm d}$ are consistent with their direct observational values. It is very difficult to simply call it a coincidence, and as systematic uncertainties are removed, it would not be a surprise if the age of the Universe turns out to be smaller than the Planck18 $\Lambda$CDM prediction.

\subsection{$\omega_b$ discrepancy}

The BBN constraints on $\omega_{\rm b}$ depend on the assumed nuclear reaction rates. The most important one for deuterium destruction relevant to BBN is the $D(p,\gamma)^3\rm He$ reaction rate which was recently measured by the LUNA  (The Laboratory for Underground Nuclear Astrophysics) experiment~\cite{Mossa:2020gjc}. They use their measurements to give the constraint $\omega_{\rm b}^{\rm LUNA}=0.02233\pm0.00036$ on the physical density
parameter of baryons. Using the same measurement by LUNA and a more theoretically guided approach for the two other important processes for deuterium destruction, Ref.~\cite{Pitrou:2020etk} has reported the constraint $\omega_{\rm b}^{\rm PCUV21}=0.02195\pm0.00022$. Compared to the CMB only prediction $\omega_{\rm b}=0.02237 \pm 0.00015$ from Planck, which increases to  $\omega_{\rm b}=0.02242 \pm 0.00014$ when BAO data is included~\cite{Planck:2018vyg}, $\omega_{\rm b}^{\rm LUNA}$ shows excellent agreement while the more theoretical value  $\omega_{\rm b}^{\rm PCUV21}$ presents some discrepancy. These are in line with the previous trend where predictions of CMB agree well with empirical approaches based on experimentally measured cross sections while more theoretical approaches are discrepant. See Ref.~\cite{Cooke:2017cwo} where the theoretical $D(p,\gamma)^3\rm He$ rate yields $\omega_{\rm b}=0.02166\pm0.00019$ whereas the empirical one yields $\omega_{\rm b}= 0.02235 \pm 0.00037$.

In Ref.~\cite{Akarsu:2021fol}, both $\Lambda$CDM and $\Lambda_{\rm s}$CDM yielded similar $\omega_{\rm b}$ values discrepant with theoretical BBN constraints, and inclusion of the BAO in the data set resulted in an exacerbation of this discrepancy for $\Lambda$CDM as in the analyses of Planck, but, it resulted in an amelioration for $\Lambda_{\rm s}$CDM. Intrigued by these results, in this paper, we also computed the tensions of both models with  both empirically and theoretically guided BBN constraints on $\omega_b$. From~\cref{tab:withMB,tab:withoutMB}, we see that both models yield $\omega_{\rm b}$ values higher than BBN constraints. While inclusion of the $M_B$ prior increases these values further, inclusion of the BAO data increases $\omega_{\rm b}$ for $\Lambda$CDM but decreases it for $\Lambda_{\rm s}$CDM pulling the extended model towards BBN constraints. When $\omega_{\rm b}^{\rm LUNA}$ is considered, both models are in excellent agreement for all data sets; when $\omega_{\rm b}^{\rm PCUV21}$ is considered, both models are moderately discrepant for all six data sets. However, it is worth noting that $\omega_{\rm b}$ values for $\Lambda_{\rm s}$CDM are lower for all data sets in better agreement with BBN constraints up to $0.5\sigma$. Also, since the $\omega_{\rm b}$ tensions within $\Lambda_{\rm s}$CDM increase with the inclusion of the $M_B$ prior and decrease with the inclusion of the full BAO data set, out of the six different discrepancies presented in~\cref{tab:tensionswithoutMB}, it is the only one that prefers relatively larger $z_\dagger$ values.

\section{Conclusion}
\label{sec:conc}
The $\Lambda_{\rm s}$CDM model is based on the recent conjecture that the Universe went through a spontaneous AdS to dS transition characterized by a sign-switching cosmological constant ($\Lambda_{\rm s}$) at ${z\sim2}$~\cite{Akarsu:2019hmw,Akarsu:2021fol}. This conjecture was inspired by the promising observational findings on the gDE model that showed the gDE, which smoothly transitions from negative to positive energy densities, can simultaneously ameliorate the $H_0$ and Ly-$\alpha$ discrepancies by preferring a rapid transition at ${z\sim2}$, and it was further compelled by some theoretical advantages of $\Lambda_{\rm s}$ over the gDE~\cite{Akarsu:2019hmw}. In this paper, we consider the simplest $\Lambda_{\rm s}$CDM model, constructed simply by promoting the usual cosmological constant $\Lambda$ of the standard $\Lambda$CDM model to an abrupt sign-switching cosmological constant $\Lambda_{\rm s}$, which we treat as an idealized description of a rapid transition (may or may not be smooth) from an AdS vacuum provided by $\Lambda_{\rm s}=-\Lambda_{\rm s0}$ to a dS vacuum provided by $\Lambda_{\rm s}=\Lambda_{\rm s0}$, or DE models such as gDE, that can mimic this behavior. This model has been recently proposed in Ref.~\cite{Akarsu:2021fol} and explored theoretically and observationally. It was found that, when $\Lambda_{\rm s}$CDM is guaranteed to be consistent with the CMB data at the background level, it predicts a higher $H_0$ value compared to $\Lambda$CDM and agrees with the Ly-$\alpha$ data for $z_\dagger\lesssim2.3$. In the robust observational analyses, it was able to simultaneously ameliorate the $H_0$, $M_B$, and $S_8$ tensions along with the Ly-$\alpha$ and $\omega_{\rm b}$ anomalies. However, while the CMB data alone was consistent with any $z_\dagger$ value for $z_\dagger\gtrsim1.5$; when a compilation of BAO data was combined with the CMB data, the constraint on $z_\dagger$ turned out to be $z_\dagger\sim2.4$, compromising the success of the model in ameliorating the tensions. This compromise was attributed to the opposition of galaxy BAO to lesser $z_\dagger$ values thereby preventing the model from achieving $z_\dagger\sim2$ required for complete removal of the tensions under consideration, or equivalently, it was attributed to the discordance of low-redshift and high-redshift BAO within $\Lambda_{\rm s}$CDM (that is also present within $\Lambda$CDM).

In this paper, we have constrained the parameters of $\Lambda_{\rm s}$CDM and $\Lambda$CDM models with various combinations of updated and extended data compared to Ref.~\cite{Akarsu:2021fol}, with particular focus on the Pantheon SNIa data set with and without the SH0ES $M_B$ prior. The extended analyses in the present paper let us asses how $\Lambda_{\rm s}$CDM performs in the light of this extended set of cosmological observations, and further investigate the constraints on $z_\dagger$ without the full BAO data set, which exhibits internal conflicts within $\Lambda_{\rm s}$CDM, similar to those in the case of  $\Lambda$CDM. The results confirm the pushback from the galaxy BAO, and show that the $M_B$ prior strongly favors $\Lambda_{\rm s}$CDM as expected since the model predicts higher $H_0$ values and respects the internal consistency of the SH0ES $H_0$ measurement utilizing $M_B$. When the $M_B$ prior is present without the full BAO data set, $\Lambda_{\rm s}$CDM is very strongly favored over $\Lambda$CDM in Bayesian evidence with exceptional $\Delta \ln \mathcal{Z}$ values of $12.32$ and $7.77$ with and without the Ly-$\alpha$ data respectively. The inclusion of the \textit{completed} full BAO data set in the analysis hinders the promising features of $\Lambda_{\rm s}$CDM by pushing $z_\dagger$ to higher values, yet, $\Lambda_{\rm s}$CDM is still strongly favored over $\Lambda$CDM in Bayesian evidence; namely, we have $\Delta \ln \mathcal{Z}=3.23$ (CMB+Pan+BAO+$M_{B}$) in this case. It is important to observe the trend in the case of the $\Lambda_{\rm s}$CDM model that, inclusion of the $M_B$ prior without the full BAO data set simultaneously removes all the prominent discrepancies that prevail within the standard cosmological model (viz., the $H_0$, $M_B$, and $S_8$ tensions), as well as the $t_0$ anomaly, with strict constraints on $z_\dagger$, while its inclusion causes only minor improvements in the case of the $\Lambda$CDM model.

Generically, $\Lambda_{\rm s}$CDM performs better for all six discrepancies of $\Lambda$CDM considered in this paper (viz., $H_0$, $M_B$, $S_8$, Ly-$\alpha$, $t_0$, and $\omega_{\rm b}$ discrepancies) for all six data compilations; particularly, in the case of $z_\dagger\sim1.8$, $\Lambda_{\rm s}$CDM is remarkably successful in simultaneous alleviation of these six discrepancies. In Ref.~\cite{Akarsu:2019hmw}, the presence of an AdS to dS transition at $z_\dagger\sim2$ was argued mainly based on the Ly-$\alpha$ data preferring negative DE densities at their effective redshifts greater than 2. In Ref.~\cite{Akarsu:2021fol}, for CMB+BAO data, it was indeed the Ly-$\alpha$ data that insisted on $z_\dagger\lesssim2.3$ despite the opposition of the galaxy BAO to lower $z_\dagger$ values. Pleasantly, the results in this paper show that the presence of the $M_B$ prior finds excellent constraints of $z_\dagger\sim2$ ($z_\dagger\sim1.8$ when the full BAO data is not included) even when the Ly-$\alpha$ data is not included, and the consequent predictions of the $\Lambda_{\rm s}$CDM model efficiently address the tensions of $\Lambda$CDM that are considered in this work.

The inclusion of the full BAO data in the data set hinders the success of $\Lambda_{\rm s}$CDM due to the galaxy BAO, as was the case in Ref.~\cite{Akarsu:2021fol}. However, upon careful inspection, \cref{tab:BAO_measurements} hints that
$\Lambda_{\rm s}$CDM is only discrepant with the BAO data at $z_{\rm eff}<0.5$ when the $D_H(z)/r_{\rm d}$ measurements are considered, but these discrepancies carry over to $D_M(z)/r_{\rm d}$ measurements from BAO at higher effective redshifts since deviations in $H(z)$ at small redshifts also cause deviations in $D_M(z)/r_{\rm d}$ at all redshifts. If the $\Lambda_{\rm s}$CDM model is to describe the present full BAO data, a wiggly modification to its Hubble function at low redshifts as suggested in Ref.~\cite{Akarsu:2022lhx} (see also references therein)
is in order. From an alternative point of view, the present full BAO data seem to have internal conflicts between low- and high-redshift BAO data within both the $\Lambda$CDM and $\Lambda_{\rm s}$CDM models; if these are due to systematics in the galaxy BAO measurements that are to be resolved in the future bringing BAO data to concordance within $\Lambda_{\rm s}$CDM, this could allow $\Lambda_{\rm s}$CDM to have excellent fit to all of the data considered in our analyses without suffering from the serious to mild tensions within $\Lambda$CDM ($H_0$, $M_B$, $S_8$, Ly-$\alpha$, and $t_0$), in contrast, if these are due to systematics in the high-redshift BAO measurements that are to be resolved in the future bringing BAO data to concordance within $\Lambda$CDM, $\Lambda$CDM would still be discrepant with multitude of cosmological observations.

Further analyses of $\Lambda_{\rm s}$CDM can be carried out by including additional data related to structure formation such as weak lensing and redshift-space distortion from Kilo-Degree Survey (KiDS)~\cite{Heymans:2020gsg} and Dark Energy Survey (DES)~\cite{DES:2021vln}, to robustly determine the model's consistency with regards to amplitude and growth of structures, and/or the most recent CMB data from the Atacama Cosmology Telescope (ACTPol)~\cite{ACT:2020gnv} and the South Pole Telescope (SPT-3G)~\cite{SPT-3G:2021wgf} along with the Planck data. In addition, the recent Pantheon+~\cite{Scolnic:2021amr} sample includes SNIa from the Cepheid-host galaxies whose distances are calibrated by SH0ES; thus, $\Lambda_{\rm s}$CDM can be analyzed using Pantheon+ in combination with the SH0ES distance measurements instead of using the Pantheon sample along with the SH0ES $M_B$ prior. Our analyses with the $M_B$ prior suggest that, in this case, $\Lambda_{\rm s}$CDM would perform better compared to $\Lambda$CDM; and thanks to the model's submission to the internal consistency of the SH0ES $H_0$ measurement, this better performance would also manifest itself in a high $H_0$ prediction in agreement with the SH0ES value. It is worth noting that a recent study reinforces this expectation by suggesting that Pantheon+ data set itself shows the presence of negative DE density at high redshifts~\cite{Malekjani:2023dky}.

Other future works may investigate extensions of $\Lambda_{\rm s}$CDM itself. A straightforward extension of the model can be achieved by allowing nonzero spatial curvature. This scenario is particularly interesting due to the preference of positive spatial curvature on top of $\Lambda$CDM by the CMB data; since positive spatial curvature mimics cosmic strings with negative energy density in the Friedmann equation, whether this preference of a closed space by the CMB data (i.e., the $\Omega_{k}$ anomaly that is closely related to the $A_{\rm L}$ anomaly due to the degeneracy between the two) still exists within $\Lambda_{\rm s}$CDM, which already incorporates a negative DE density at large redshifts, is worthy of investigation~\cite{LSCDMcurvaturelens}. Alternatively, considering that $\Lambda_{\rm s}$CDM's struggle with galaxy BAO data appears to be the main factor preventing it from simultaneously fitting excellently to the variety of the high precision data considered in the present work, one may extend the model by introducing wiggles (see Ref.~\cite{Akarsu:2022lhx} and references therein) to its Hubble function at low redshifts (that can accommodate the full BAO data) without excessive number of free parameters.

The apparently spontaneous nature of the $\Lambda_{\rm s}$, or a DE density mimicking it, and also the fact that it shifts to a larger value, in particular from negative to positive, may render finding a concrete physical mechanism underlying this scenario challenging~\cite{Akarsu:2019hmw,Akarsu:2021fol}. However, the phenomenological success of the $\Lambda_{\rm s}$CDM model despite its simplicity (particularly, when the abrupt sign-switching $\Lambda_{\rm s}$ is considered), is highly encouraging to look for possible physical mechanisms underlying this scenario as well as their specific imprints in the sky. We treat the abrupt sign-switching transition of $\Lambda_{\rm s}$ defined in \cref{{eq:ssdef}} as an idealized description of a rapid transition (may or may not be smooth) from an AdS vacuum provided by $\Lambda_{\rm s}=-\Lambda_{\rm s0}$ to a dS vacuum provided by $\Lambda_{\rm s}=\Lambda_{\rm s0}$ at/around a certain redshift, $z_\dagger$, in the late Universe---or DE models such as gDE, that can mimic/approximate this behavior. However, this begs the question of why this transition occurs at/around a certain time instance $t=t_\dagger$, corresponding to $z=z_\dagger$, in the history of the Universe. The way this question is answered may have far-reaching theoretical and even observational implications. For instance, if we take $\Lambda_{\rm s}$ as an approximation to a smoothly evolving dynamical DE, whose density rapidly changes sign around $z_\dagger$, then the time instance DE density passes from negative to positive, $t_\dagger$, is not different from any other time in the time evolution of the DE (determined by the continuity equation according to the EoS parameter that characterizes it), and the concerns regarding spontaneity are mitigated; in this case, the sign change in the DE density occurs simultaneously across the entire Universe. On the other hand, if we take $\Lambda_{\rm s}$ as a transition phenomenon (such as phase transition, spontaneous symmetry breaking, spontaneous emission, phenomena related to catastrophe theory), subtler points arise. First of all, it becomes crucial to address what critical event/condition (could be external) triggers the sign switch. While the answer would be mechanism-dependent, it is conceivable that the sign-switch occurs when a critical local energy level is reached. Assuming such a critical energy level, in a universe with perfect spatial uniformity (i.e., in a universe perfectly described by the RW spacetime metric), every point in space would reach the critical energy level at the same cosmic time instance leading to a simultaneous sign-switch at every point in space. But in reality, the Universe is not exactly uniform (spatially), but almost-exactly uniform (cf. the CMB temperature anisotropies are $\Delta T/T\sim10^{-5}$ level, over a wide range of angular scales), therefore the sign switch must have occurred at/around slightly different comoving time instances in the slightly overdense and underdense regions (on cosmological scales), and also, in some overdense regions (viz., the regions where the structures have grown), the $\Lambda_{\rm s}$ must have never transitioned to the dS phase and remained in AdS phase. This could have observable consequences in the sky, which in turn can allow for new tests of $\Lambda_{\rm s}$CDM and its possible underlying mechanisms. For instance, the asynchronization in $t_\dagger$ and the possibility that sign switch has never occurred in some regions may lead to specific imprints in the CMB, the clustering of galaxies etc.---taking this at face value, the effects of the sign-switch on the CMB anomalies by itself is an intriguing topic. Finally, let us comment on one more interesting point; if the sign-switching transition of the cosmological constant is triggered when, e.g., the local energy level reaches a critical value, then it may be possible to relate $t_\dagger$ (or $z_\dagger$) to some other cosmological parameters, which in turn leads to the possibility of reducing the free parameters of $\Lambda_{\rm s}$CDM to that of the base $\Lambda$CDM model, a possibility that may crown the success of the $\Lambda_{\rm s}$CDM model in light of the currently available observational data.

\begin{acknowledgments}
The authors are grateful to the referee for valuable comments and suggestions. \"{O}.A. acknowledges the support by the Turkish Academy of Sciences in the scheme of the Outstanding Young Scientist Award  (T\"{U}BA-GEB\.{I}P), and the COST Action CA21136 (CosmoVerse). \"{O}.A. is supported in part by TUBITAK Grant No. 122F124. S.K. gratefully acknowledges support from the Science and Engineering Research Board (SERB), Govt. of India (File No.~CRG/2021/004658).  E.\"{O}.~acknowledges the support by The Scientific and Technological Research Council of Turkey (T\"{U}B\.{I}TAK) in scheme of 2211/A National PhD Scholarship Program. J.A.V. acknowledges the support provided by FOSEC SEP-CONACYT Investigaci\'on B\'asica A1-S-21925, Ciencias de Frontera No. CONACYT-PRONACES/304001/202 and No. UNAM-DGAPA-PAPIIT IN117723. A.Y. is supported by Junior Research Fellowship (CSIR/UGC Ref. No. 201610145543) from University Grants Commission, Govt. of India. 
 \end{acknowledgments}
 
 \newpage
 
  \appendix
\section{Triangle Posteriors}
\label{sec:Appendix}
In this appendix, we present the
one- and two-dimensional (at 68\% and 95\% CLs) marginalized distributions of the model parameters for both models. We do not see strong correlations between $z_\dagger$ and the six baseline parameters, but these exist among $z_\dagger,\,H_0,\,M_B,\,S_8$, and $\Omega_{\rm m}$. Thus,  triangular plots showing the joint posteriors between the parameters present extra complementary information to the tables in the main text.

\begin{figure*}[htbp]
 \centering
\includegraphics[width=15cm]{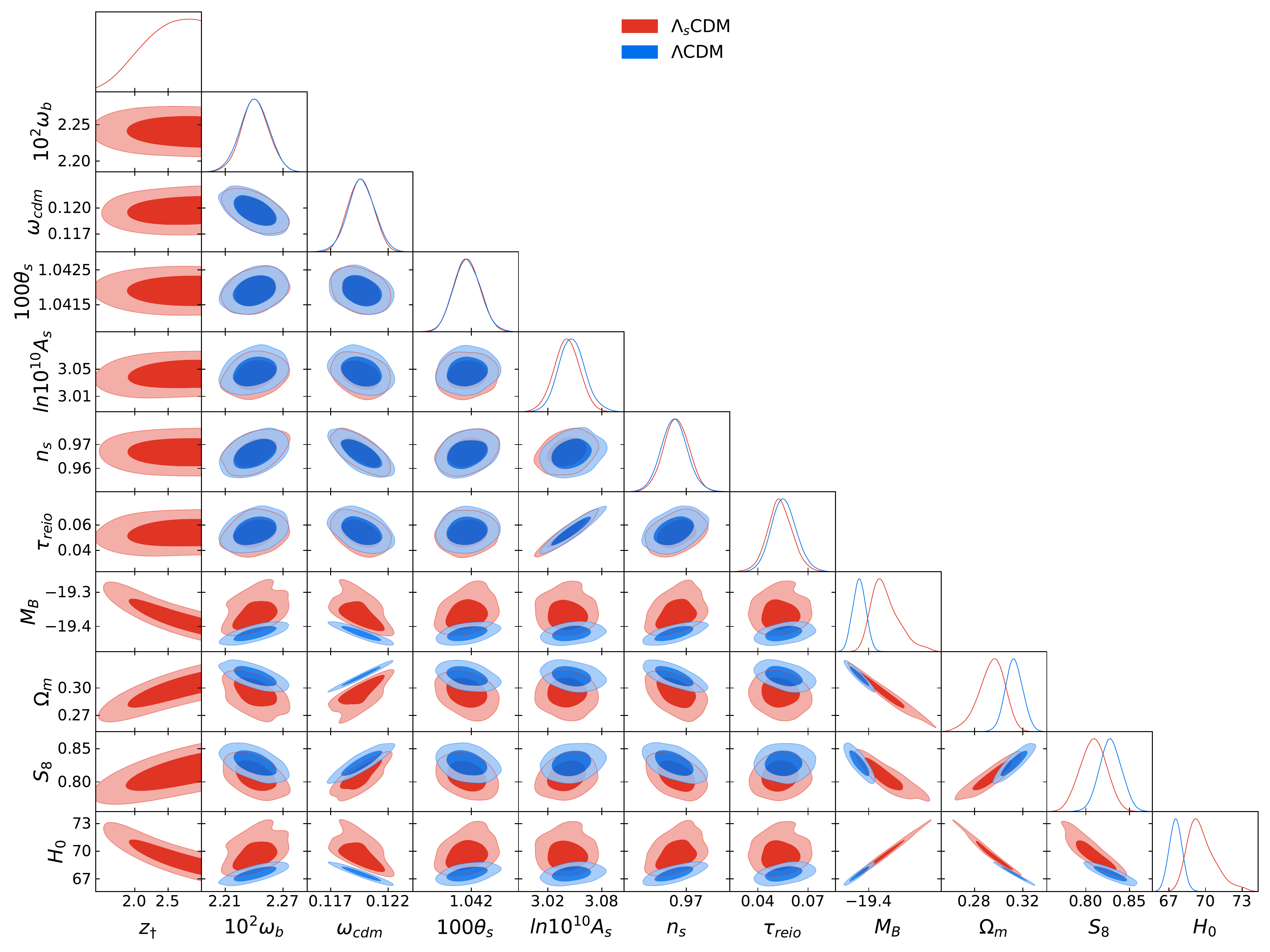}
\caption{One- and two-dimensional (68\%, 95\% CLs) marginalized distributions of the model parameters from CMB+Pan without $M_{B}$ prior.} 
\label{fig:nomb}
 \end{figure*}
 \begin{figure*}[htbp]
\centering
\includegraphics[width=15cm]{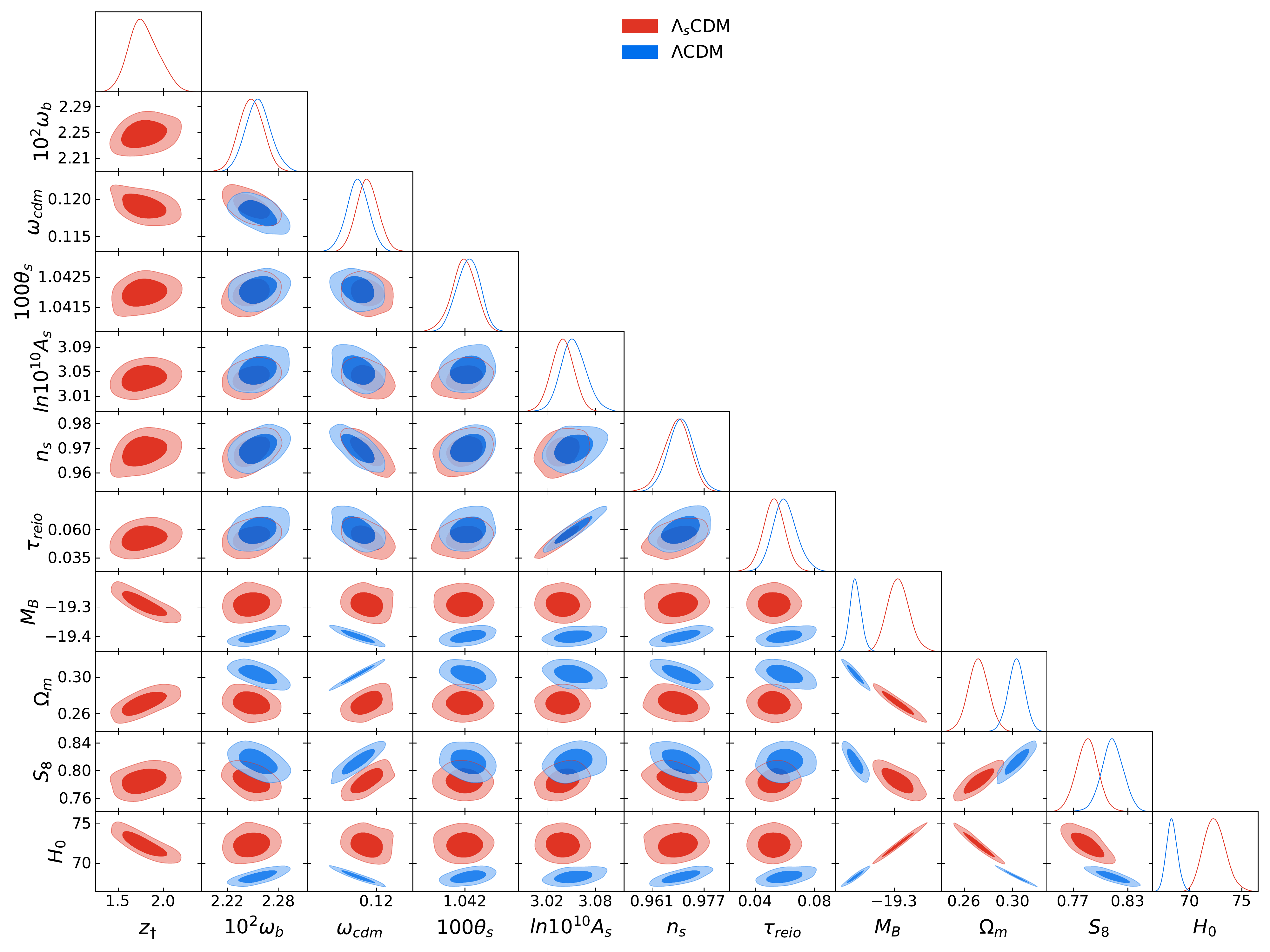}
\caption{One- and two-dimensional (68\%, 95\% CLs) marginalized distributions of the model parameters from CMB+Pan with $M_{B}$ prior.} 
\label{fig:mb}
 \end{figure*}
 
   \begin{figure*}[htbp]
 \centering
\includegraphics[width=15cm]{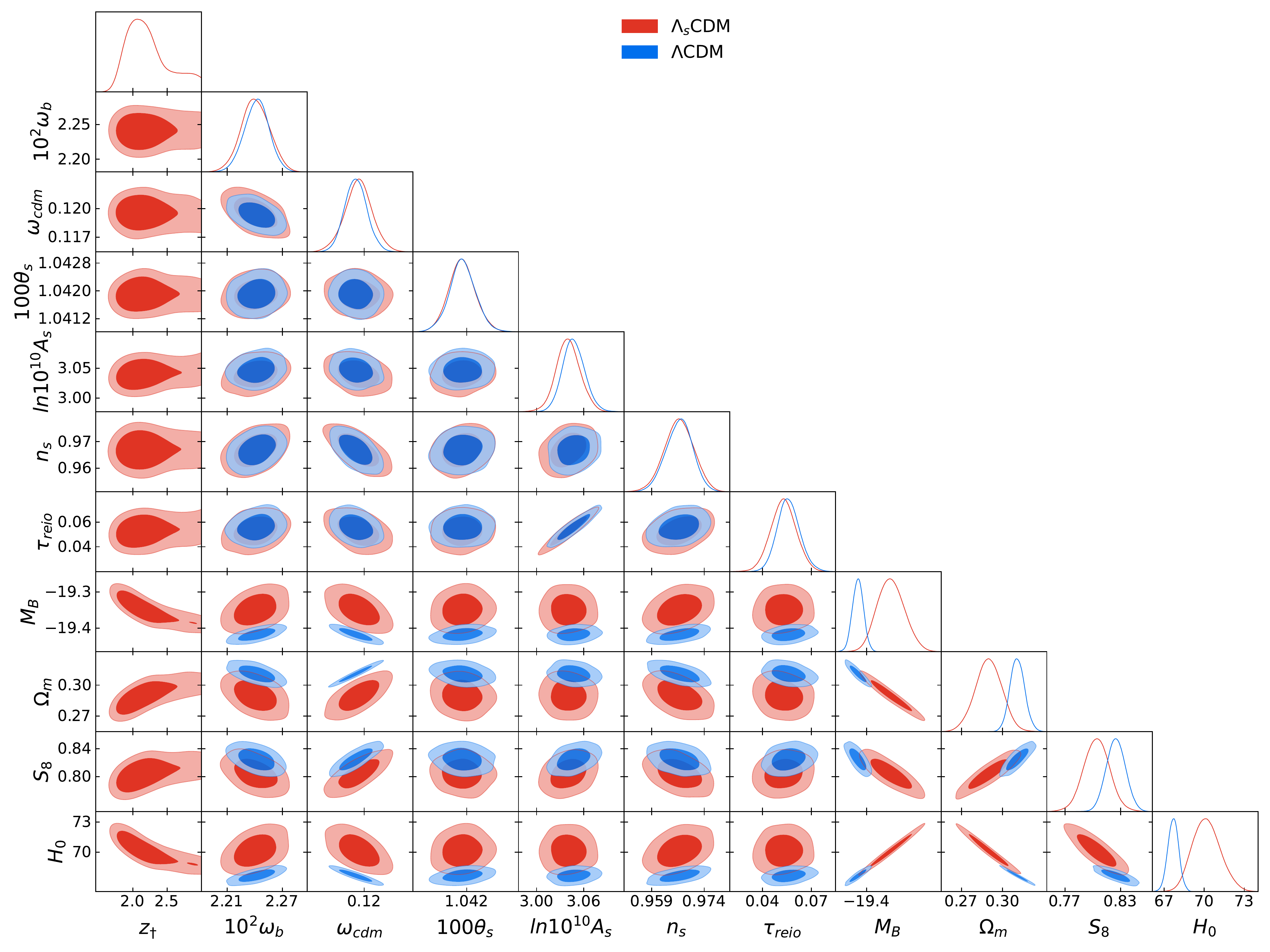}
\caption{One- and two-dimensional (68\%, 95\% CLs) marginalized distributions of the model parameters from CMB+Pan+Ly-$\alpha$ without $M_{B}$ prior.} 
\label{fig:nomblya}
 \end{figure*}
 \begin{figure*}[htbp]
 \centering
\includegraphics[width=15cm]{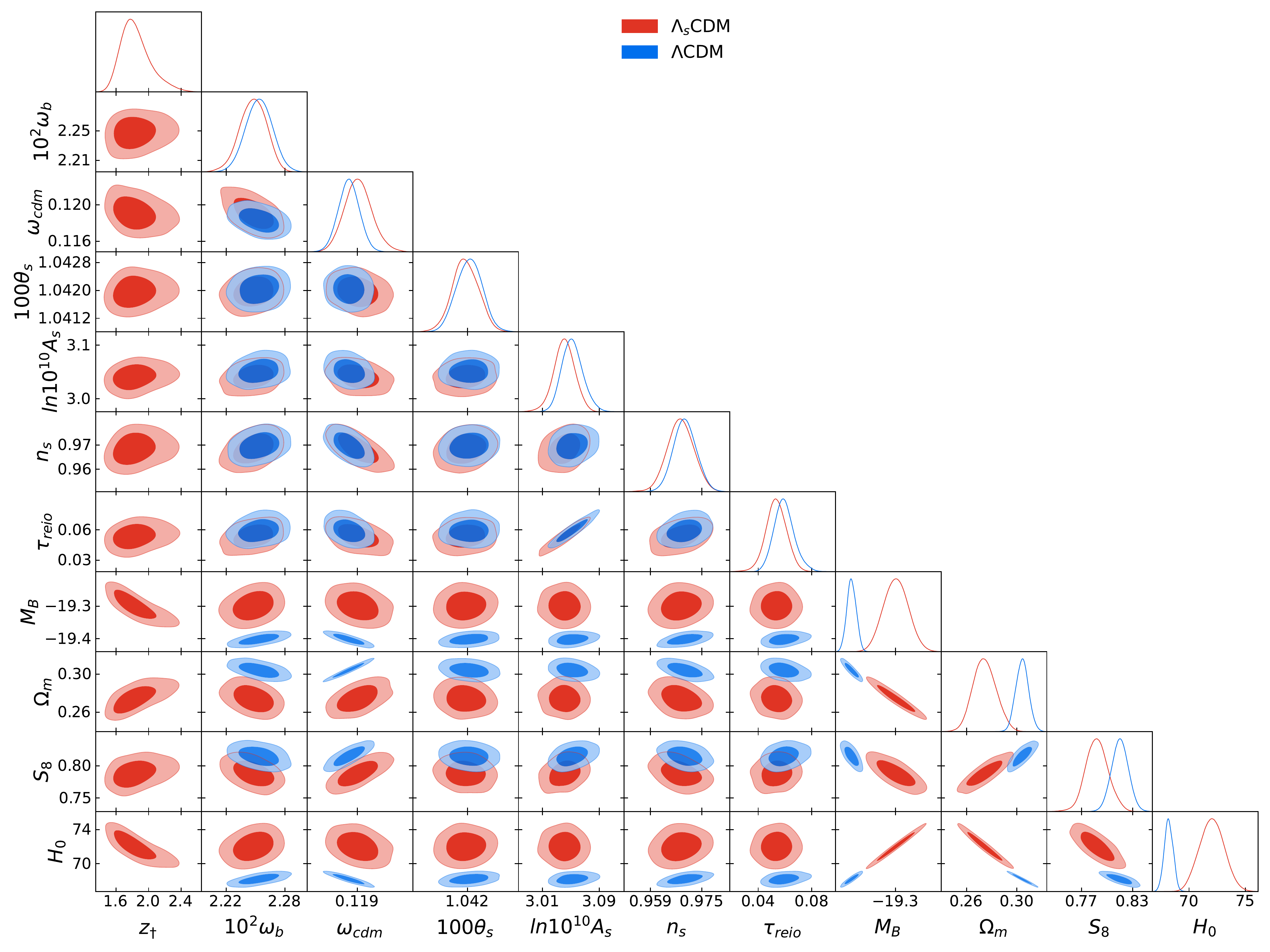}
\caption{One- and two-dimensional (68\%, 95\% CLs) marginalized distributions of the model parameters from CMB+Pan+Ly-$\alpha$ with $M_{B}$ prior.} 
\label{fig:mblya}
 \end{figure*}
 
   \begin{figure*}[htbp]
\includegraphics[width=15cm]{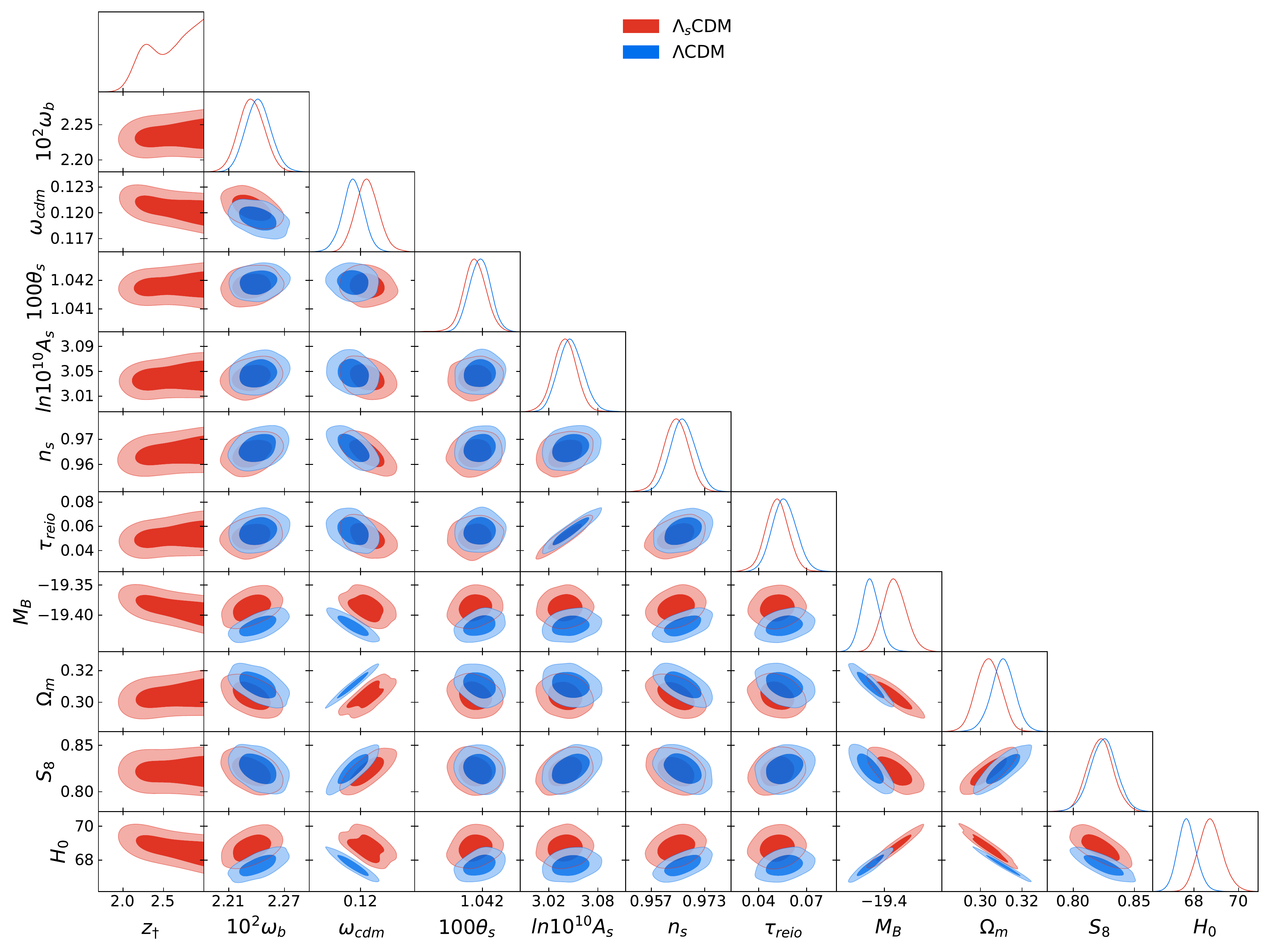}
\caption{One- and two-dimensional (68\%, 95\% CLs) marginalized distributions of the model parameters from CMB+Pan+BAO without $M_{B}$ prior.}
\label{fig:nombbao}
 \end{figure*}
 \begin{figure*}[htbp]
\centering
\includegraphics[width=15cm]{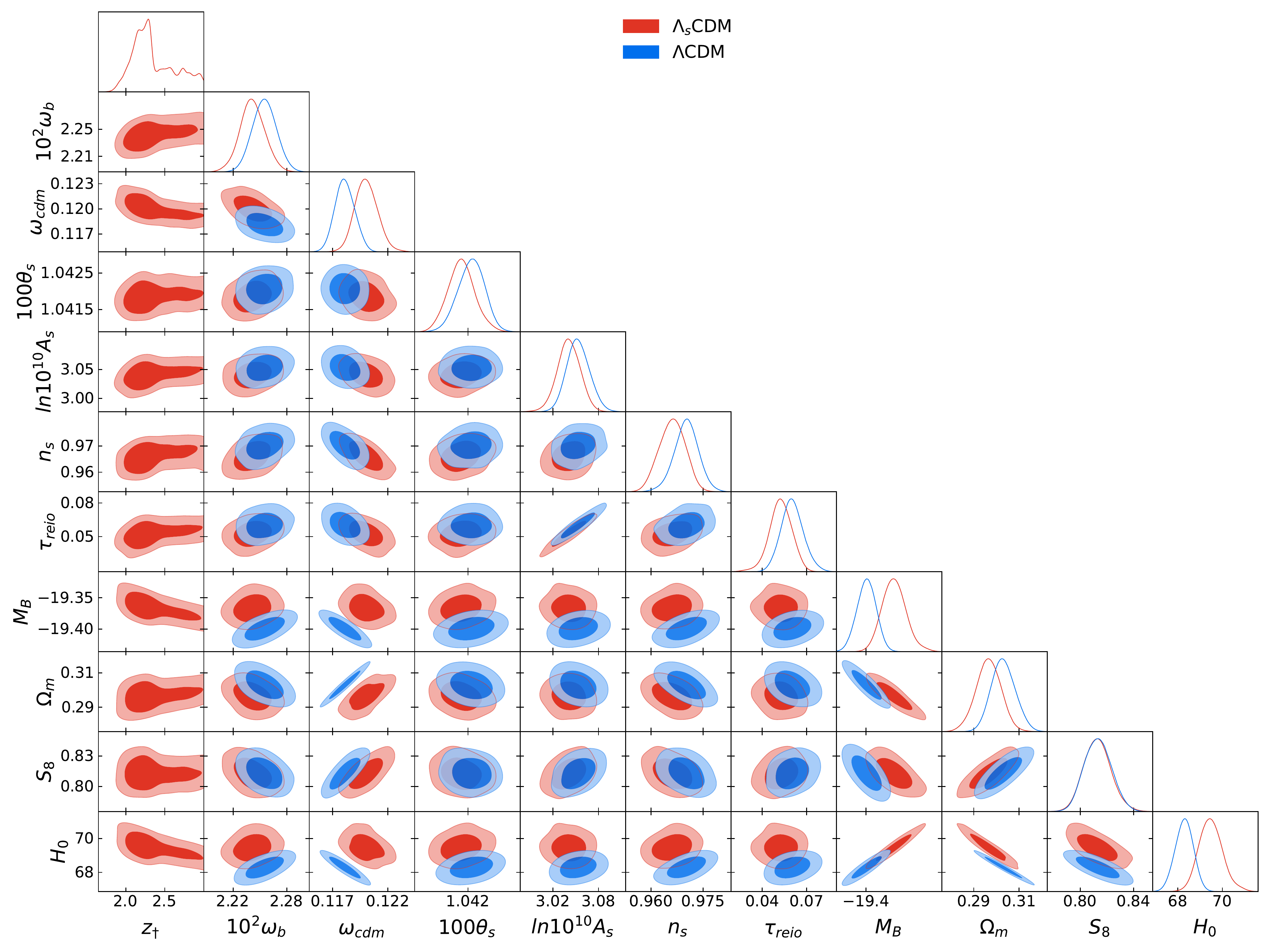}
\caption{One- and two-dimensional (68\%, 95\% CLs) marginalized distributions of the model parameters from CMB+Pan+BAO with $M_{B}$ prior.}
\label{fig:mbbao}
\end{figure*}

\end{document}